\documentclass[preprint,showpacs,showkeys,preprintnumbers,amsmath,amssymb,nofootinbib,aps]{revtex4}

\usepackage{amssymb}
\usepackage{amsmath}
\usepackage{graphicx}% Include figure files
\usepackage{dcolumn}% Align table columns on decimal point
\usepackage{bm}% bold math
\usepackage{epsfig}
\usepackage{hyperref}
\newcommand\beq{\begin{eqnarray}}
\newcommand\eeq{\end{eqnarray}} 
\newcommand\eqn[1]{\label{eq:#1}} 
\newcommand\eq[1]{Eq. (\ref{eq:#1})} 
\newcommand\figwidth{.48\textwidth}
\newcommand\Eq[1]{Eq.~\ref{eq:#1}}
\newcommand\Fig[1]{Fig.~\ref{fig:#1}}
\newcommand\Sec[1]{Sec.~\ref{sec:#1}}

\newcommand\Ap[1]{Appendix~\ref{sec:#1}}
\newcommand\Tab[1]{Table~\ref{tab:#1}}
\newcommand\bfx{\mathbf x}
\newcommand\bfn{\mathbf n}
\newcommand\bfj{\mathbf j}

\newcommand\bfp{\mathbf p}
\newcommand\bfq{\mathbf q}
\newcommand\bfC{\mathbf C}
\newcommand\bfr{\mathbf r}

\newcommand\calB{\mathcal B}
\newcommand\calK{\mathcal K}

\newcommand\calC{\mathcal C}
\newcommand\calO{\mathcal O}
\newcommand\calT{\mathcal T}
\newcommand\calV{\mathcal V}
\newcommand\calD{\mathcal D}
\newcommand\calH{\mathcal H}
\newcommand\calP{\mathcal P}

\begin{document}

\title{Lattice Monte Carlo calculations for unitary fermions in a harmonic trap}

\begin{abstract}
We present a new lattice Monte Carlo approach developed for studying large numbers of strongly interacting nonrelativistic fermions and apply it to a dilute gas of unitary fermions confined to a harmonic trap. In place of importance sampling, our approach makes use of high statistics, an improved action, and recently proposed statistical techniques.  We show how improvement of the lattice action can remove discretization and finite volume errors systematically.  For $N = 3$ unitary fermions in a box, our errors in the energy scale as the inverse lattice volume, and we reproduce a previous high precision benchmark calculation to within our $0.3\%$  uncertainty; as additional benchmarks we reproduce precision calculations of $N = 3, . . . , 6$ unitary fermions in a harmonic trap to within our $\sim 1\%$ uncertainty. We then use this action to determine the ground state energies of up to 70 unpolarized fermions trapped in a harmonic potential on a lattice as large as $64^3 \times 72$. In contrast to variational calculations we find evidence for persistent deviations from the thermodynamic limit for the range of $N$ considered.
\end{abstract}

\author{Michael G. Endres}
\email{endres@riken.jp}
\affiliation{Physics Department, Columbia University, New York, NY 10027, USA}
\affiliation{Theoretical Research Division, RIKEN Nishina Center, Wako, Saitama 351-0198, Japan}
\author{David B. Kaplan}
\email{dbkaplan@uw.edu}
\author{Jong-Wan Lee}
\email{jwlee823@u.washington.edu}
\author{Amy N. Nicholson}
\email{amynn@u.washington.edu}
\affiliation{Institute for Nuclear Theory, University of Washington, Seattle, WA 98195-1550, USA}


\preprint{CU-TP-1198}
\preprint{INT-PUB-11-024}
\preprint{RIKEN-QHP-3}

%\received{XX/XX/XXXX}


\date{\today}

\maketitle

\section{Introduction}
\label{sec:introduction}

Developing a predictive understanding of strongly interacting many-body systems is one of the most difficult and potentially rewarding challenges in physics.  A paradigm for this problem  in perhaps its purest form is to determine the behavior of a gas of unitary fermions (for a brief overview, see \cite{2007NatPh...3..469H}). These are nonrelativistic fermions with zero range interactions tuned such that the two-body s-wave scattering length diverges.  Thus the s-wave phase shift satisfies $\delta(k)=\pi/2$ for all $k$ and the field theory describing the many-body system is at a conformal fixed point\footnote{Since the underlying theory is conformal,  at nonzero chemical potential $\mu$ and $\hbar=1$, all dimensionful quantities, such as the ground-state energy and pairing gap $\Delta$, are given as pure numbers times the function of $\mu$ and the fermion mass $M$ combined to give the corresponding dimension.};   in 1998 it was suggested that unitary fermions could serve as the starting point for an effective field theory expansion for nuclear physics \cite{Kaplan:1998we, Kaplan:1998tg}.  Since then the unitary fermion gas has been created and studied experimentally by trapping atoms tuned to a Feshbach resonance by means of an applied magnetic field, exhibiting collective effects interpolating between the well understood phenomena of BCS pairing and Bose-Einstein condensation  \cite{O'Hara13122002,PhysRevLett.90.230404,PhysRevA.68.011401,Bourdel:2003zz,PhysRevLett.89.203201,PhysRevLett.92.040403,PhysRevLett.92.150402,2003Natur.424...47R,PhysRevLett.92.120403,PhysRevLett.92.120401}.  The nonperturbative nature of the strongly coupled interaction between unitary fermions poses a nontrivial challenge for theory, and numerical simulation has played an essential role in making progress.  A large body of recent theoretical work exists for unitary fermions, both analytical \cite{2007PhRvA..75d3605A,2004cond.mat.12764T,2006PhRvL..97e0403N,2007PhRvD..76h6004N,2006PhRvA..74e3604W,2006AnPhy.321:197S,Braaten:2008bi} and numerical 
\cite{2005PhRvC..72b4006L,Bulgac:2005pj,Lee:2005is,Lee:2005it,Lee:2008xsa,2008PhRvC..77c2801G,2008PhRvA..78b3625B,2010arXiv1011.2197M,2007JPhA...4012863P,2004PhRvL..93t0404A,2011PhRvA..83d1601G,PhysRevLett.91.050401,2010PhRvA..82e3621G,PhysRevLett.96.160402,2007PhRvA..76d0502B,2005PhRvA..72d1603P,2011arXiv1104.2102B,Abe:2007ff,PhysRevLett.95.060401,Gezerlis:2007fs,PhysRevLett.99.233201,Chang:2007zzd,Chang:2004zz}.

In this paper we describe a new lattice approach for simulating unitary fermions, and determine the ground state energies for up to 70 unitary fermions in a harmonic trap on lattices as large as $64^3\times 72$, allowing for an extrapolation to the infinite volume limit. This significantly extends  preliminary findings 
published in the Lattice 2010 conference proceedings \cite{Endres:2010sq,Lee:2010qp,Nicholson:2010ms}, building on the lattice construction  of \cite{Chen:2003vy}. In addition, for this work we have made several improvements, including the use of a Galilean-invariant interaction \footnote{By Galilean invariant, we mean that the interaction is only a function of the transferred  three-momentum between interacting particles, although that momentum is necessarily discrete and periodic on the finite volume lattice. Momentum dependent separable interactions, for example, would not be Galilean invariant.} for tuning to unitarity and reducing time discretization errors in the implementation of the harmonic oscillator; these are outlined in \Sec{lattice_construction}.

Our approach differs from previous numerical studies in several ways:
\begin{itemize}

\item The theory is defined on a four dimensional Euclidian  lattice,  and fermion-fermion interactions are induced by an auxiliary scalar  field $\phi$. We compute $N$-fermion correlators in the background $\phi$ field, then average observables over an ensemble of these fields -- in much the same way one computes the hadron spectrum in lattice QCD.  Unlike some approaches \cite{Chang:2004zz,Chang:2007zzd, PhysRevLett.99.233201}, our computation is not variational in nature, and so our result for the ground state energy does not depend on an accurate parametrization of the many-body ground state wavefunction. In practice, however, using good sources and sinks for the correlators is necessary to achieve this goal, blurring the  boundary between unconstrained and variational calculations when $N$ is large.

\item We formulate the lattice action in such a way  that the fermion determinant is independent of the auxiliary field $\phi$ so that the so-called ``quenched approximation" is exact, greatly simplifying the computation.  This requires open boundary conditions in the temporal direction, and we can therefore only study properties at zero temperature.

\item We do not use importance sampling (that is, we do not include the correlator we are trying to compute as part of the measure for $\phi$).  
Instead, our $\phi$ ensemble consists of random $Z_2$ valued variables living on the time-like lattice links, and therefore is extremely cheap to generate (see \cite{Endres:2010sq} for a detailed discussion of the scaling of our algorithm with volume and number of fermions). The price we pay is that we face a serious distribution overlap problem that cannot be overcome simply by increasing statistics \footnote{%
Due to an unfortunate choice of nomenclature, the ``overlap problem'' commonly refers to one of two unrelated problems, both of which concern us here.
The first is the poor overlap between the true ground state and the choice of interpolating operators, whereas the latter is the poor overlap between the path-integral probability measure and the dominant part of the operator being estimated.
We will refer to the former as a ``interpolating operator overlap problem'' and the second as a ``distribution overlap problem.''
}.

\item The distribution overlap problem is identified as arising from heavy-tailed distributions for our correlators, similar to what is seen for conductance electrons in disordered media near the Anderson localization transition.  We have developed a statistical method for greatly ameliorating the problem, as discussed  in a separate paper, Ref.~\cite{Endres:2011jm}.

\item We use a greatly improved lattice action that exactly reproduces single particle dispersion relations up to a momentum cutoff related to the inverse lattice spacing as well as the first several two-particle energy levels in a box with zero lattice spacing.  
We show that the  volume dependence we find for the energies of two-body states are consistent with  fermions having the 
first  four or five terms in the effective range expansion tuned to zero.
Thus our fermions are much closer to the unitarity limit than have ever been studied before for $N>3$ particles, and as a result we have small discretization errors and do not have to extrapolate our results to zero range, as do most simulations.

\end{itemize}
\noindent
We have formulated this theory both for unitary fermions in a box (``untrapped") or in a harmonic potential (``trapped").  In this paper we will present only the results for trapped fermions, leaving the untrapped results for future publication \cite{Endres:2011tba}, although we use   results for two and three untrapped fermions to help establish the validity of our method.

The organization of this paper is as follows.
In \Sec{lattice_construction} we describe the theoretical details of our lattice construction, including notational conventions, lattice parameter tuning methods, and an analysis of discretization errors.
In \Sec{analysis_and_results} we present ensemble details and measurement results for the ground state energies of up to 70 unpolarized unitary fermions confined to a harmonic trap.
We conclude in \Sec{conclusion} with a summary of results and a discussion of possible future applications of our lattice construction.
% A
More technical details are provided in appendices:  \Ap{tuning} gives details about tuning the lattice interaction;
% B
 \Ap{lattice_construction_observables} describes how we construct our multi-fermion correlators which incorporate pairing correlations; 
 % C
 \Ap{measurement_strategy} explains our strategy for extracting accurate estimates of the multi-fermion energies using cumulant expansion techniques of Ref.~\cite{Endres:2011jm}; 
 % D
  \Ap{simulation_details}  provides details of our simulation, including various numerical checks performed in order to verify the correctness of our code.

\section{Lattice Construction}
\label{sec:lattice_construction}

\subsection{Action, notation and conventions}
\label{sec:lattice_construction.action_notation_and_conventions}

The starting point for our construction is a highly improved variant of the nonrelativistic Euclidean-time lattice action proposed in \cite{Chen:2003vy}:
\begin{eqnarray}
S = b_\tau b_s^3 \sum_{\tau,\bfx} \left[ \bar\psi_{\bfx,\tau}  (\partial_\tau \psi)_{\bfx,\tau}  - \frac{1}{2M} \bar\psi_{\bfx,\tau} (\nabla^2 \psi)_{\bfx,\tau} + (\sqrt{C} \phi)_{\bfx,\tau} \bar\psi_{\bfx,\tau} \psi_{\bfx,\tau-1} \right]\ .
\label{eq:action}
\end{eqnarray}
This action describes two species of one-component interacting fermions $\psi = (\psi^\uparrow, \psi^\downarrow)$ with equal mass $M$ defined on a $T\times L^3$ lattice, with the temporal and spatial lattice spacings given by $b_\tau$ and $b_s$, respectively.
For convenience, we work primarily in lattice units, where $b_s=b_\tau=1$, however in some sections we restore the lattice spacings in order to discuss temporal and spatial discretization errors.
Throughout this work, we consider a lattice with open boundary conditions in the time direction with time labeled by integers $\tau\in[0,T-1]$, and periodic boundary conditions in the spatial directions with position labeled by integers $x_j\in[-L/2,L/2-1]$, for $j=1,2,3$.
As a result of using open temporal boundary conditions, the utility of our lattice action is limited to studies at zero temperature.
In addition, this choice of boundary conditions forbids the introduction of a chemical potential and  we work in the canonical, rather than grand-canonical ensemble.

The derivative operator $\partial_\tau$ appearing in \Eq{action} represents a backward difference operator in time, i.e., $(\partial_\tau\psi)_{\bfx,\tau} = \psi_{\bfx,\tau}-\psi_{\bfx,\tau-1}$, whereas $\nabla^2$ represents a lattice gradient operator defined so as to give a perfect continuum-like single particle dispersion relation for free fermions.
This kinetic term is highly nonlocal, although as will be described below, the nonlocality poses no challenge in a numerical simulation of \Eq{action}.

A four-fermion contact interaction is achieved via the introduction of a stochastic  auxiliary scalar field $\phi_{\bfx,\tau}$ associated with the time-like links of the lattice.
This field is chosen to satisfy the conditions
\beq
\langle \phi_{\bfx,\tau} \rangle = 0\ ,\qquad \langle \phi_{\bfx,\tau} \phi_{\bfx',\tau'}\rangle = \delta_{\bfx,\bfx'} \delta_{\tau,\tau'}
\eeq
where the expectation value represents ensemble averaging over $\phi$, and in this work the $\phi$ distribution is taken to either be unit-variance Gaussian or $Z_2$.
The point-split character of the interaction ensures that scattering propagates fermions forward in time by one unit.
This choice, along with the absence of fermion propagation in the negative temporal direction and open boundary conditions in time,  ensures that no closed fermion loop depends on $\phi$.  A consequence is that the fermion determinant is $\phi$-independent  and has no effect on the measure for $\phi$, greatly simplifying numerical simulation of \Eq{action}. 

The operator $C_{\bfx\bfx'}=C(\bfx-\bfx')$ acts only in space and is taken to be real, symmetric, local, and invariant under lattice translations; it can be thought of as a differential operator acting on $\phi$ which allows the interaction between fermions induced by $\phi$ exchange to depend on the transfer momentum.  Not only does this give us a momentum-dependent interaction we can tune to attain unitarity, but it is also  Galilean invariant in that  it depends   only on the difference between the ingoing and outgoing fermion momenta.  This is important, since tuning a non-Galilean invariant interaction to give unitarity in one frame would lead to non-unitary fermions in another, and boosted pairs of particles would see an interaction which did not correspond to unitarity.
Integrating out the auxiliary field $\phi$ 
yields the four-fermion interaction
\begin{eqnarray}
(\sqrt{C} \phi)_{\bfx,\tau} (\bar\psi\psi)_{\bfx,\tau} \rightarrow  (  \bar\psi\psi)_{\bfx,\tau} ( C \bar\psi\psi)_{\bfx,\tau} \ ,
\label{eq:int_out}
\end{eqnarray}
where $(\bar\psi\psi)_{\bfx,\tau} = \bar\psi_{\bfx,\tau} \psi_{\bfx,\tau-1}$, and we have used the Hermiticity of $C$.
We may express \Eq{action} succinctly as $S=\bar\psi K \psi$, where the time components of the fermion matrix $K$ are given in block-matrix form by:
\begin{eqnarray}
K = \left( \begin{array}{cccccc}
D      & -X(T-1)   & 0         & 0         & \ldots & 0      \\
0      & D         & -X(T-2)   & 0         & \ldots & 0      \\
0      & 0         & D         & -X(T-3)   & \ldots & 0      \\
0      & 0         & 0         & D         & \ldots & 0      \\
\vdots & \vdots    & \vdots    & \vdots    & \ddots & -X(0) \\
0      & 0         & 0         & 0         & \ldots & D      \\
\end{array} \right)\ ,
\end{eqnarray}
with
\begin{eqnarray}
D = 1 - \frac{\nabla^2}{2M} \ ,\qquad X(\tau) = 1 - \sqrt{C} \Phi(\tau) \ .
\label{eq:d_and_x_operators}
\label{eq:Kblocks}
\end{eqnarray}
Note that the $L^3 \times L^3$ matrices $D$, $X$, $C$ and $\Phi(\tau)$ act only in space and that $\Phi(\tau)$ is a diagonal matrix with statistically independent random elements $\phi_\bfx(\tau)$.

We choose to realize the lattice Laplacian in such a way that $D$ has the following form in momentum space (\cite{2008PhRvA..78b3625B}):
\begin{eqnarray}
 D_{ \bfp \bfp'}  =  \delta_{\bfp,\bfp'} \times  \left\{ \begin{array}{ll}
e^{\bfp^2/(2M)} & |\bfp| < \Lambda \\
\infty  & |\bfp| \ge \Lambda
\end{array} \right. \ ,
\label{eq:d_operator}
\end{eqnarray}
where $p_j=2\pi m_j/L$ for integers $m_j\in[-L/2,L/2-1]$ and $j=1,2,3$.
The parameter $\Lambda=\pi\times (1-10^{-5})$ is a hard momentum cutoff imposed on the fermions; a small shift away from $\pi$ has been introduced in the cutoff in order to avoid inclusion of momenta lying on the very edge of the Brillouin zone (BZ).  For free fermions, $X=1$ and the propagator is just a transfer matrix, which in momentum space has the form
\beq
\left[K^{-1}_\text{free}(0,\tau)\right]_{\bfp\bfp'} =\left[ D^{-\tau} \right]_{\bfp\bfp'} \equiv \delta_{\bfp,\bfp'}e^{-E(\bfp) \tau} \theta(\Lambda-|\bfp|)
\eeq
and yields the exact one-particle energy, $E(\bfp)= \bfp^2/2M$.  So we see that the choice \eq{d_operator} is designed to give the exact one-particle dispersion relation up to a momentum cutoff $|\bfp|=\Lambda$, beyond which the fermions do not propagate. Imposing the $\Lambda$ cutoff just within the Brillouin zone boundary was necessary to reconcile the exact continuum dispersion relation with the periodicity of the reciprocal lattice.

For the interaction we take in momentum space
\begin{eqnarray}
 C_{ \bfp \bfp'}  =  \delta_{\bfp,\bfp'} \times  \left\{ \begin{array}{ll}
C(\bfp) & |\bfp| < \Lambda \\
C(\Lambda)  & |\bfp| \ge \Lambda 
\end{array} \right. \ ,
\label{eq:c_operator}
\end{eqnarray}
where below $\Lambda$, $C(\bfp)$ is an analytic function of $\bfp^2$ which we adjust   to construct the desired continuum phase shift for two-particle scattering (for example, the constant $\delta=\pi/2$ phase shift for unitary fermions).
How we tune $C$ is discussed in \Sec{lattice_construction.transfer_matrix_formalism} and \Sec{lattice_construction.parameter_tuning}.

In order to simulate the partition function defined by \Eq{action}, it is necessary to first integrate out the fermionic degrees of freedom, yielding an effective action involving only the auxiliary field.
The resulting partition function is given by
\begin{eqnarray}
Z = \int [d\phi] \rho(\phi) \det{K}\ ,
\end{eqnarray}
where 
\beq
\rho(\phi) =\begin{cases}  \prod_\bfx e^{-\frac{1}{2}\phi_\bfx^2}\ , & \text{Gaussian} \cr 
\prod_\bfx ( \delta_{\phi_\bfx,1} + \delta_{\phi_\bfx,-1} )\ ,&Z_2 \end{cases} 
\eeq
The corresponding expectation value of an arbitrary operator $\calO(\psi,\bar\psi)$ is given by:
\begin{eqnarray}
\langle \calO(\psi,\bar\psi) \rangle = \frac{1}{Z} \int [d\phi] \rho(\phi) \det{K} \, \tilde\calO(K^{-1}) \ ,
\end{eqnarray}
where $\tilde\calO(K^{-1})$ is some new calculable operator that depends implicitly on $\phi$ through the propagator $K^{-1}$.
Both $\calO$ and $\tilde\calO$ may have explicit dependence on $\phi$ as well.
Since $K$ is an upper triangular block matrix, its determinant is given by the product of determinants of its diagonal blocks,   $\det K =( \det D)^T$, which is independent of the auxiliary field. Therefore the full numerical simulation of the partition function with action given in \Eq{action} is equivalent to a quenched simulation, with expectation values given by:
\begin{eqnarray}
\langle \calO(\psi,\bar\psi) \rangle = \frac{1}{Z_\text{quenched}}\int [d\phi] \rho(\phi) \tilde\calO(K^{-1}) \ ,
\end{eqnarray}
where $Z_\text{quenched} = \int [d\phi] \rho(\phi)$ is the quenched partition function.
Note that the absence of a nontrivial probability measure for the auxiliary field ensures that the path integral is free of the sign problem.

Because $K$ is upper triangular in form, interacting fermion propagators measured from time slice zero to time slice $\tau$ may be expressed exactly as a sequence of applications of $D^{-1}$ and $X$ operators, resulting in a simple recursive formula:
\begin{eqnarray}
K^{-1}(\tau;0) &=& D^{-1} X(\tau-1) K^{-1}(\tau-1;0)\ ,
\label{eq:prop}
\end{eqnarray}
with $K^{-1}(0;0) = D^{-1}$.
The form of this result is evident from the fact that there are no time-like closed fermion loops, which is a consequence of using open boundary conditions and from the absence of anti-particles in the nonrelativistic  theory.
Inversion of the nonlocal $D$ operator and application of the $X(\tau)$ operator may be performed efficiently with fast Fourier transforms (FFTs); it is this feature that allows us to use the perfect dispersion relation and momentum dependent interaction defined in \Eq{d_operator} and \Eq{c_operator}.
 
\subsection{Transfer matrix formalism}
\label{sec:lattice_construction.transfer_matrix_formalism}

Multi-fermion correlation functions $\calC(\tau)$ are obtained from an ensemble average of direct products of propagators
\begin{eqnarray}
\calK^{-1}(\tau;0) = \underbrace{ K^{-1}(\tau;0) \otimes \ldots \otimes K^{-1}(\tau;0) }_N \ ,
\label{eq:correlator}
\end{eqnarray}
which are sandwiched between properly antisymmetrized $N$-fermion initial and final states (i.e., interpolating fields associated with time slices zero and $\tau$, respectively). We will refer to the initial and final states as sources and sinks, respectively. 
We may translate our lattice action in \Eq{action} into Hamiltonian language by noting that the expectation value of $\calK^{-1}(\tau;0) $ is just the Euclidean time evolution operator for a system of $N$ particles.
Since the single particle propagator $K^{-1}$ is itself a product of uncorrelated random matrices (because the auxiliary field probability measure is separable in time), the multi-fermion correlation function will factor into a matrix product of ensemble averages.
If we define the matrix:
\begin{eqnarray}
\calT = {\calD}^{-1/2} (1-\calV) {\calD}^{-1/2}\ ,
\label{eq:Nfermion_tmatrix}
\end{eqnarray}
where
\begin{eqnarray}
\calD = \underbrace{ D \otimes \ldots \otimes D }_N  \ ,
\end{eqnarray}
and
\begin{eqnarray}
(1-\calV) = \underbrace{ \langle  X(\tau) \otimes \ldots \otimes X(\tau)\rangle }_N \ ,\quad\textrm{for every $\tau$}
\label{eq:interaction}
\end{eqnarray}
are $V^N$ dimensional matrices, then the $N$-fermion correlator may be written in the highly suggestive form:
\begin{eqnarray}
\langle \calK^{-1}(\tau;0) \rangle   = \calD^{-1/2} (\calT)^\tau \calD^{-1/2}\ ,
\end{eqnarray}
and we may identify $\calT$ as a transfer matrix and $\calH=-\ln \calT$ as a Hamiltonian for the $N$-fermion system, provided $\calT$ is Hermitian and positive\footnote{This is a stronger condition than necessary; if $\calT$ is hermitian but not positive then $\calT\calT=\calT^\dagger\calT$ is Hermitian and positive, guaranteeing  that a sensible definition of the Hamiltonian will exist with a time step of $2b_\tau$.}.

A general expression for the multi-particle interaction $\calV$ may be computed analytically from \Eq{d_and_x_operators} and \Eq{interaction} by explicit integration of the auxiliary fields.
The expression is somewhat complicated for large numbers of particles and will therefore not be explicitly derived here.
Observe, however, that although the auxiliary field interaction $X(\tau)$ involves a square root of the operator $C$, the multi-particle interaction $\calV$ is in fact an analytic function of momenta.
This is 
due to the presence of momentum conserving delta functions which ensure that $\sqrt{C}$ always comes in pairs; in terms of Feynman diagrams, there are identical factors of $\sqrt{C}$ at each end of the $\phi$ propagator,   only depending on the magnitude of the momentum flowing through that propagator.
This property is generally true for any $N$-particle system since only an even number of insertions of the interaction survive integration over the auxiliary fields; it is also evident from the right-hand-side of \Eq{int_out}.

In the case of two fermions, where $N_\downarrow=N_\uparrow=1$, the transfer matrix defined by \Eq{Nfermion_tmatrix} may be evaluated in momentum space and is given by:
\begin{eqnarray}
\langle {\bfq^\downarrow} {\bfq^\uparrow} | \calT | \bfp^\downarrow \bfp^\uparrow \rangle = \frac{ \delta_{\bfq^\downarrow,\bfp^\downarrow} \delta_{\bfq^\uparrow,\bfp^\uparrow} + \frac{1}{L^3} C(\bfp^\downarrow-\bfq^\downarrow)\delta_{\bfq^\downarrow+\bfq^\uparrow,\bfp^\downarrow+\bfp^\uparrow} }{ e^{( {\bfq^\downarrow}^2+{\bfq^\uparrow}^2 + {\bfp^\downarrow}^2 + {\bfp^\uparrow}^2)/(4M)}}\ , 
\label{eq:tmatrix}
\end{eqnarray}
for momenta below the cutoff $\Lambda$.
$C(\bfp)$ is a periodic function of the operator $\bfp$ for $|\bfp|<\Lambda$ which we choose to expand in a convenient basis of local functions:
\begin{eqnarray}
 C(\bfp) = \frac{4\pi}{M} \sum_{n=0}^{N_{\calO}-1} C_{2n} \calO_{2n}(\bfp)\ ,
\eqn{cexp}
\end{eqnarray}
with unknown coefficients $C_{2n}$ to be determined from scattering data.
Our choice of basis  functions is:
\begin{eqnarray}
\calO_{2n}(\bfp) = M_0^n\times\begin{cases} \left(1-e^{-\bfp^2/M_0}\right)^n & |\bfp|\le\Lambda\ ,\cr  \left(1-e^{-\Lambda^2/M_0}\right)^n  & |\bfp|>\Lambda\end{cases}
\label{eq:ops}
\end{eqnarray}
for $\bfp$ within the first Brillouin zone, and periodic from one Brillouin zone to the next.   The basis functions behave as $O_{2n}(\bfp) \approx \bfp^{2n}$ for small $\bfp^2\ll M_0$ and tend to a constant for $\bfp^2>M_0$; this basis  was chosen to approximate continuum 2-body contact interactions with $2n$ derivatives for low transfer momentum, while not getting excessively big for momenta at the edge of the Brillouin zone.   Throughout this work we take $M_0 = M$, and both to be $O(1)$ in lattice units.

In the special case where $N_{\calO}=1$ the only operator in the sum \Eq{cexp} is $\calO_0$ which is  constant, and the two-fermion transfer matrix may be diagonalized analytically on the finite volume lattice.
All nonzero total momentum eigenstates of \Eq{tmatrix} correspond to plane waves, whereas the zero total momentum eigenstates are given by
\beq
\langle \bfp^\downarrow \bfp^\uparrow | \Psi_k \rangle \propto \frac{e^{p^2/2M}}{e^{-E_k+ p^2/M }-1} \delta_{\bfp^\downarrow+\bfp^\uparrow,0}
\end{eqnarray}
where $p = |\bfp^\downarrow|=|\bfp^\uparrow|$.
The corresponding energy eigenvalues $E_k$ are given by solutions to the integral equation
\begin{eqnarray}
\frac{M}{4\pi} \frac{1}{C_0} = \frac{1}{L^3} \sum_{\bfp<\Lambda} \frac{1}{e^{-E+ p^2/M}-1}\ ,
\eeq
which, for every value of $p^2$, admits a single bound state for any value of $C_0>0$ at finite volume. This negative energy state becomes a scattering state in the infinite volume limit for $0<C_0<C_\text{crit}$ and a bound state for $C_\text{crit} < C_0$, where $C_\text{crit}$ is an $M$-dependent critical value; tuning $C_0\to C_\text{crit}$ yields a zero energy bound state at infinite volume, corresponding to unitarity and the continuum limit of the lattice theory.

In the case where $N_{\calO}>1$,  even semi-analytic solutions for the $C_{2n}$ coefficients are not feasible, but they
 may be determined numerically by explicit diagonalization of \Eq{tmatrix}.
It is helpful to restrict the transfer matrix to the zero center-of-momentum subspace, thus reducing the dimensionality of the matrix from $L^6$ down to a more manageable size of $L^3$.
A further reduction in the dimensionality of \Eq{tmatrix} may be achieved by projecting the zero center-of-momentum part of the transfer matrix onto appropriate representations of the octahedral group $O_h$ (e.g., in the case of s-wave scattering, the trivial representation $A_1^+$).
Performing such a projection makes numerical diagonalization feasible for lattices at least as large as $L=64$, which is the maximum lattice size we consider in our numerical studies.

\subsection{Parameter tuning}
\label{sec:lattice_construction.parameter_tuning}

Unitary fermions in the continuum are a conformal system, while a lattice simulation necessarily involves finite lattice spacing and volume, both breaking conformal symmetry.  Critical to a numerical simulation is the ability to tune the interactions to unitarity and control the systematic errors.  In contrast to  chiral symmetry in lattice QCD, for example, there is no phase transition associated with unitarity, despite the enhanced symmetry, and so there is no general feature in the $N$-body spectrum that allows one to easily evaluate how far one is from unitarity.  It is important therefore to collect as many results as possible about unitary fermions in the continuum that are known exactly or to high numerical precision in order to facilitate the tuning of the lattice action and to control systematic errors. 

What is known exactly about  unitary fermions in the continuum is (i) the spectrum of two unitary fermions in a box of size $L$ \cite{Luscher:1985dn,Luscher:1986pf,Luscher:1990ux,Beane:2003da};
%, which follows from 
%L\"uscher's  formula  \cite{Luscher:1985dn,Luscher:1986pf,Luscher:1990ux,Beane:2003da}
%and  the vanishing of $p\cot\delta$;   
(ii) the spectrum of two and three unitary fermions in a harmonic trap \cite{2006PhRvA..74e3604W}; (iii) the scaling dimension of local composite operators involving unitary fermions\footnote{The scaling of  two-body operators was determined in Ref.~\cite{Kaplan:1998tg,Kaplan:1998we} (see also \cite{Birse:2010fj}); the scaling of low dimension three-body operators was first analyzed by Griesshammer 
\cite{Griesshammer:2005ga, Griesshammer:2005sj}, and a beautiful general analysis was subsequently supplied by Nishida and Son  \cite{2007PhRvD..76h6004N}.}.
Not known exactly but determined  to high numerical accuracy are (iv) the few lowest energy levels for three unitary fermions in a box, extrapolated from a lattice Hamiltonian diagonalization very close to the continuum limit, with lattice size up to $L = 50$ \cite{2007JPhA...4012863P}; and (v)
the ground state energies for 4, 5, 6 unitary fermions in a harmonic trap, obtained by solving the Schr\"odinger equation  \cite{Blume201186}. The ground state energy for $N=4$ fermions in a box has also recently been precisely studied by several methods in Ref.~ \cite{2011arXiv1104.2102B}, 
but involves extrapolation to the continuum from very small lattices, $L\le 8$, which makes the evaluation of potential systematic errors difficult.

Our strategy for utilizing this information to tune our lattice action and estimate the size of systematic errors is to adjust our $C_{2n}$ coefficients to correctly reproduce the low-lying two-particle spectrum in a box in the continuum, subsequently showing that we can reproduce the correct volume scaling relations of measured  energies, as well as the precisely known ground state energies for 3-fermions in a box or 3-6 trapped fermions.  Here we discuss the tuning and energy levels of two and three untrapped fermions; our results for few-body trapped fermions are discussed in \Sec{analysis_and_results}

\subsubsection{Tuning and scaling of  low-lying 2-body  untrapped energy levels}
\label{sec:lattice_construction.parameter_tuning.two_body}

The two-particle  energies $E$ for $s$-wave particle pairs in a box with zero net momentum and phase shift $\delta_0$ are given by the solutions to 
\begin{eqnarray}
p \cot \delta_0 = \frac{1}{\pi L} S(\eta) \ ,\quad S(\eta) = \lim_{\Lambda\to\infty} \left[ \sum_{|\bfj|<\Lambda} \frac{1}{\bfj^2-\eta} - 4\pi\Lambda \right]\ ,
\label{eq:Sfunc}
\end{eqnarray}
where $\bfj$ is an integer  three-vector, $\eta = (pL/2\pi)^2$, and $p$ is related to the energy by   $E=p^2/M$ \cite{Luscher:1985dn,Luscher:1986pf,Luscher:1990ux,Beane:2003da}. If scattering is due to short range interactions, then $p\cot\delta_0$ is analytic in $p^2$ at sufficiently low $p$ and one has the effective range expansion,
\begin{eqnarray}
p\cot\delta_0 = -\frac{1}{a} + \frac{1}{2} r_0 p^2 + r_1 p^4\ldots\ ,
\end{eqnarray}
where $a$ is the scattering length, $r_0$ is the effective range, and $r_1$, with dimension of volume, is what we will call the shape parameter.  By means of \Eq{Sfunc}, knowledge of the energy eigenvalues for the low-lying two-particle modes in a box can be used to determine effective range expansion parameters.  Conversely,  given a target set of effective range expansion parameters, we can tune our operator coefficients $C_{2n}$ in  \eq{cexp} of our lattice theory until we attain the correct low-lying energy eigenvalues.  This general tuning procedure was introduced in \cite{Lee:2007ae}.  For unitary fermions in the continuum we set $p\cot\delta_0=0$ on the lefthand side of \eq{Sfunc} and find the solutions $\eta^*_k$ to the equation $S(\eta^*_k)=0$.  The function $S(\eta)$ is shown in Fig.~\ref{fig:zeta3d}, and the roots $\eta^*_k$ correspond to the points where the function crosses the $\eta$ axis.  The first 27 solutions are listed in \Tab{S_roots} \footnote{To compute the $\eta_k^*$ it is very helpful to recognize that the number of integer three vectors $\bfj$ with equal norm is given by the coefficient of $x^{|\bfj |^2}$ in the Taylor expansion of $\left[\theta_3(0,x)\right]^3$, where $\theta_3(u,x)$ is one of the Jacobi theta functions.}. 

On the lattice the energy eigenvalues are defined from $\lambda = e^{- b_\tau E}$, where $\lambda$ are the eigenvalues of the two-particle transfer matrix discussed above and $b_\tau$ is  the temporal lattice spacing.  Spatial discretization effects make it impossible to exactly reproduce the continuum $\eta^*_k$ on the lattice. For one thing, there are an infinite number of $\eta^*_k$ while the lattice transfer matrix has only a finite number of eigenvalues. Furthermore, since the lattice restricts how easily fermions can get close to each other --- effectively creating a repulsive interaction ---  the phase shift for lattice unitary fermions necessarily falls below $\pi/2$ for large lattice momenta, and $p\cot\delta_0$ as computed from \eq{Sfunc} gets large. So the best one can do is tune a number $N_\calO$ of the $C_{2n}$ coefficients to reproduce the  lowest $N_\calO$  solutions  $\eta^*_k$. Details of how this tuning was performed numerically are provided in \Ap{tuning}.  In  \Tab{C_vals}. we give as an example the results for tuning operators for an $L=32$ lattice with mass $M=5$.

\begin{figure}
\includegraphics[width=\figwidth]{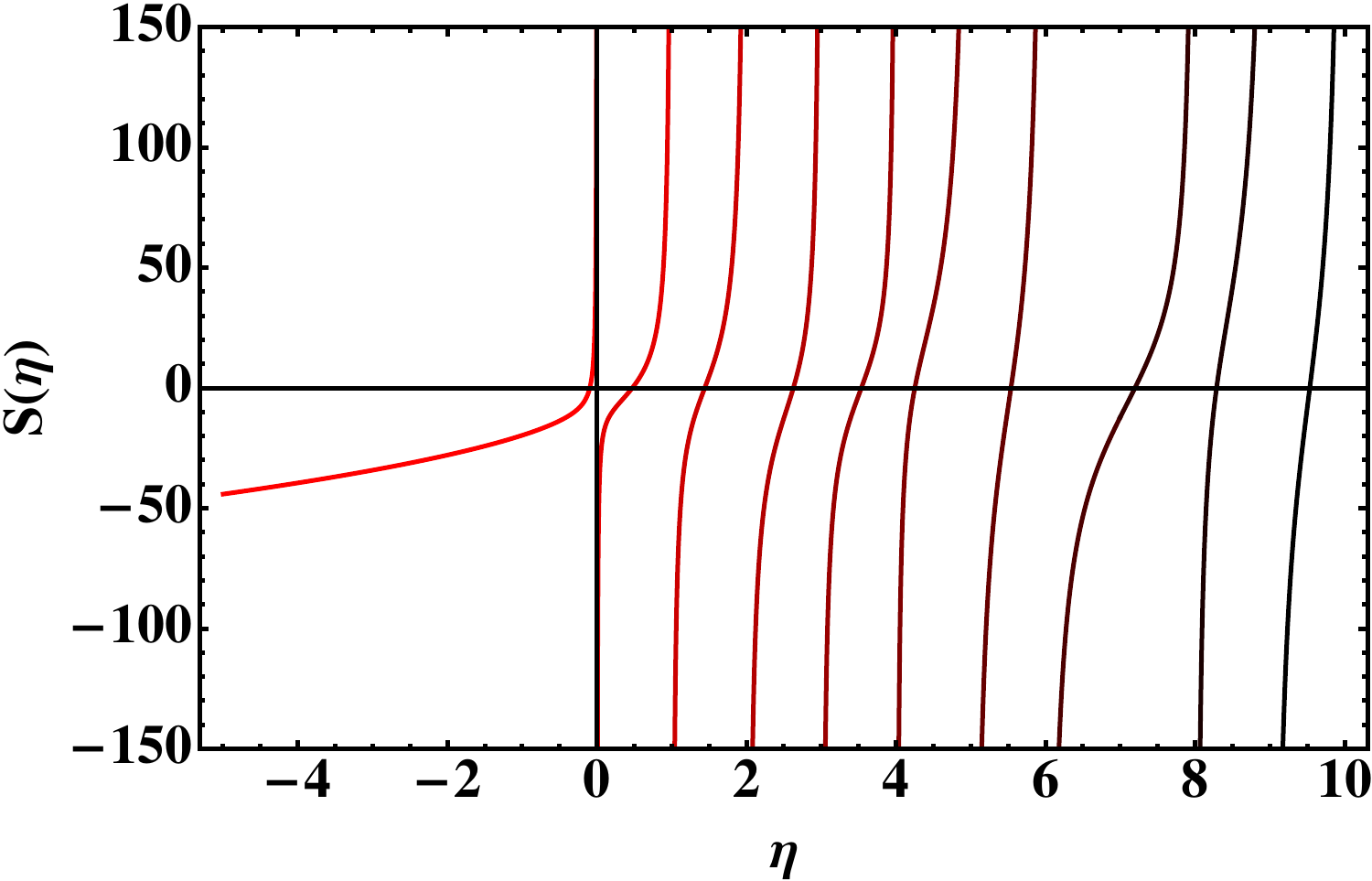}
\caption{\label{fig:zeta3d}A plot of the three-dimensional $\zeta$-function $S(\eta)$.}
\end{figure}

\begin{table}
\caption{\label{tab:S_roots} First 27 roots $\eta_k^*$ ($k=1,\ldots,27$) of $S(\eta)$.}
\begin{ruledtabular}
\begin{tabular}{cc|cc|cc|cc}
$k$ & $\eta_k^*$ &  $k$ & $\eta_k^*$ &  $k$ & $\eta_k^*$ &  $k$ & $\eta_k^*$\\
\hline
 1  & -0.0959007 & 8  &  7.1962633 & 15 & 15.3537376 & 22 & 23.0194729 \\
 2  &  0.4728943 & 9  &  8.2879537 & 16 & 16.1218254 & 23 & 24.3306210 \\
 3  &  1.4415913 & 10 &  9.5345315 & 17 & 17.5325416 & 24 & 25.3016129 \\
 4  &  2.6270076 & 11 & 10.5505341 & 18 & 18.6053932 & 25 & 26.6803601 \\
 5  &  3.5366200 & 12 & 11.7014958 & 19 & 19.5186394 & 26 & 27.8780020 \\
 6  &  4.2517060 & 13 & 12.3102392 & 20 & 20.4033187 & 27 & 29.6156511 \\
 7  &  5.5377008 & 14 & 13.3831152 & 21 & 21.6944179 &    &             \\
\end{tabular}
\end{ruledtabular}
\end{table}

\begin{table}[t]
\caption{\label{tab:C_vals} Results for tuning $N_\calO$ $C_{2n}$ coefficients for an $L=32$, $M=5$ lattice. Uncertainties  in the coefficients reflect a numerical uncertainty in $\eta_k^*$ at $\calO(10^-7)$.}
\begin{ruledtabular}
\begin{tabular}{c|llll%l
}
$N_\calO$ & $C_0$ &  $C_2$ &  $C_4$  &  $C_6$ %&  $C_8$
\\
\hline
 1& 0.6815346(1)&--&--&--%&--
 \\
2&0.466516(2)& 0.0856007(8)&--&--%&--
\\
3&0.489085(8) & 0.00853(2)& 0.020778(6)&--%&--
\\
4&0.50142(5)& 0.00958(3)& 0.00350(8)& 0.00430(2)%&--
\\
%5&0.4515(3)& 0.02782(6)& -0.01885(1)& 0.03432(9)& -0.00628(3)\\
\end{tabular}
\end{ruledtabular}
\end{table}

Once we have tuned the $C_{2n}$ operator coefficients, we can compute all eigenvalues of the 2-particle transfer matrix relevant for continuum $s$-wave scattering and use \Eq{Sfunc} to determine $p\cot\delta_0$. \Fig{tuned_pcotd} shows the result of this exercise for the successive tunings of \Tab{C_vals}. In the left panel we show that $p\cot\delta_0 \ll 1$ over a wide range of momenta, extending well beyond that of the $\le 4$ lowest eigenvalues we used to tune the $C_{2n}$. 

\begin{figure}[b]
\includegraphics[width=\figwidth]{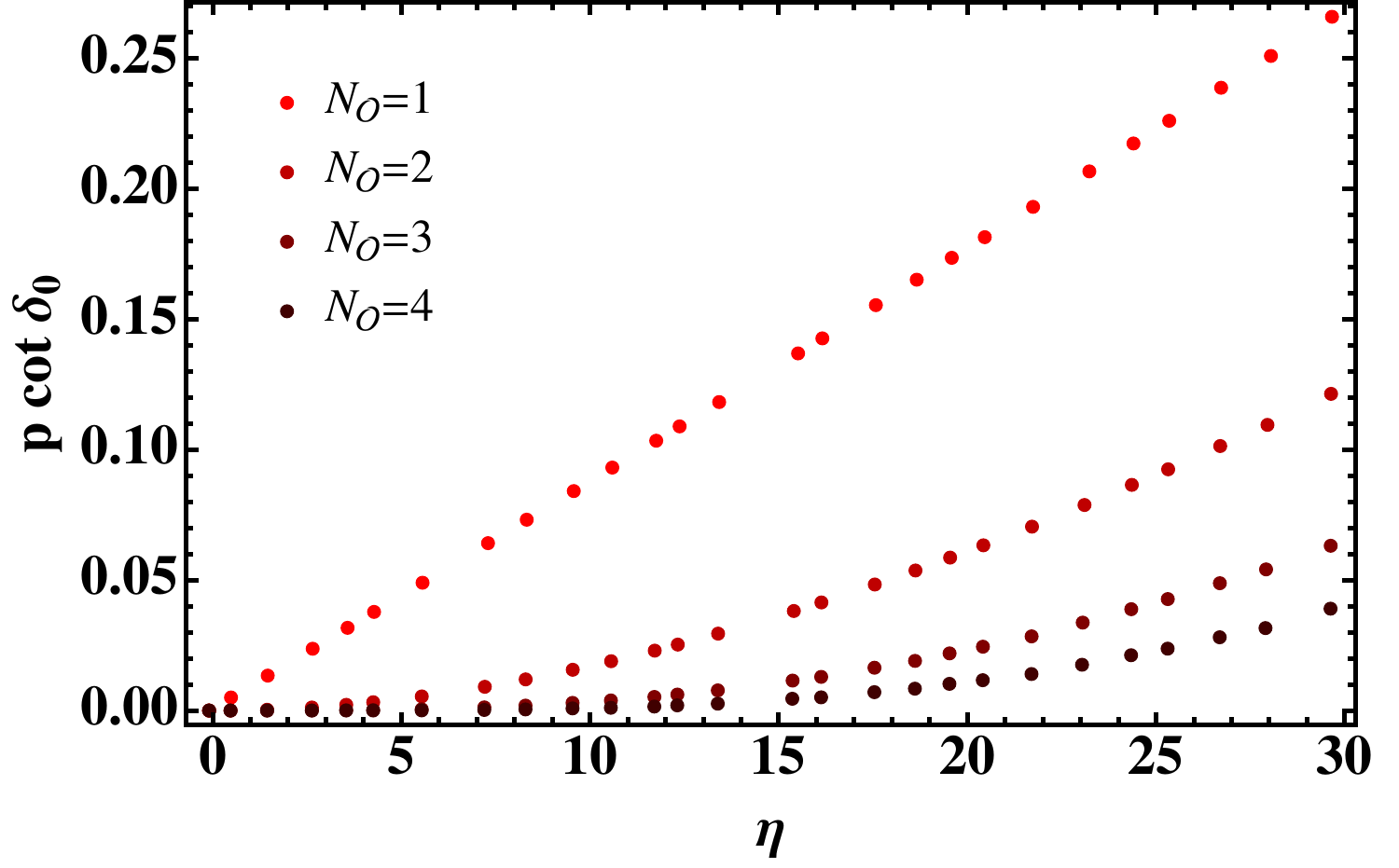}
\includegraphics[width=\figwidth]{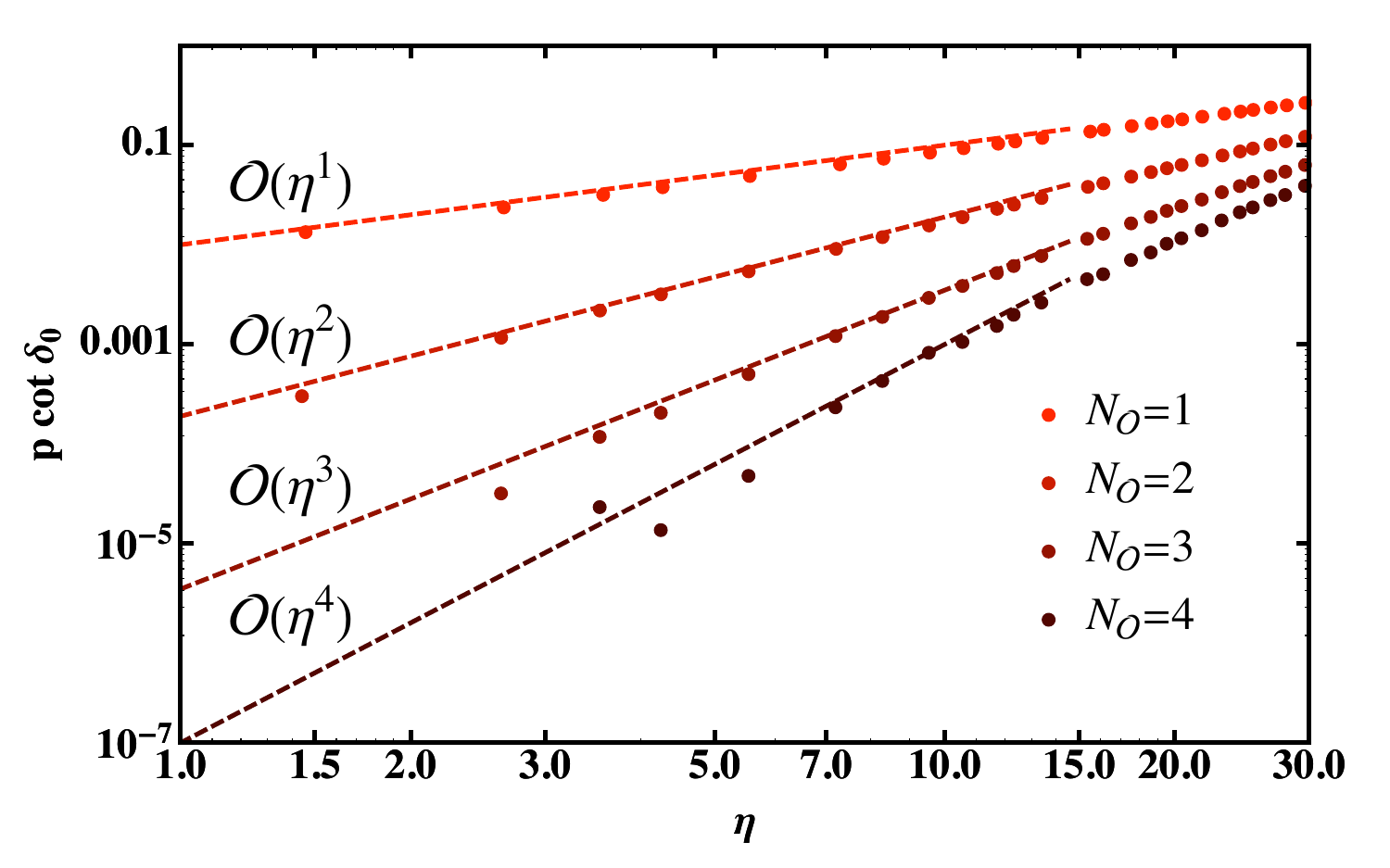}
\caption{%
\label{fig:tuned_pcotd}%
Left: $p\cot\delta_0$ as computed from exact lattice 2-particle energy eigenvalues using L\"uscher's formula, with the first $N_\calO$ terms in the effective range expansion tuned to zero for $N_\calO=1,\cdots,4$. Right: Same data on a ln-ln plot along with expected $\eta$ scaling (dashed lines) for various $N_{\cal O}$. Data is from an $L=32$ and $M=5$ lattice.
}
\end{figure}

Having $p\cot\delta_0$ look progressively flatter with each tuning is only a qualitative indication that we are attaining unitarity with improvement at each order.  It is not advisable to try to fit this curve with a polynomial to extract effective range expansion coefficients; the reason is that the lattice function is only defined at discrete points, and one expects a finite -- but unknown -- radius of convergence for the effective range expansion.  As a result it is possible to extract wildly different effective range coefficients from a polynomial fit, depending on the order of the fit and its momentum range.    
The situation is clarified in the right panel of \Fig{tuned_pcotd} which plots $p\cot\delta_0$ on a ln-ln plot.  This plot shows clear evidence that with each successive tuning we are setting successive terms in the effective range expansion to zero.  Furthermore, the convergence of the dashed lines in the plot at $\eta\sim 30$ demonstrates that the radius of convergence for the effective range expansion is $\eta\sim 30$, with deviations of the plotted points from the dashed lines indicating  significant breakdown of the expansion at $\eta\gtrsim 15$, or $|p|\sim 0.76/b_s$.  Note that for free fermions, $\eta$ is an integer that denotes the energy shell, and that a degenerate fermi gas filled to the $\eta=15$ shell would contain 251 fermions of each spin,  far above the number of fermions we actually are able to study \footnote{The scattered behavior of the lowest $\eta$ points in the right panel of \Fig{tuned_pcotd} seem to indicate the difficulties with our procedure when we attempt to tune too many $C_{2n}$ parameters.}.

\begin{figure}[t]
\includegraphics[width=\figwidth]{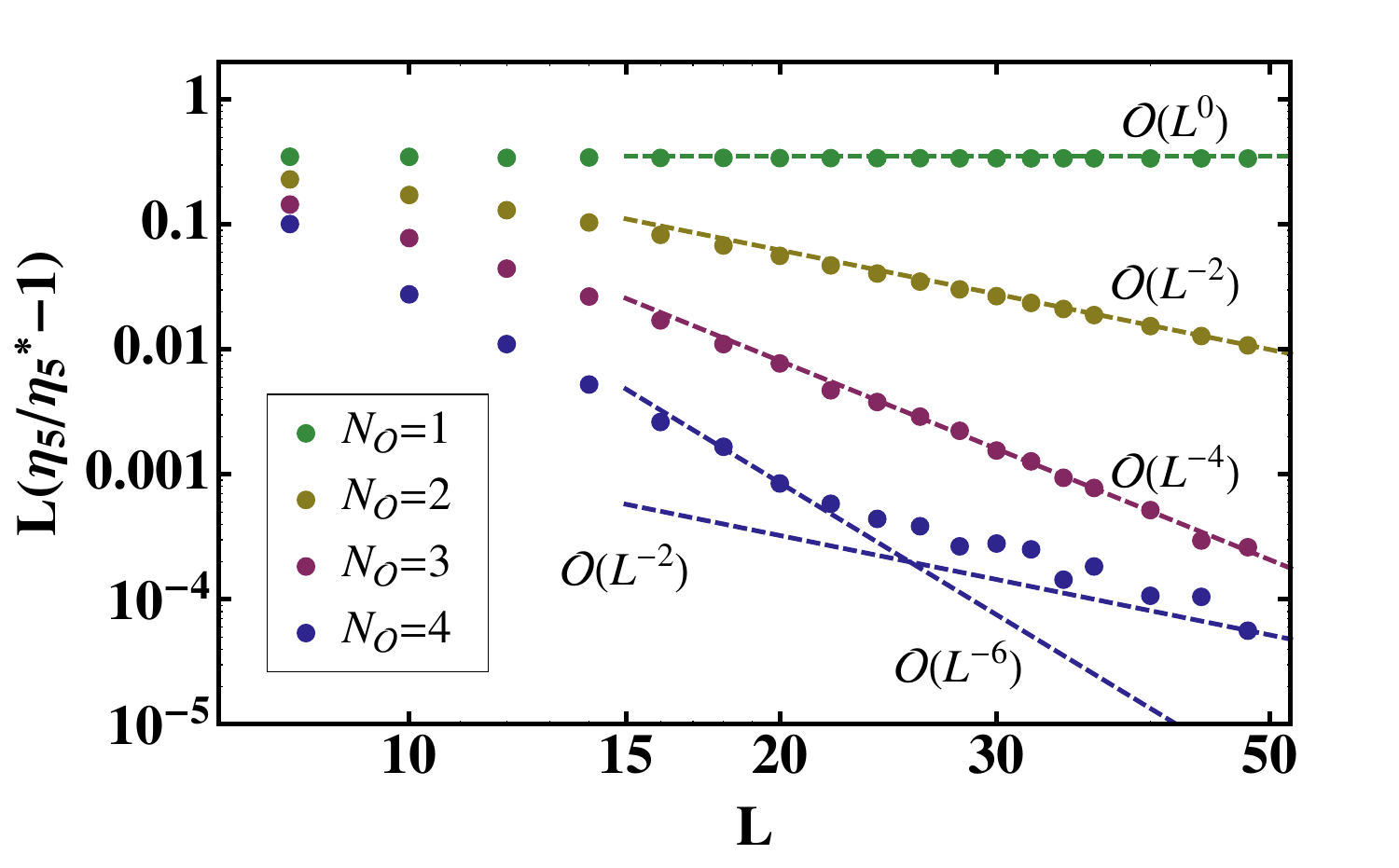}
\includegraphics[width=\figwidth]{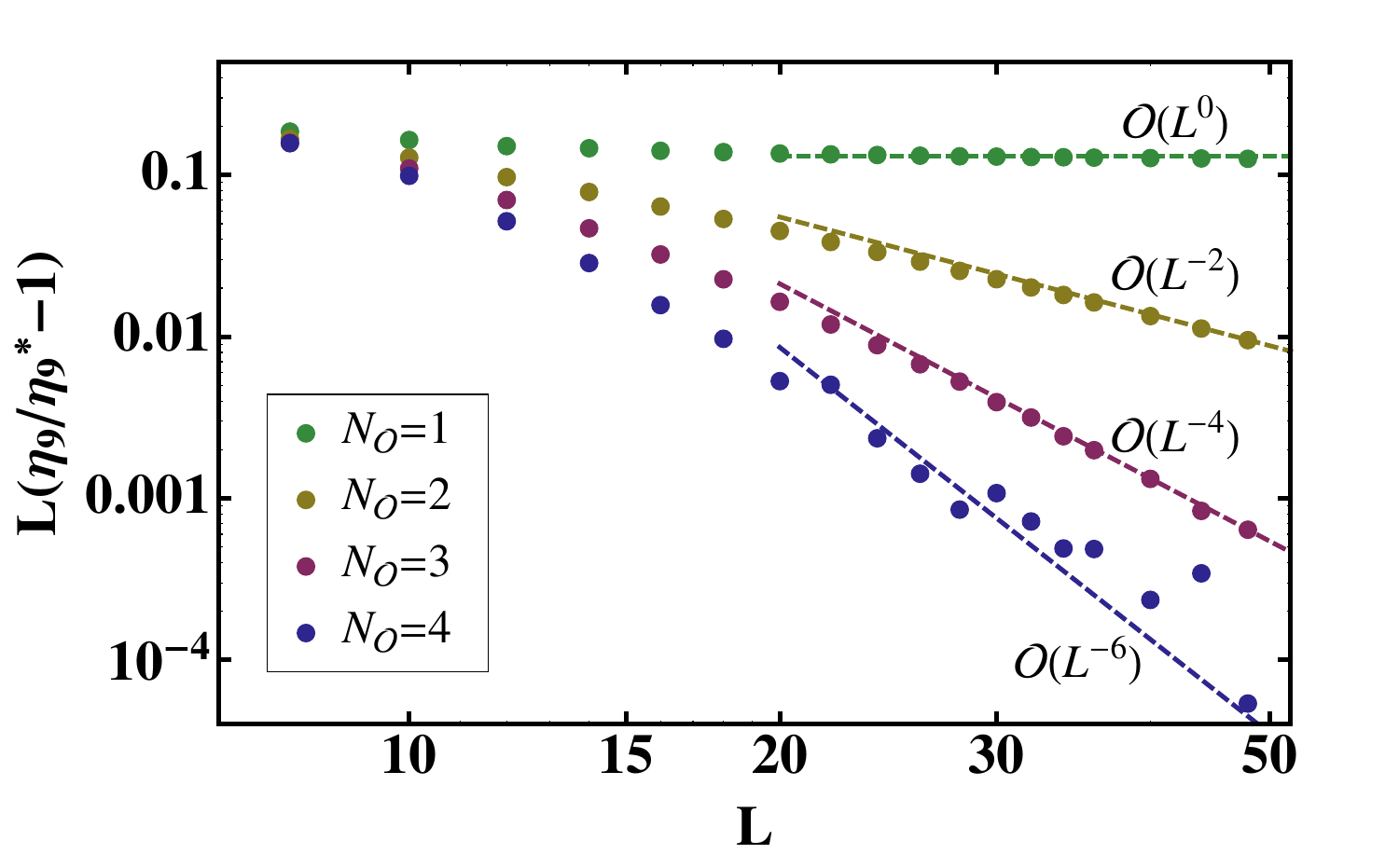}
\caption{%
\label{fig:etaLdep}%
Succesful tuning of effective range parameters may be seen in the $L$ dependence of individual energy eigenvalues for two particles in a box.  Here we see agreement with \Eq{etaLdep} for the $L$-dependence (in lattice units) of  levels $\eta_5$ and $\eta_9$ which were not tuned.
}
\end{figure}

Another way to see if the tuning procedure is successful is to look at the $L$-dependence of the low-lying energy eigenmodes on the lattice.  Assume that we have tuned $p\cot\delta_0$ so that the leading term in the effective range expansion is 
\beq
\pi L p\cot\delta_0 \sim \pi L r_{n-1} p^{2n} =\frac{1}{2} (2\pi)^{2n+1} L^{1-2n} r_{n-1} \eta^n \ ,
\eeq
where $r_{n-1}$ has dimensions $(\text{length})^{2n-1}$,  and that $\eta_k$ are the solutions to $S(\eta)=\pi L p\cot\delta_0$, while as before, the $\eta^*_k$ are the unitary limit solutions to  $S(\eta)=0$.  For sufficiently small $\eta_k-\eta_k^*$ we have $S(\eta_k)\simeq c_k(\eta_k-\eta^*_k)$ where $c_k$ are the slopes of $S$ where it intersects the $\eta$-axis in \Fig{zeta3d}.  Thus we find
\beq
\frac{1}{2} (2\pi)^{2n+1} L^{1-2n} r_{n-1} (\eta^*_k)^n\simeq c_k(\eta_k-\eta^*_k)
\eeq
or
\beq
L\left(\frac{\eta_k}{\eta^*_k}-1\right)\simeq \frac{(2\pi)^{2n+1} r_{n-1}}{2 c_k}(\eta^*_k)^{n-1}\,L^{2-2n}  
\eqn{etaLdep}\eeq
Thus the prediction is that a plot of $L\left(\frac{\eta_k}{\eta^*_k}-1\right)$ should scale like $L^{-(2n-2)}$ when $n$ terms in the effective range expansion have been tuned away.  Note that because of the $L^{-2n}$ factor in the above equation, the effects of  a small residual term at lower order in the effective range expansion will dominate at sufficiently large $L$.
We have computed the low-lying energy eigenvalues for two particles on lattices of a number of different sizes, and in \Fig{etaLdep} we plot the results for  energy levels $\eta_5$ and $\eta_9$, both at higher shells than were used in our tuning procedure.  The scaling of \Eq{etaLdep} is evident in these plots:  at each successive tuning we see that the $L$ dependence is steepened by an additional factor of $L^{-2}$.   An interesting exception is for $\eta_5$ with four parameters tuned and $L\gtrsim 22$; there we see points flattening out to perhaps an $L^{-2}$ slope, suggesting that  a small residual shape parameter $r_1$ is beginning to dominate at that point.  We can use this deviation, \eq{etaLdep}, the value of $\eta_5^*$ from \Tab{S_roots}, and a calculation that gives $c_5 \simeq  96$ to estimate an upper bound on the residual shape parameter, $r_1\lesssim 10^{-3} $ in lattice units.

\subsubsection{ 3-body untrapped ground state energy }
\label{sec:lattice_construction.three.body}

As a nontrivial test of the precision of our lattice method we have computed the lowest energy of three unitary fermions in a zero total momentum eigenstate; the energies of this state and higher eigenstates were computed to high accuracy by Pricoupenko and Castin in Ref.~\cite{2007JPhA...4012863P}.  We performed the calculation for lattice sizes $L=8,10,12,14,16$, tuning the coefficients of four $\calO_{2n}$ operators for the $L=8$ lattice, and five for the other lattices; for each lattice we used $1.5-1.9\times 10^8$ scalar configurations.  With a perfect one-body dispersion relation and this many two-body s-wave operators tuned, the leading $L$ dependence of our result for the $N=3$ energy will be due to the untuned two-derivative two-body p-wave operator at $\mathcal{O}(L^{-3})$; subleading scaling  would be due to the lowest dimension three-body operator, scaling as $L^{-4.72}$, followed by the four derivative p-wave and d-wave two-body operators, scaling as $L^{-5}$; for more details see \cite{Endres:2011tba}. In Fig. 4 we have plotted our results versus $L^{-3}$ --- the leading scaling behavior expected ---  including combined statistical and fitting systematic errors.  Evidently the $L\ge 10$ numbers exhibit  $L^{-3}$ scaling nicely, while the $L=8$ result is off, suggesting that $L=8$ is too small a lattice to see the asymptotic scaling behavior. The red lines in Fig. 4 give the range of two-parameter fits of the $L \geq10 $ data to $c_1+c_2/L^3$, which reflects the uncertainty in our data, while the black line is the fit of the central values of the data using the same fit function. At $L\rightarrow\infty$, the energy we obtain is $0.3735^{+0.0014}_{-0.0007}$ in units of the energy of three noninteracting fermions. As a result we find that our lattice action reproduces the Pricoupenko-Castin result to within our $0.3\%$ uncertainty.

\begin{figure}[t]
\includegraphics[width=\figwidth]{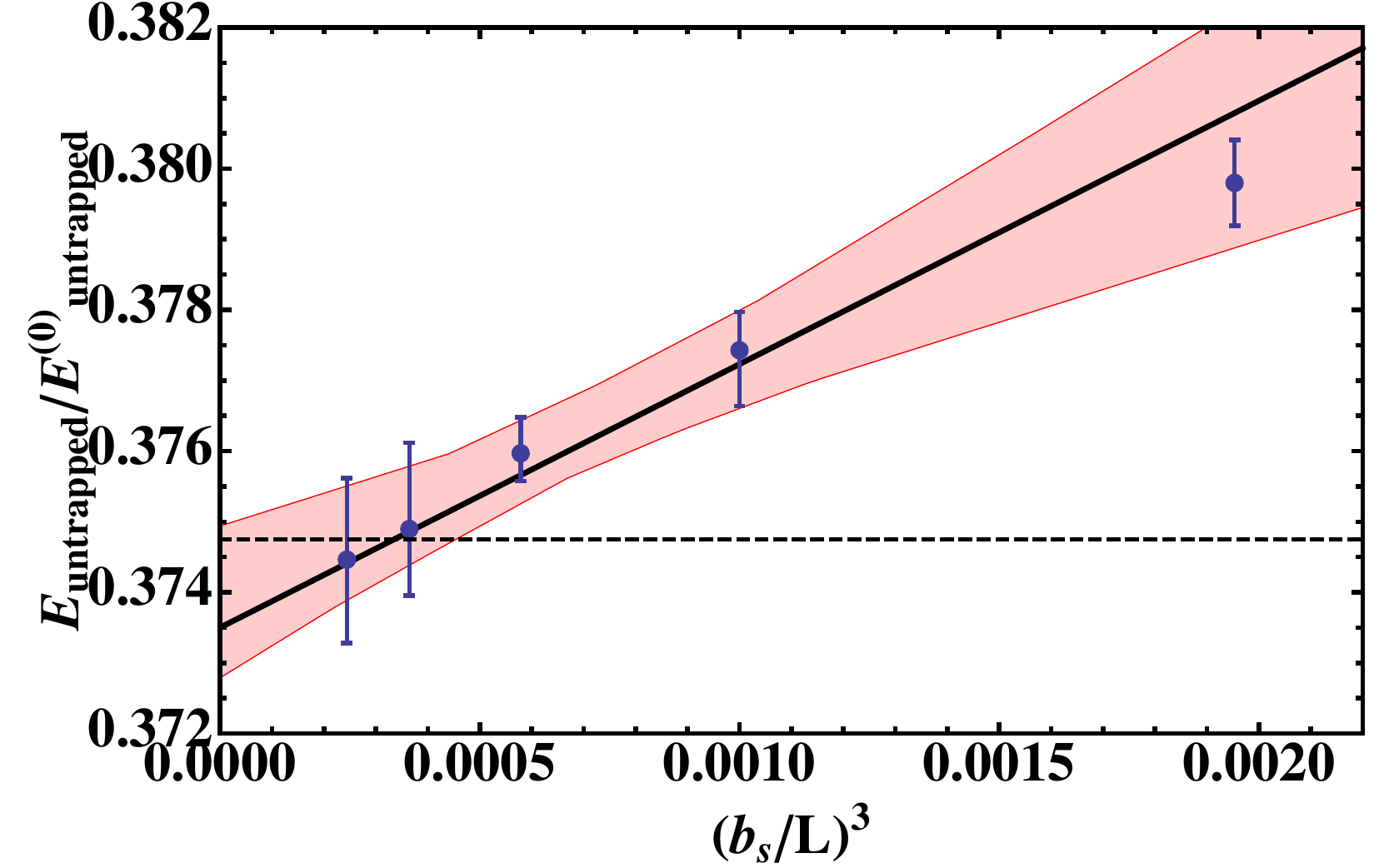}
\caption{%
\label{fig:N3}%
Energy of three untrapped unitary fermions in a zero total momentum eigenstate in units of the energy of three noninteracting fermions, $E^{(0)}_{untrapped}=2\times(2 \pi/L)^2/(2M)$, plotted versus $(b_s/L)^3$ for $L/b_s=8, 10, 12, 14, 16$. The error bars include statistical and fitting systematic errors (for a discussion of these errors, see \Sec{analysis_and_results.unitary_fermions_in_a_finite_trap.extraction_of_ground_state_energies}). The red band represents all possible two-parameter fits of the $L/b_s \geq 10$ data to the function $c_1+c_2/L^3$, reflecting both statistical and fitting systematic errors in our measurements, while the black line is the fit to the central values. The dashed line is the precise Pricoupenko-Castin result, Ref. ~\cite{2007JPhA...4012863P}, with which we agree to within our $\sim 0.3\%$ uncertainty.
}
\end{figure}

\subsection{External potentials}
\label{sec:exttice_construction.ternal_potentials}

Until now, we have concentrated on a system of interacting nonrelativistic fermions in the absence of an external potential.
An external potential $U$ may be introduced in a natural way by replacing the single particle interaction operator $X$ defined in \Eq{Kblocks} with:
\begin{eqnarray}
X(\tau) \rightarrow  e^{-U/2} X(\tau) e^{-U/2}\ ,
\label{eq:potential_discretization}
\end{eqnarray}
where the $L^3\times L^3$ matrix $U$ is given by $U_{\bfx\bfx'} = U(\bfx) \delta_{\bfx,\bfx'}$.
In the case of a harmonic trap, we use a potential of the form $U(\bfx)= \frac{1}{2}\kappa \bfx^2$ centered about $\bfx=0$, and with simple harmonic oscillator (SHO) spring constant $\kappa$.
For fermions of mass $M$, the characteristic trap size is given by $L_0 = (\kappa M)^{-1/4}$, and the oscillator frequency by $\omega = \sqrt{\kappa/M}$.

In the absence of interactions, the single fermion transfer matrix for our lattice theory is given by
\begin{eqnarray}
\calT_\text{SHO} = e^{-\bfp^2/4M\, b_\tau} e^{-U\, b_\tau}  e^{-\bfp^2/4M\, b_\tau}\ ,
\label{eq:lattice_SHO}
\end{eqnarray}
which may be recognized as Trotter's product formula with $\calO(b_\tau^2)$ time discretization errors\footnote{The relation $\calT(-b_\tau) = \calT^{-1}(b_\tau)$ ensures that the energy can only suffer from corrections even in $b_\tau$.}.
Specifically, temporal discretization errors are controlled by the dimensionless quantity $(\omega b_\tau)^2$, and are eliminated in the limit that $\omega\to0$ in lattice units.

Finite volume errors the other hand, are controlled by the dimensionless ratio $L/L_0$.
In the continuum limit, finite volume errors for the noninteracting system may be computed analytically, since the SHO potential is separable.
A plot of the energy dependence of the SHO on $L/L_0$ is shown in \Fig{sho_finite_vol} for several low energy single fermion states; at large $L/L_0$, the energies in units of $\omega$ are just an integer plus the zero point energy $3/2$ for a three-dimensional SHO.
However, for very small volumes, the harmonic potential plays no role and the system is effectively a free particle in a finite box, with energies increasing proportional to $ \frac{1}{2} \left(\frac{2\pi }{L/L_0}\right)^2 $  with decreasing $L/L_0$.
The dashed lines in \Fig{sho_finite_vol} indicate this limiting behavior for several SHO states.

\begin{figure}
\includegraphics[width=\figwidth]{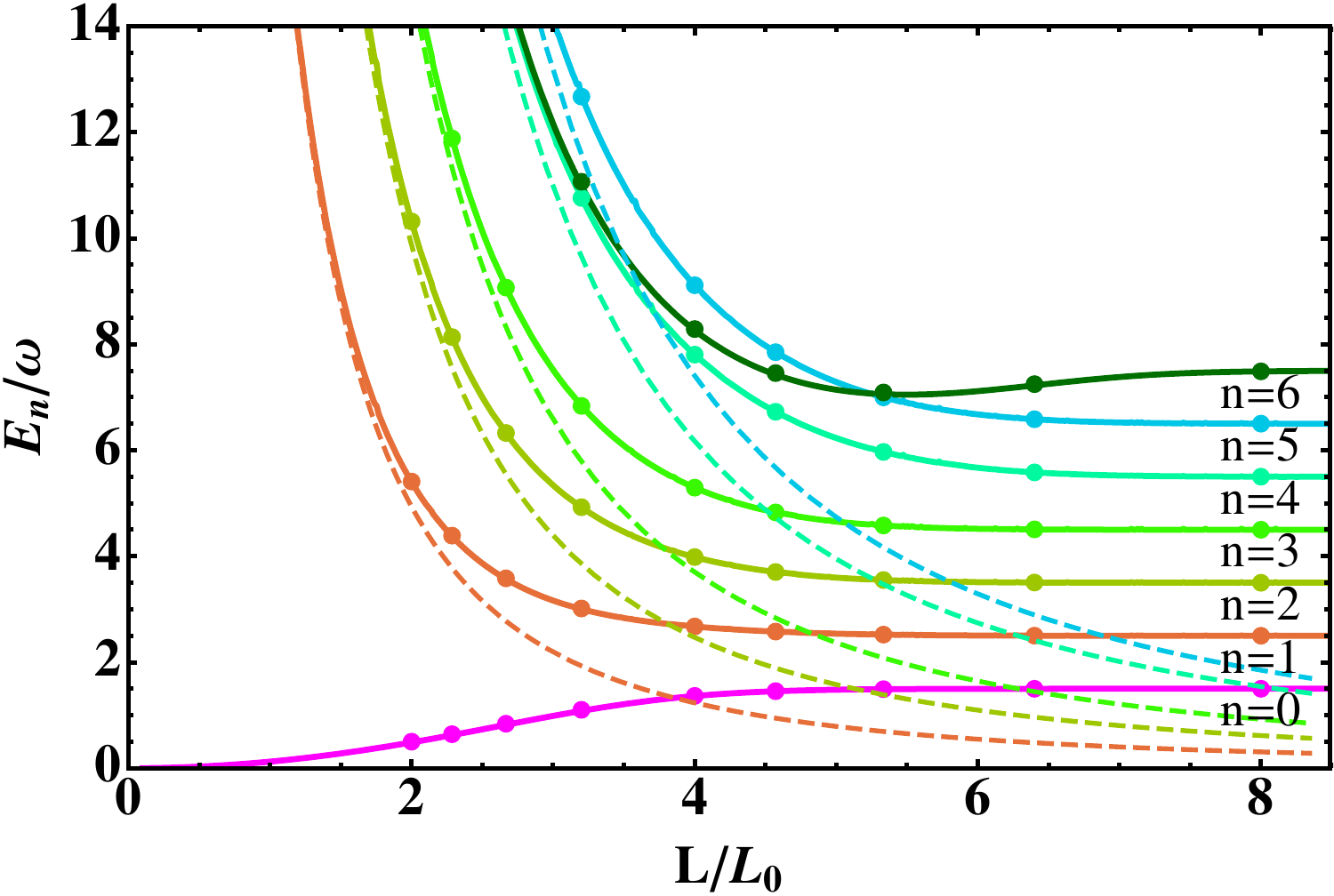}
\caption{%
\label{fig:sho_finite_vol}%
$L/L_0$ dependence of the SHO energies $E_n$ ($n = \sum_j n_j$) corresponding to the single fermion states $\bfn = (0,0,0), (1,0,0), (1,1,0), (1,1,1), (2,1,0), (3,1,1)$, and $(4,2,0)$.
Solid lines indicate an exact continuum limit calculation, whereas the data-points indicate simulation results for $\omega=0.005$ and $L_0\ge2$.
Dashed lines correspond to free fermions in a finite box (small $L/L_0$ limit).
}
\end{figure}

When tuned interactions are turned on, both temporal and spatial discretization errors are controlled in part by how the couplings are chosen.
As was demonstrated in the previous section, by tuning the couplings one may completely eliminate both sources of discretization errors in the low end of the spectrum for two unitary fermions. Writing the transfer matrix for  untrapped  unitary fermions as
\beq
\calT_\text{untrapped} = e^{- b_\tau\bfp^2/4M}(1-b_\tau\calV) e^{-b_\tau\bfp^2/4M}  = e^{-b_\tau \calH}
\eeq
where $\calH$ is assumed to have been tuned free of discretization errors, then
the transfer matrix for unitary fermions in a harmonic trap is given by
\begin{eqnarray}
\calT_\text{trapped} &=&  e^{- b_\tau\bfp^2/4M} e^{-b_\tau U/2} (1-b_\tau\calV) e^{-b_\tau U/2}  e^{-b_\tau\bfp^2/4M} \cr
      &=&  e^{-b_\tau U/2 - b_\tau^2 [U/2, \bfp^2/4M] + \calO(b_\tau^3)} e^{-b_\tau \calH}  e^{-b_\tau U/2 + b_\tau^2 [U/2,  \bfp^2/4M ] + \calO(b_\tau^3)} \cr
      &=&  e^{-b_\tau (\calH+U) + \calO(b_\tau^3)}
\end{eqnarray}
where $ (\calH+U) $ is the target Hamiltonian for trapped unitary fermions. We see that in the lattice definition of the trapped lattice Hamiltonian,  $\calH_\text{trapped} \equiv -\frac{1}{b_\tau}\ln \calT_\text{trapped} $, temporal discretization errors appear at $\calO(b_\tau^2)$.

As was the case for noninteracting fermions in a harmonic trap, interacting fermions will possess spatial discretization and finite volume errors that scale as $b_s/L_0$ and $L/L_0$, respectively. These errors must be explored numerically, and will be presented in detail in \Sec{analysis_and_results}.

\section{Analysis and results}
\label{sec:analysis_and_results}

In this section, we report results for the ground state energies of up to $N=70$ unitary fermions confined to a harmonic potential.
We benchmark our method and systematic errors for up to $N=6$ against high precision solutions to the many-body Schr\"odinger equation, achieving agreement at $1\%$.
We believe this is the first microscopic study to explore $N>6$ fermions in a trap without invoking a variational principle or requiring costly importance sampling.

Numerical simulations of the trapped unitary Fermi gas have been performed with two objectives in mind: evaluation of systematic errors using known few-body ($N\le6$) results as a benchmark, and numerical calculation of ground state energies of the many-body system ($N\le70$).  We explore the  question of whether one can use the trapped fermion data to extract the Bertsch parameter, defined as  $\xi= E_\text{untrapped}/E^{(0)}_\text{untrapped}$, where $E^{(0)}_\text{untrapped}$ is the energy of noninteracting, untrapped fermions. $\xi$ is related to the ground state energies of trapped fermions via the local density approximation  \cite{2005PhRvA..72d1603P}
\beq
E_\text{trapped} = E^{(0)}_\text{trapped} \sqrt{\xi}\left( 1 - 4\sqrt{2\xi} \pi^2\left(c_1-\frac{9}{2}c_2\right)  (3 N)^{-2/3} + O(N^{-4/3})\right)
\eqn{Etrapped}
\eeq
where $c_1$ and $c_2$ are unknown phenomenological constants and $ E^{(0)}_\text{trapped}$ is the energy of $N$ noninteracting trapped fermions,
\beq
 E^{(0)}_\text{trapped} = \frac{(3N)^{4/3}}{4}\,\omega\ .
 \eqn{Efreetrapped}
 \eeq

Note that if  $\left(c_1-\frac{9}{2}c_2\right)\simeq 1$, then   for $N=70$ and $\xi\simeq0.4$ one finds the subleading term in the expansion \eq{Etrapped} to be the same size as the leading term, suggesting that  $N = 70$ is not enough particles for the trapped system to be considered near the thermodynamic limit. In fact, that is what we find:  we see significant shell structure all the way up to $N=70$ and conclude that we are not yet in the thermodynamic limit.  This is in contrast with what we find in the untrapped case, where shell structure disappears at much lower $N$ \cite{Endres:2011tba}. At $N=70$ the system has not yet reached the thermodynamic limit, we are not able to extract the value of $\xi$ or the unknown parameters $c_1$ and $c_2$ from the trapped data. Our data does, however, give information about possible differences in how the trapped and untrapped systems approach the thermodynamic limit.

\subsection{Extraction of ground state energies}
\label{sec:analysis_and_results.unitary_fermions_in_a_finite_trap.extraction_of_ground_state_energies}

The energies of multi-fermion systems may be extracted from correlation functions using conventional techniques.
Given a correlator $\calC(\tau)$ describing the Euclidean time evolution of some N-fermion initial state (source) at time slice zero into some final state (sink) at time slice $\tau$, a generalized effective mass may be defined as
\begin{eqnarray}
m_{eff}(\tau) = \frac{1}{\Delta \tau} \log{ \left[ \frac{ \calC(\tau) }{  \calC(\tau+\Delta\tau) } \right]   }\ ,
\label{eq:effm}
\end{eqnarray}
which satisfies $\lim_{\tau\to\infty} m_{eff}(\tau) = E_{0}$, where $E_0$ is the ground state energy of the system.
At late times, energies are given by a plateau in the effective mass, with excited state contamination falling off exponentially in the energy difference between lowest and first excited states.
For noisy correlators, a stride of $\Delta\tau>1$ may be used to facilitate detection of the time window over which a plateau appears. 

For large numbers of fermions, the standard effective mass exhibits a distribution overlap problem (see Appendix ~\ref{sec:measurement_strategy}). For this reason, we utilize the effective mass defined using the cumulant expansion truncated at $\calO (N_{\kappa}$),
\begin{eqnarray}
m_{eff}^{(N_\kappa)}(\tau) = \frac{1}{\Delta \tau} \sum_{n=1}^{N_\kappa} \frac{ 1}{n!} \left[  \kappa_n(\tau) -  \kappa_n(\tau + \Delta\tau)  \right]\ ,
\label{eq:cumulant_effm}
\end{eqnarray}
where $\kappa_n(\tau)$ is the $n$th cumulant of $\log(\calC(\tau))$. Details of this technique may be found in Appendix ~\ref{sec:measurement_strategy}, and details of the particular strategy used for systems of trapped fermions will be discussed in \Sec{analysis_and_results.unitary_fermions_in_a_finite_trap.many-body_results.statistics}. 

To extract the energies of the system, we perform correlated $\chi^2$ fits to the plateau region of the effective mass associated with the $N$ fermion correlator.
Statistical error estimates are obtained by resampling the data using the bootstrapping technique. Fitting systematic errors are found by varying the endpoints of the fitting interval. For small $N \lesssim 8$, contamination from excited states persists to very large Euclidean times. Because of this, the data we fit is quite noisy and determining the plateau region becomes difficult. For this reason, we vary the endpoints of our fits by $\delta \tau = \pm 10$ to account for any systematic error due to the choice of fitting region.
For large $N$ we find that it is sufficient to vary the endpoints of the fit region by an amount $\delta \tau=\pm 2$ to determine our fitting systematic errors. Because the plateaus are well-resolved for many time steps we do not find significant deviations in the error bars by considering larger variations of the endpoints.

\subsection{Ensembles and parameters}
\label{sec:analysis_and_results.unitary_fermions_in_a_finite_trap.ensembles_and_parameters}

A complete analysis of the systematic errors due to finite volume and lattice spacing artifacts requires performing scans in the parameters $L$, $L_0$ and $\omega$.
Since performing such scans would be prohibitively costly for large numbers of fermions, we have instead chosen to generate two sets of ensembles that allow us to address these questions in a cost-effective manner.

The first set of ensembles consists of a series of scans in the aforementioned parameters, while restricting the number of particles to values $N\le6$.
Restricting the number of fermions in this way greatly reduces the computational resources required, and also permits a higher resolution in the parameter scans.
Few fermion ensembles of size $N_{conf}=1M$ were generated for $L=48$ and $L=64$ lattices using trap sizes $L_0=3,4,5,6, 6.5, 7, 7.5$ and $8$.
Scans were primarily performed at $\omega=0.005$, however, several simulations were also performed at $\omega=0.01$.
The temporal extent for all of the few-body lattices was $T=80$.

With guidance from our analysis of the systematic errors of the few-body system, we then performed a more targeted set of simulations for up to $N=70$ fermions, using parameter choices $L=48, 54$ and $64$, $L_0=7$ and $8$, and $\omega=0.005$.
The parameter choices used in our $N\le70$ simulations are detailed in \Tab{trapped_parameters}.
For our simulations of up to $N=70$ fermions, we have generated approximately one million configurations for each value of the volume and trap size, using a total of less than one million CPU hours.
In all of the trapped fermion studies, we have used $N_\calO=4$ tuned couplings for the interaction.

\begin{table}
\caption{%
\label{tab:trapped_parameters}%
Many-fermion simulation parameters for trapped fermion using the pairing wave function given by \Eq{trapped_pair_source} with $\beta=1/(\sqrt{2}L_0)$.
}
\begin{ruledtabular}
\begin{tabular}{cccc|cccc|cc}
  &   &   &    & \multicolumn{4}{c|}{$C_{2n}$ ($N_{\calO}=4$)}                  &            &           \\
$L$ & $T$ & $\omega$  & $L_0$ & $n=0$ & $n=1$ & $n=2$ & $n=3$  & $N_{conf}$ & $N_\calB$ \\
\hline
48 & 60 & 0.005 & 7.5 & 0.556104 & 0.0182354 & 0.0023426 & 0.01116874 & 1M   & 200 \\
48 & 60 & 0.005 & 8.0 & 0.582780 & 0.0221117 & 0.0016339 & 0.01659503 & 1M   & 200 \\
54 & 60 & 0.005 & 7.5 & 0.554506 & 0.0175868 & 0.0074880 & 0.00953156 & 600K & 200 \\
54 & 60 & 0.005 & 8.0 & 0.581951 & 0.0216195 & 0.0053643 & 0.01527565 & 600K & 200 \\
64 & 60 & 0.005 & 7.5 & 0.555115 & 0.0180441 & 0.0049583 & 0.01031768 & 400K & 200 \\
64 & 60 & 0.005 & 8.0 & 0.582084 & 0.0218977 & 0.0041476 & 0.01568453 & 400K & 200 \\
\end{tabular}
\end{ruledtabular}
\end{table}

Details of our construction of multi-fermion correlation functions are given in \Ap{lattice_construction_observables}. Following \cite{PhysRevLett.91.050401}, we use a modified Slater determinant \Eq{slater2} and \Eq{propagator2} to include pairing correlations. The sinks are constructed from the two-particle wave functions defined in \Eq{trapped_pair_source}.
For all correlation functions, the free parameter appearing in \Eq{trapped_pair_source} was chosen as $\beta=1/\sqrt{2}L_0$.
Multi-fermion sources where constructed from free SHO single particle wave functions $|\bfn^\sigma_i\rangle$ with $\sigma=(\downarrow,\uparrow)$ provided in \Tab{sources}, with $i\le N/2$.
The sources involving odd $N$ for our few fermion studies were obtained by removing a single fermion from the highest shell, as described at the end of \Sec{lattice_construction_observables}.

\subsection{Few-body Results}
\label{sec:analysis_and_results.unitary_fermions_in_a_finite_trap.few-body_results}

To reach the continuum and infinite volume limits we require $b_s \ll L_0 \ll L$, and $b_{\tau} \ll 1/\omega$.
To balance the need for small temporal discretization errors with the computational cost associated with the number of time steps required to reach the ground state, we have chosen $\omega b_{\tau} = 0.005$ for this study.
For small $N$, we find that the discrepancies in the energies for $\omega b_{\tau}$ in the range $0.005 - 0.01$ are about $0.5\%$, and are within our error bars.

For a given box size, the choice of $L_0$ must take into account both discretization errors and finite volume errors.
The expectation is that for small $L_0/b_s$ spatial discretization errors will dominate.
The discretization is implemented as a hard cutoff in momentum space, which may be interpreted as an infinite potential at the edge of the Brillouin zone. Sensitivity of the state to this infinite potential results in an increase in the associated energy.
Conversely, for large $L_0/b_s$ and fixed volume, finite volume errors will dominate.
The periodic boundary conditions in space result in attractive interactions from image particles, causing a decrease in energy.
Thus, measurement of the ground state energies as a function of $L_0/b_s$ and $L/L_0$ is necessary to determine at which value we can minimize both types of error.

\Fig{SysErrors} presents our findings for the ground state energies of $N=3,4,5$ and $6$ fermions, with $L_0/b_s$ ranging from $3-8$ and fixed $L/b_s$=48.
Also indicated in this figure are the ground state energies for unitary fermions in a trap quoted in \cite{Blume201186}, which were obtained by numerically solving the multi-fermion Schr\"odinger equation using the correlated gaussian (CG) method.
Using the results of \cite{Blume201186} as a benchmark, we find that for $L_0/b_s \lesssim 7.0$ our discretization errors are significant.
Above this value, however, we find that the extracted energies are independent of $L_0/b_s$, indicating negligible discretization errors in this regime\footnote{%
In \cite{Nicholson:2010ms}, we found that our results agreed with those of \cite{Blume201186} for values of $L_0\approx4$.
However, it became evident that this agreement resulted from a delicate cancellation between temporal and finite volume errors, and that each source of error was individually rather significant.
In this work, we have reduced the temporal discretization errors with an improved form of the potential; this improvement results in temporal errors appearing at an order higher in $\omega b_{\tau}$.
We have also chosen a smaller value for $\omega b_\tau$, and checked that the results are consistent for both the smaller ($\omega b_\tau=0.005$) and larger ($\omega b_\tau=0.01$) values.
}.

\begin{figure}
\includegraphics[width=0.6\textwidth]{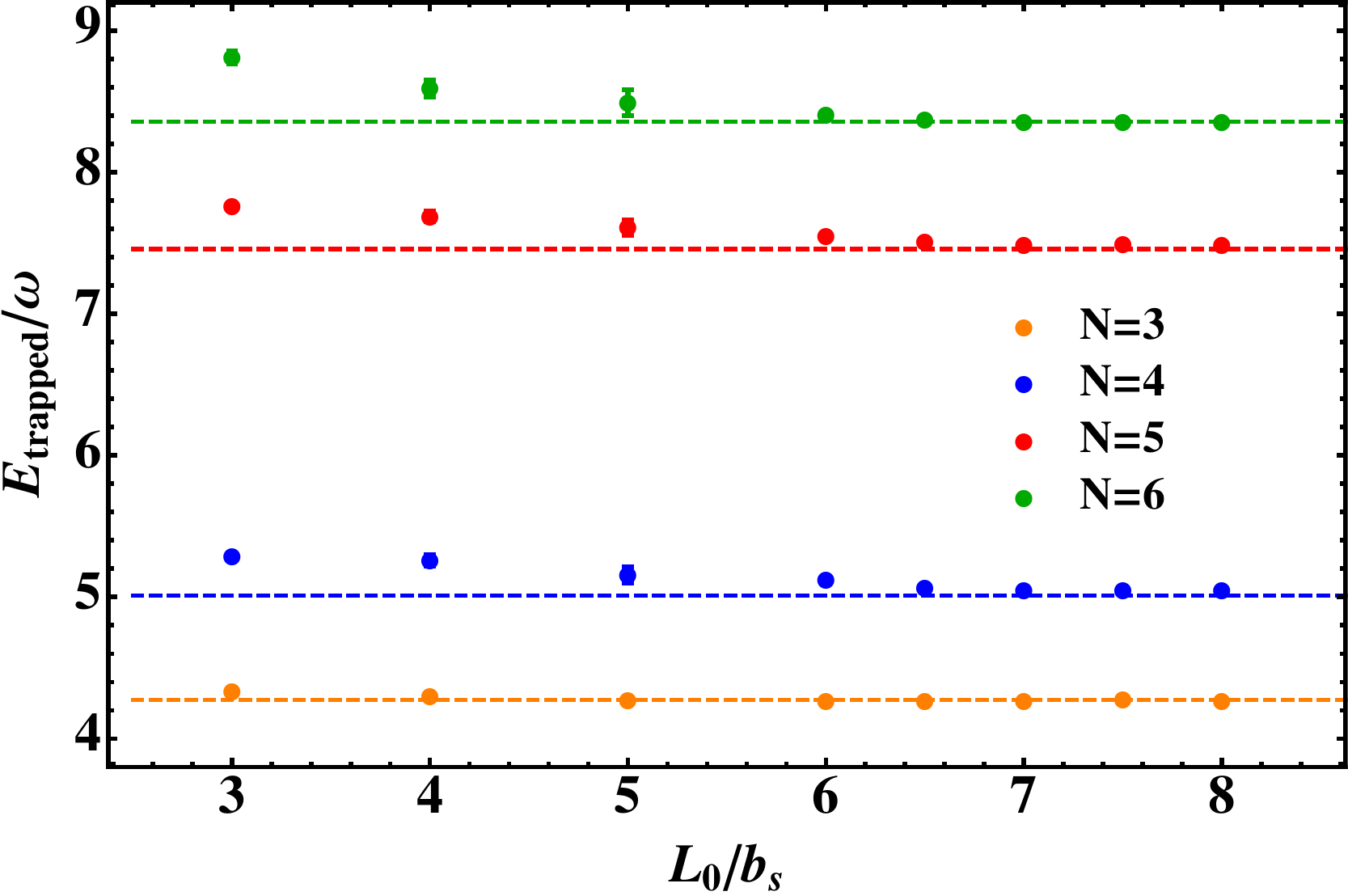}
\caption{%
\label{fig:SysErrors}%
Ground state energies (in units of $\omega$) as a function of $L_0/b_s$ at fixed $L/b_s=48$ for various values of $N$.
Dashed lines are results from \cite{Blume201186}.
}
\end{figure}

In \Fig{SmallN}, we present the $L/L_0$ dependence of the energies for $N=3,4,5$ and $6$, with $L/b_s=48$ and $64$ and $L_0\ge7$.  
\begin{figure}
\includegraphics[width=\figwidth]{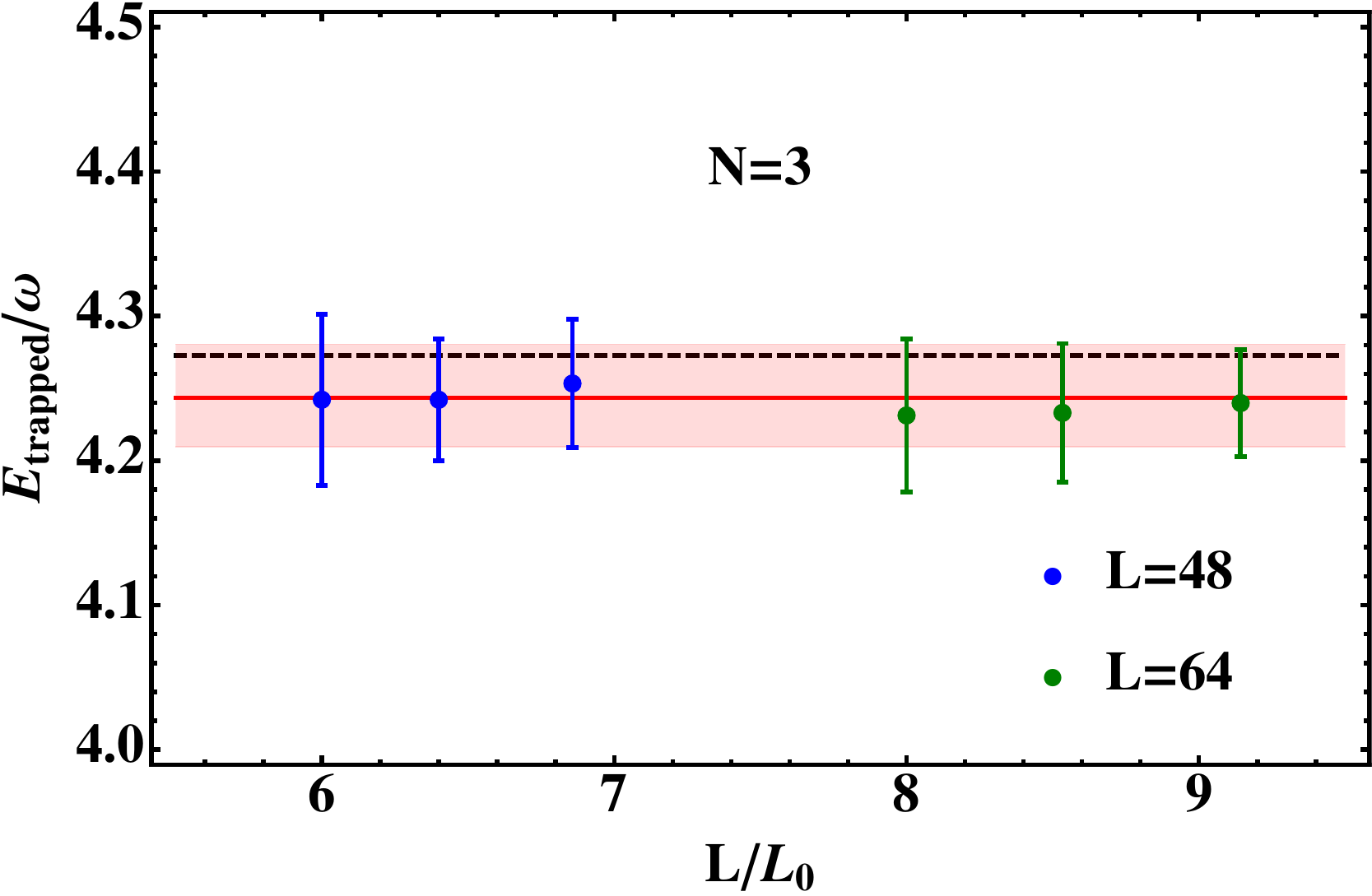}
\includegraphics[width=\figwidth]{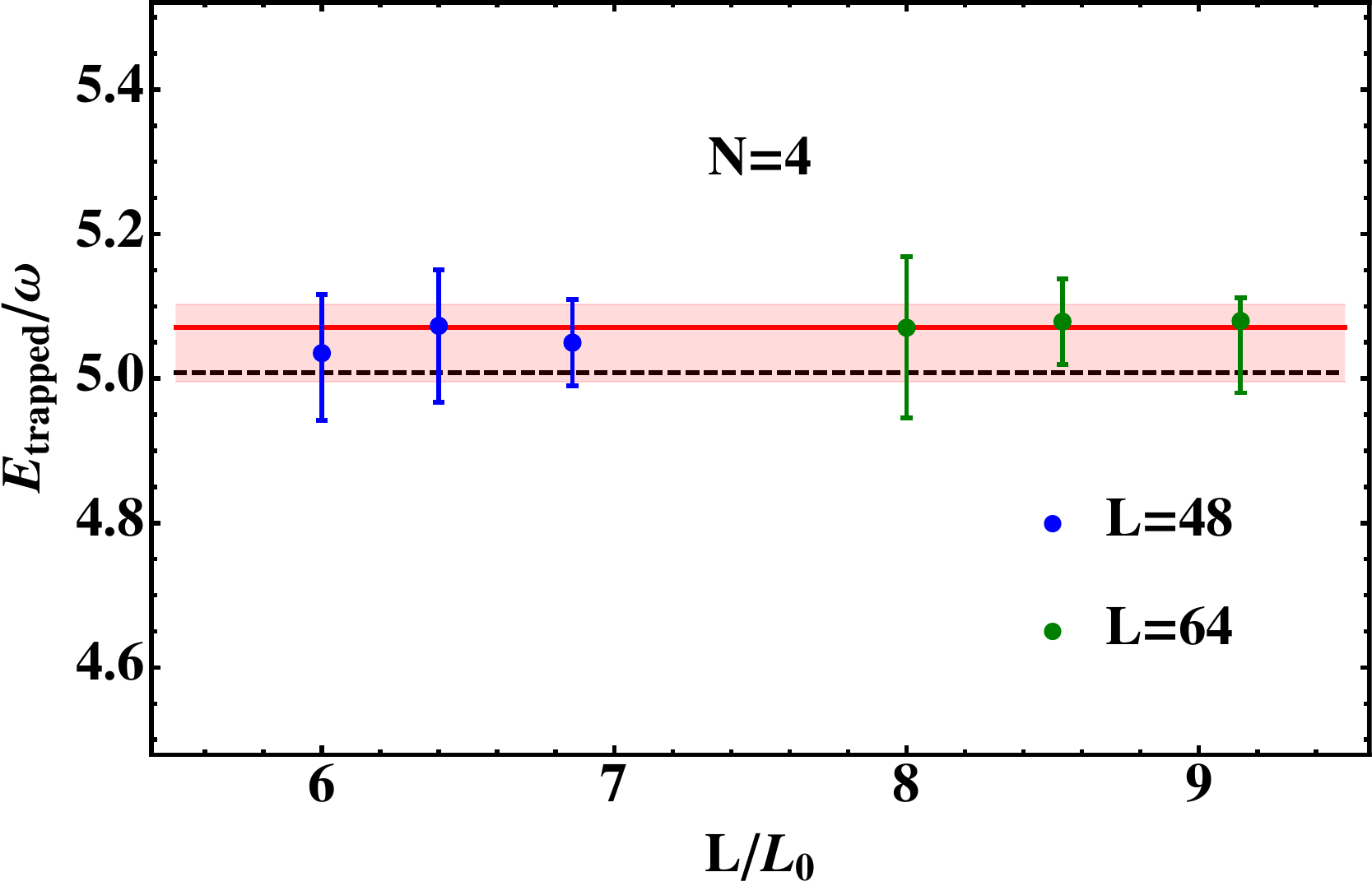}
\includegraphics[width=\figwidth]{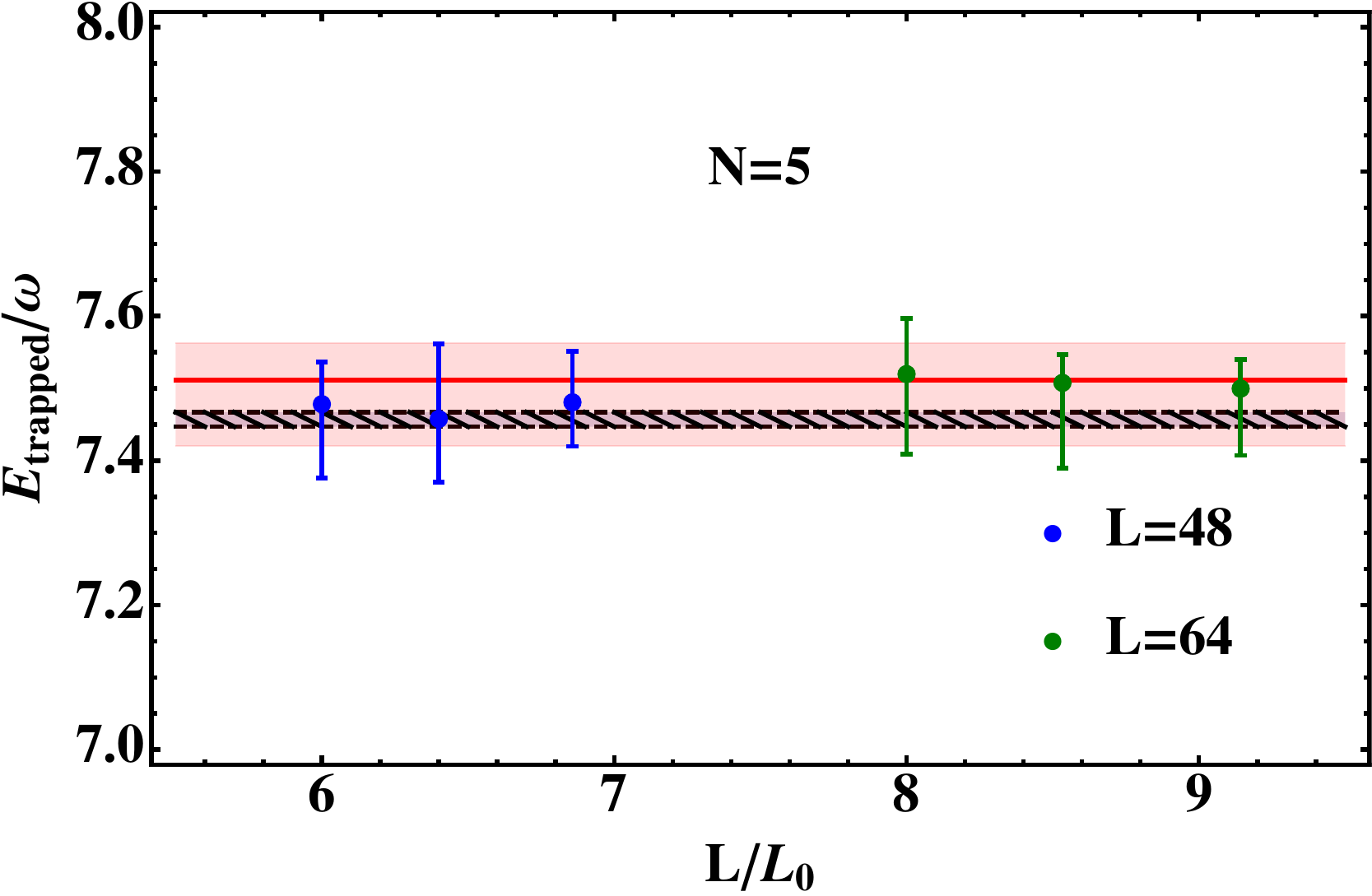}
\includegraphics[width=\figwidth]{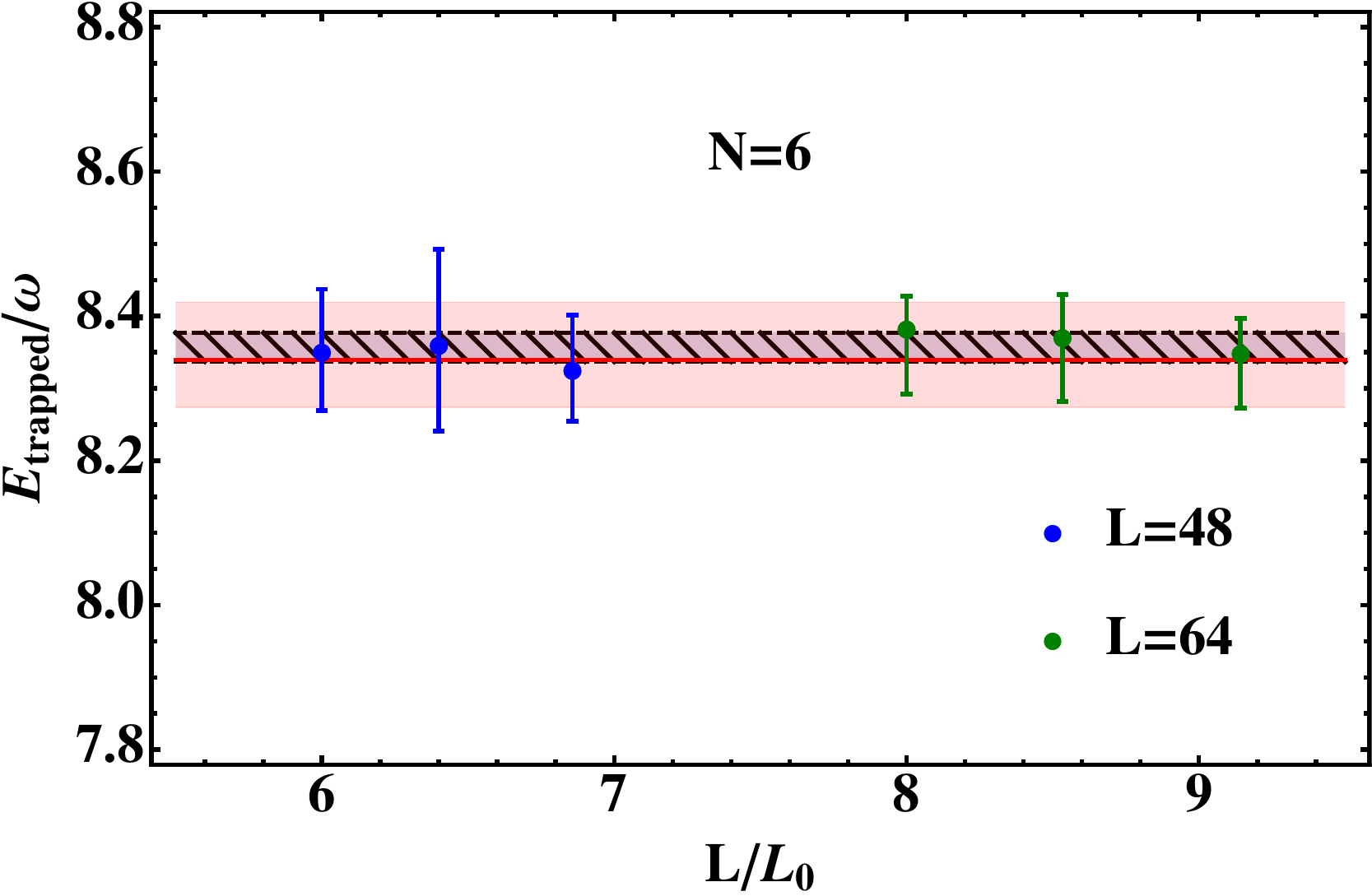}
\caption{%
\label{fig:SmallN}%
Fit results for the ground state energies (in units of $\omega$) for $N\leq6$ as a function of $L/L_0$ for two volumes, $L=48,64$, and three trap sizes, $L_0=7.0,7.5,8.0$.
The result of a correlated fit in $N,\tau$ of all data is shown as a red line, with a red band showing the combined statistical and systematic errors. The results from \cite{2004cond.mat.12764T} (N=3) and \cite{Blume201186} (N=4-6) are given by dashed lines, with any associated error bars shown by hatched regions.
}
\end{figure}
The consistency of the results between the different volumes indicates that finite volume errors are negligible within statistical uncertainties for $L/L_0$ ranging from $6-9$. The good agreement of our $L_0\ge7$ data in \Fig{SmallN} with the benchmark energy values indicates the absence of any residual errors.
We have performed a constant correlated fit using all the data in \Fig{SmallN} to obtain infinite volume, vanishing lattice spacing results for the few particle energies. 
Our final fit results for $N\leq6$ are indicated in \Fig{SmallN}, and presented along with a comparison to an exact result for $N=3$ \cite{2004cond.mat.12764T} and high-precision Hamiltonian results of \cite{Blume201186} for $N=4-6$ in \Tab{SmallN}.

\begin{table}
\caption{%
\label{tab:SmallN}%
Results for $E_\text{trapped}/\omega$ for $N \leq 6$, including combined statistical and fitting systematic errors (first row). For comparison we give the exact $N=3$ result \cite{2004cond.mat.12764T} and results of Ref. \cite{Blume201186} (second and third rows).}
\begin{ruledtabular}
\begin{tabular}{|c|c|c|c|c|}
 & 3  & 4 & 5 & 6\\
\hline
this work & $4.243^{+0.037}_{-0.034} $ & $5.071^{+0.032}_{-0.075} $ & $7.511^{+0.051}_{-0.091} $ & $8.339^{+0.080}_{-0.066} $  \\
exact, Ref. \cite{2004cond.mat.12764T}  & 4.2727 & - & - & - \\
from Ref. \cite{Blume201186}  & 4.273(2)  &  5.008(1) & 7.458(10) & 8.358(20) \\
\end{tabular}
\end{ruledtabular}
\end{table}

For $N>6$, it is likely that both discretization and finite volume errors will grow, since we expect the wave function to spread out in both position and momentum space when more particles are added to the system.
Numerical evidence suggests that extrapolations become necessary for $N\gtrsim 20$.
More details of our analysis for larger $N$ will be presented in the next section.

\subsection{Many-body Results}
\label{sec:analysis_and_results.unitary_fermions_in_a_finite_trap.many-body_results}

\subsubsection{Statistics}
\label{sec:analysis_and_results.unitary_fermions_in_a_finite_trap.many-body_results.statistics}

Examples of an effective mass plot obtained using the conventional definition, \Eq{effm} with $\Delta\tau=1$, for $N=30$ and $N=70$, are shown in \Fig{effMassStandard}.
Although we have found good signals for most values of $N$ at short times, at later times the effective mass plot shows clear evidence of a distribution overlap problem. 

Generally speaking, since the sources and sinks used to compute these correlation functions are different (see Appendix ~\ref{sec:lattice_construction_observables}), positivity of the correlator is not guaranteed.
Furthermore, there is no reason to expect that effective masses obtained from them will decrease monotonically as a function of time.
When analyzing effective mass plots, one must therefore be capable of distinguishing between local minima in the effective mass and true plateaus in order to extract reliable ground state energies.
In cases like \Fig{effMassStandard} ($N=30$) where a plateau begins ($\tau \sim 6$) well before the onset of a distribution overlap problem ($\tau \sim 22$), one may easily make the distinction between local minima and true ground state plateaus.
However, in cases like \Fig{effMassStandard} ($N=70$) where the onset of an overlap problem and the beginning of a (potential) plateau coincide, the distinction becomes less clear.

\begin{figure}
\includegraphics[width=\figwidth]{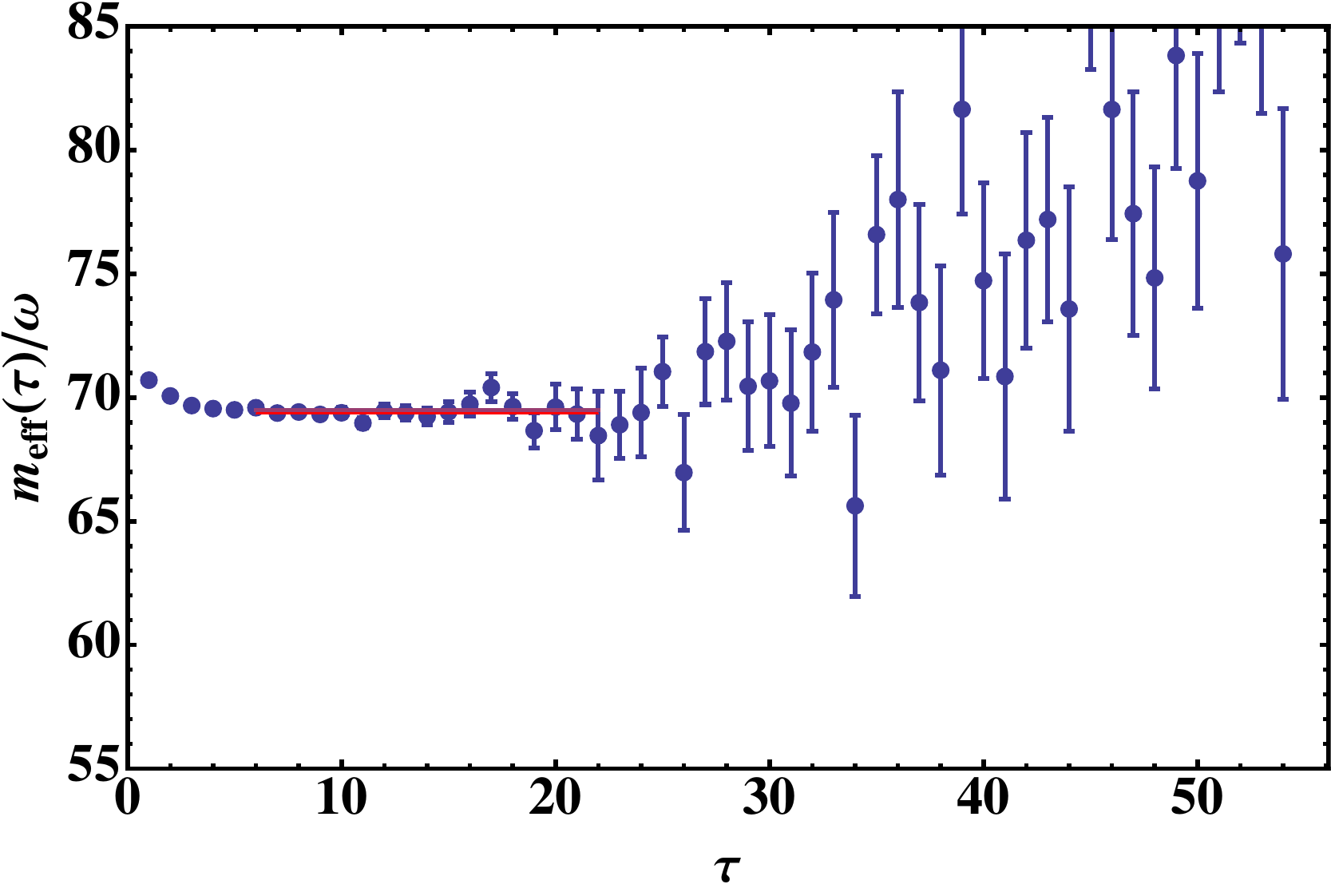}
\includegraphics[width=\figwidth]{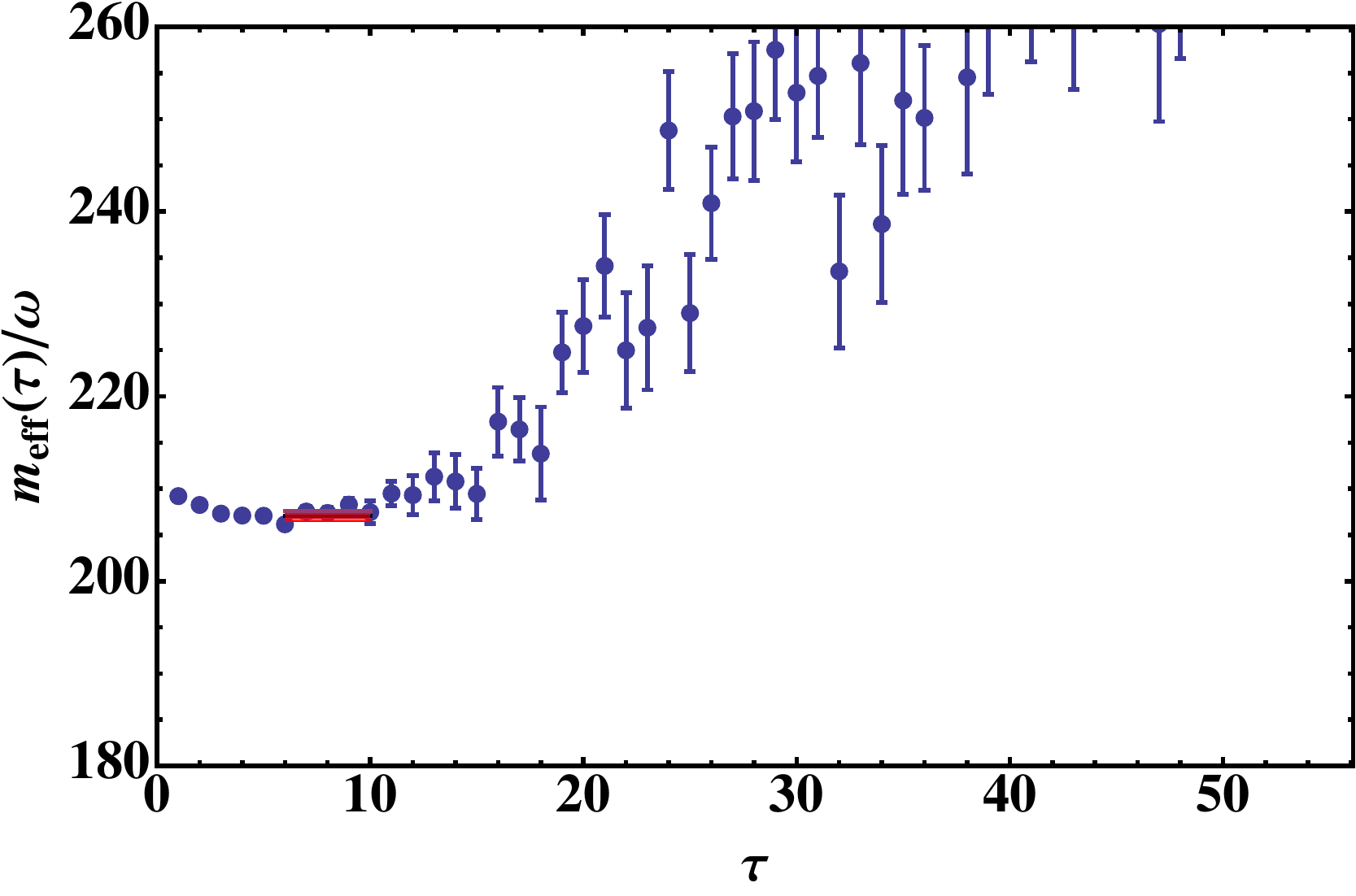}
\caption{%
\label{fig:effMassStandard}%
Effective mass as a function of $\tau$ (lattice units) for $N=30$ and $N=70$ ($L=48$, $L_0=8.0$) using the conventional definition for the effective mass \Eq{effm}.
Fits are represented by horizontal bands. 
}
\end{figure}

In light of the considerations above, effective mass calculations based on the cumulant expansion, \Eq{cumulant_effm} (for details, see Appendix ~\ref{sec:measurement_strategy}), were used to help establish and extend plateaus into the region of poor overlap.
The result of this technique is demonstrated in \Fig{effMass} for $N=70$ with $N_\kappa$ ranging from $3-5$.
Comparing the effective mass obtained from the cumulant expansion with that obtained by conventional means, we see a marked reduction in the overlap problem for $\tau\gtrsim12$, although the noise at late times increases with $N_\kappa$.
Nonetheless, the extension of the plateau region to larger times gives us confidence that we have reached the ground state, and allows us to perform fits over much longer temporal extents.

\begin{figure}
\includegraphics[width=\figwidth]{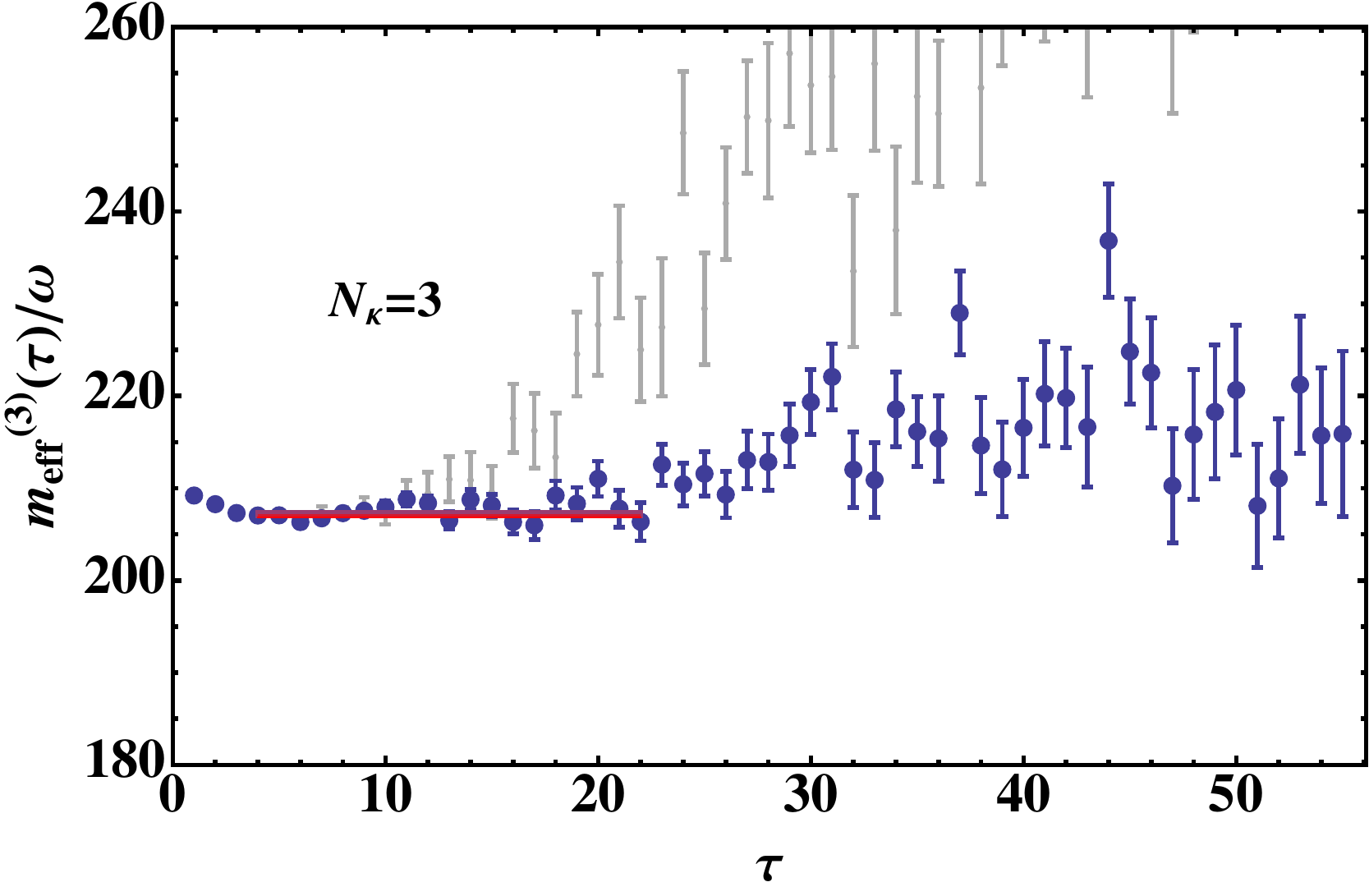}
\includegraphics[width=\figwidth]{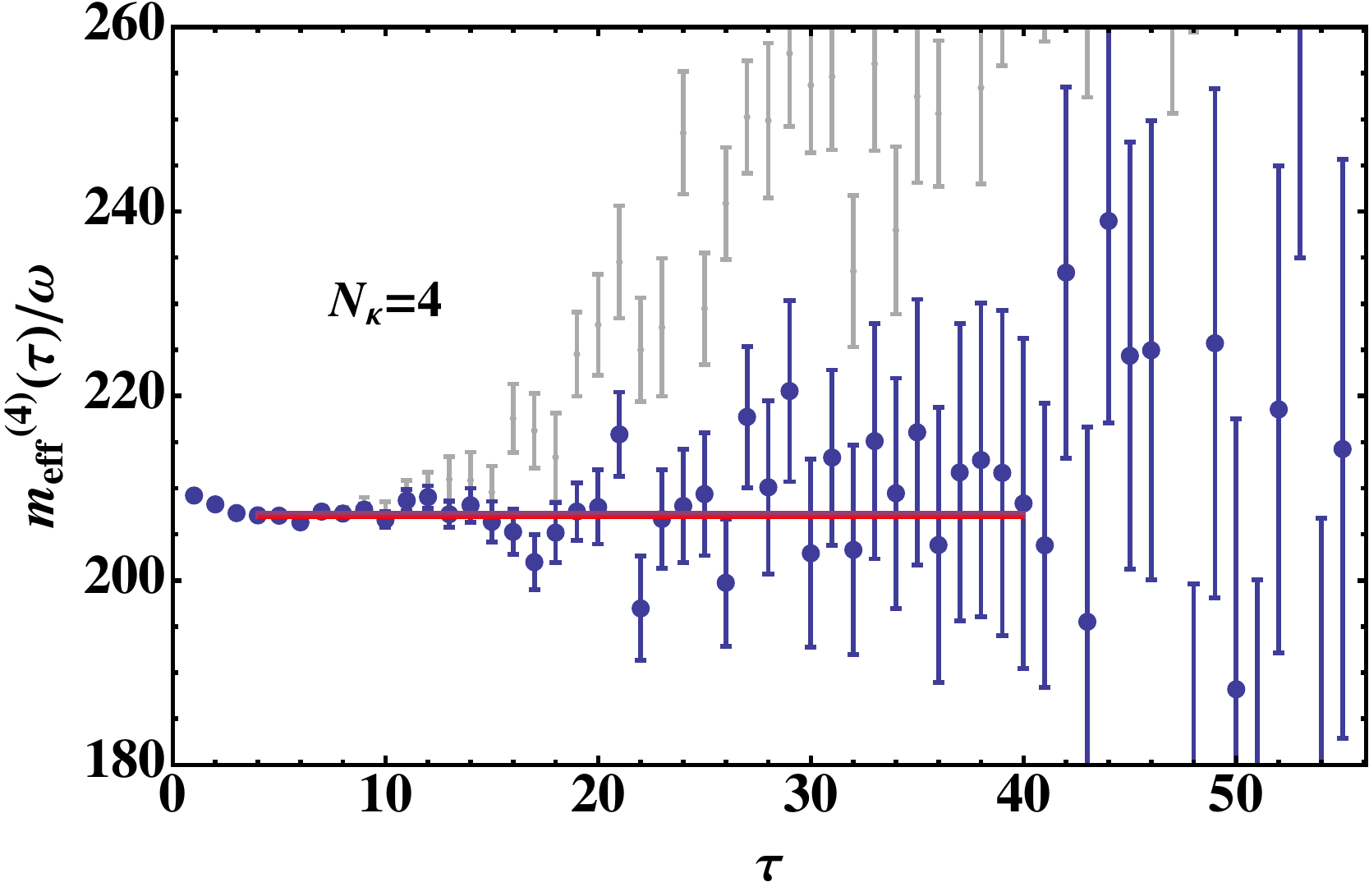}
\includegraphics[width=\figwidth]{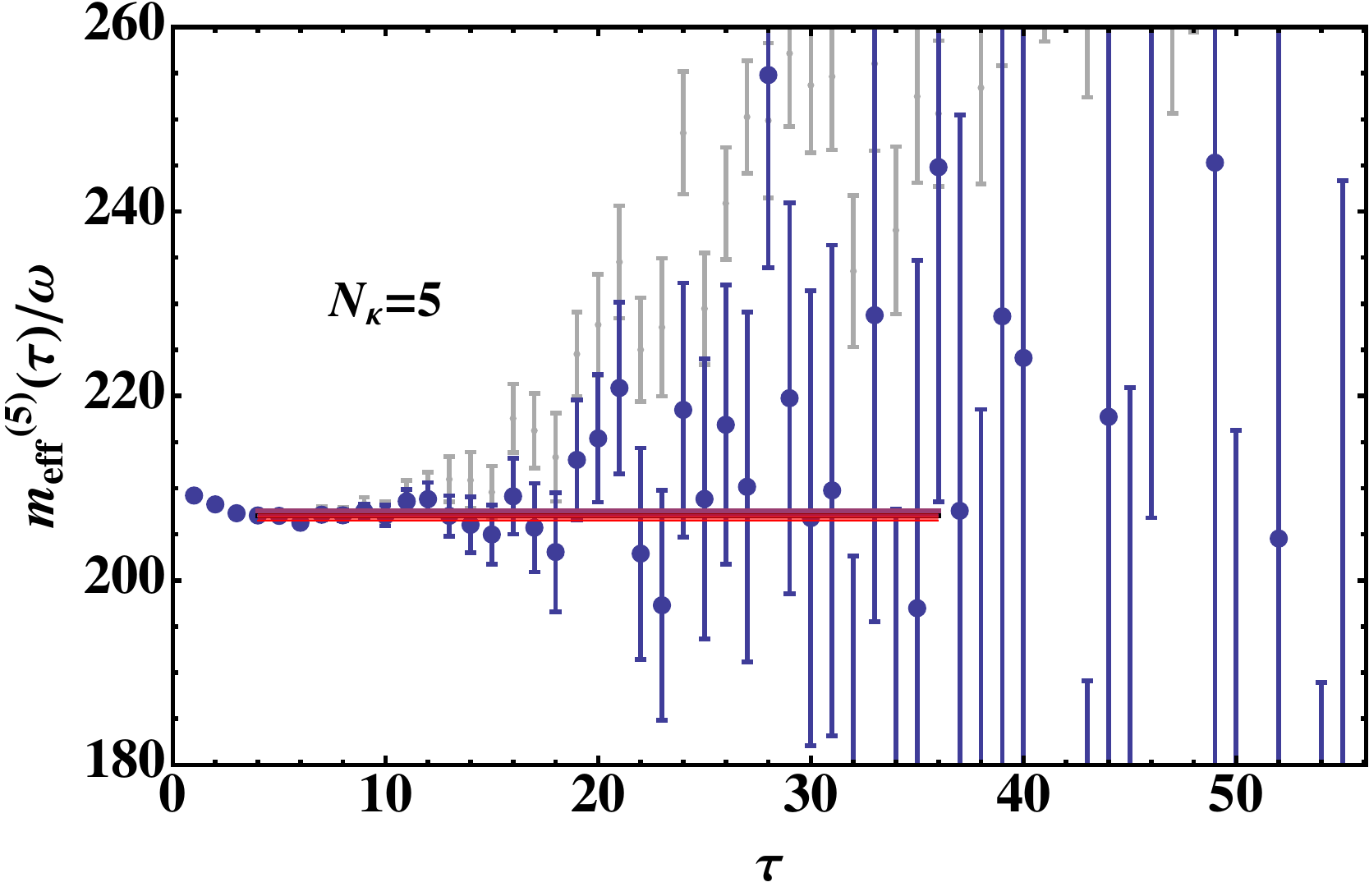}
\caption{%
\label{fig:effMass}%
Effective mass as a function of $\tau$ (lattice units) for $N=70$ ($L=48$, $L_0=8.0$) using the cumulant expansion (\Eq{cumulant_effm}) for up to $N_\kappa$ cumulants.
The fits from each are represented by horizontal bands. The conventional effective mass is shown in gray. 
}
\end{figure}

Examples of the fits obtained using the cumulant expansion for small ($N=10$), moderate ($N=30$), and large ($N=70$) numbers of fermions are shown in \Fig{untrapped_cumulant_converg}.
For small $N$, the plateaus appear at late times where we find that the cumulant expansion converges slowly.
However, because the overlap problem is less severe for small $N$ versus large $N$, we may corroborate our cumulant results with those obtained using the conventional effective mass (red bands). 
For $N=10$, convergence occurs at $N_\kappa\approx6$, whereas for all $N\gtrsim12$, convergence occurs at $N_\kappa\approx3$. 
Higher cumulants may be used to further extend the plateau region without significant growth in the error bars for the fits.
This is  presumably because the fit results are highly influenced by the early time portion of the plateau region, where the errors on the effective mass are relatively unaffected by an increase in the number of cumulants. The nearly exact agreement between results obtained using $N=3,4,5$ leads us to conclude that systematic errors introduced by truncation of the cumulant expansion are negligible.

\begin{figure}
\includegraphics[width=\figwidth]{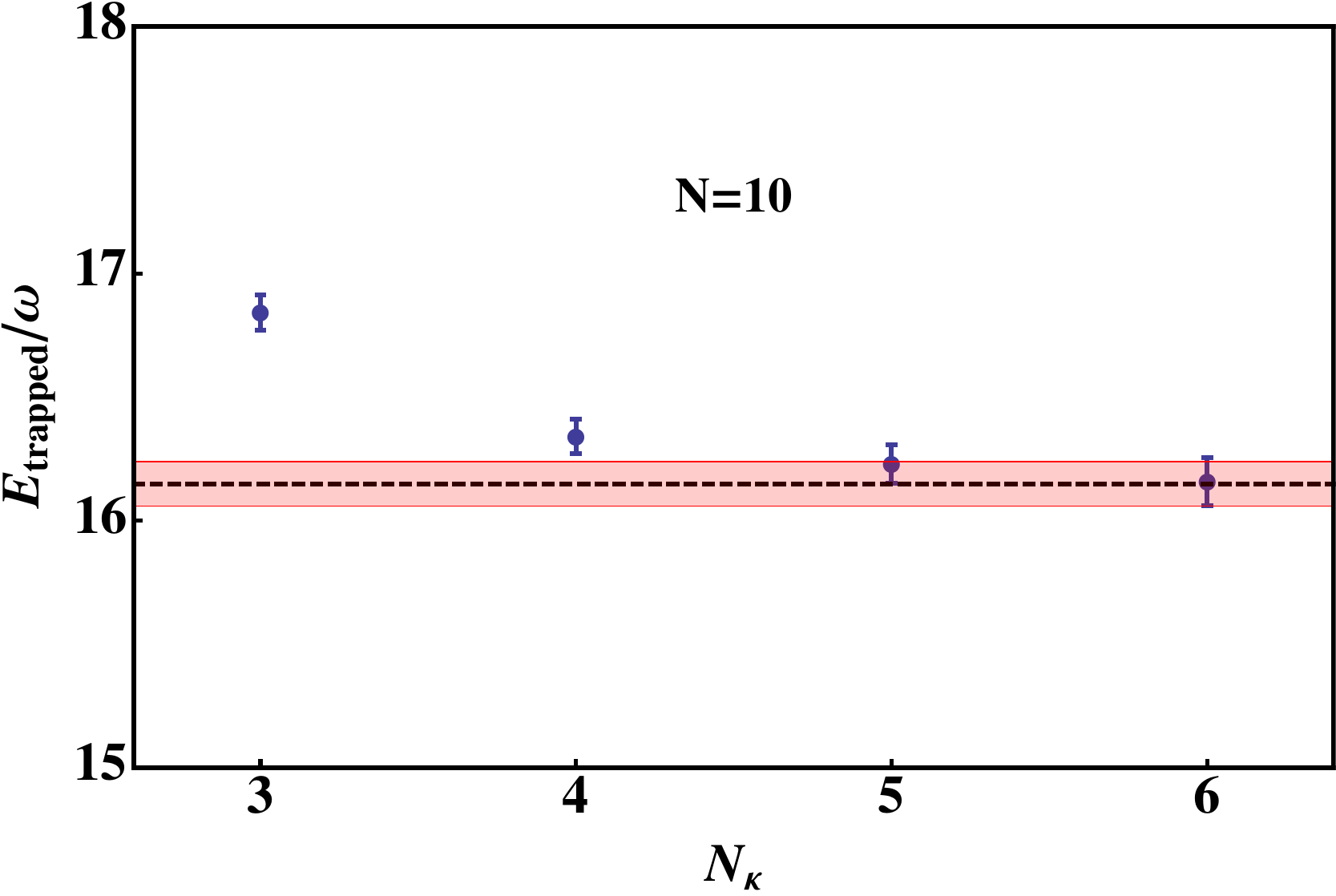}
\includegraphics[width=\figwidth]{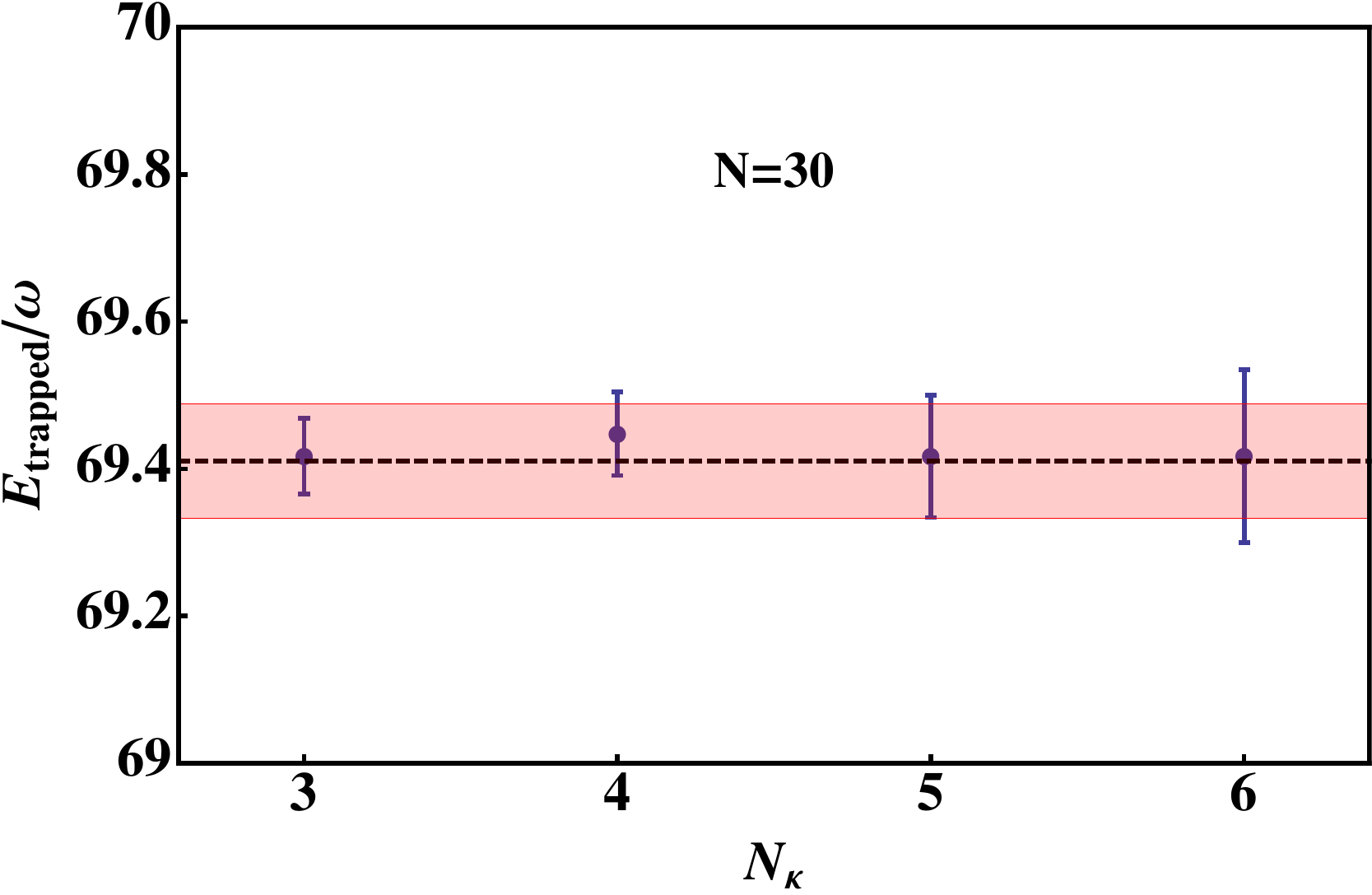}
\includegraphics[width=\figwidth]{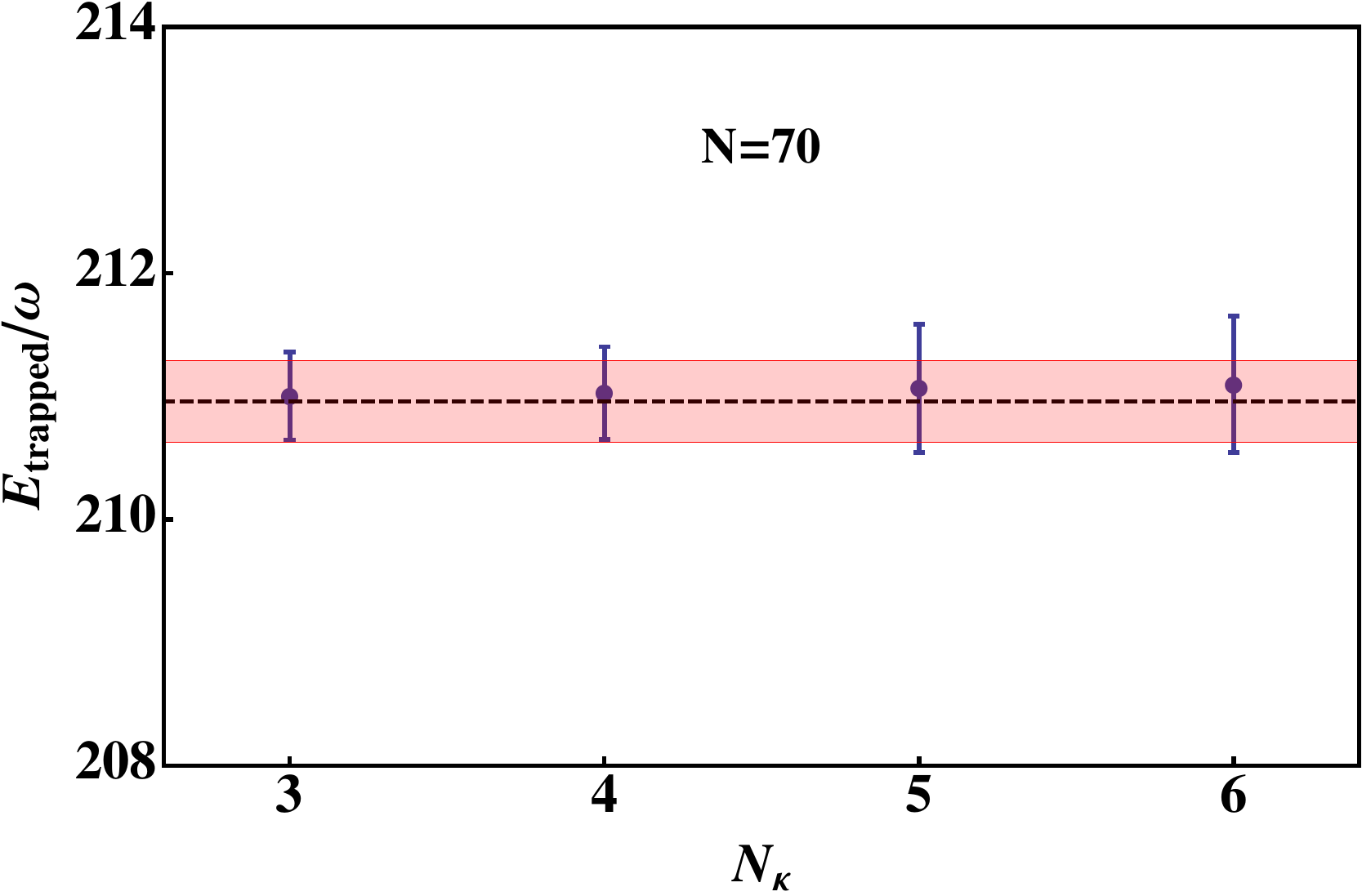}
\caption{%
\label{fig:untrapped_cumulant_converg}
Fit results for the ground state energies (in units of $\omega$) for small, moderate, and large $N$ ($L=54$, $L_0=8.0$) using the cumulant expansion including $N_{\kappa}$ cumulants (\Eq{cumulant_effm}).
The ground state energy extracted from the conventional effective mass is given as a red band.
}
\end{figure}

\subsubsection{Systematics}
\label{sec:analysis_and_results.unitary_fermions_in_a_finite_trap.many-body_results.systematics}

\begin{figure}[h!]
\includegraphics[width=\figwidth]{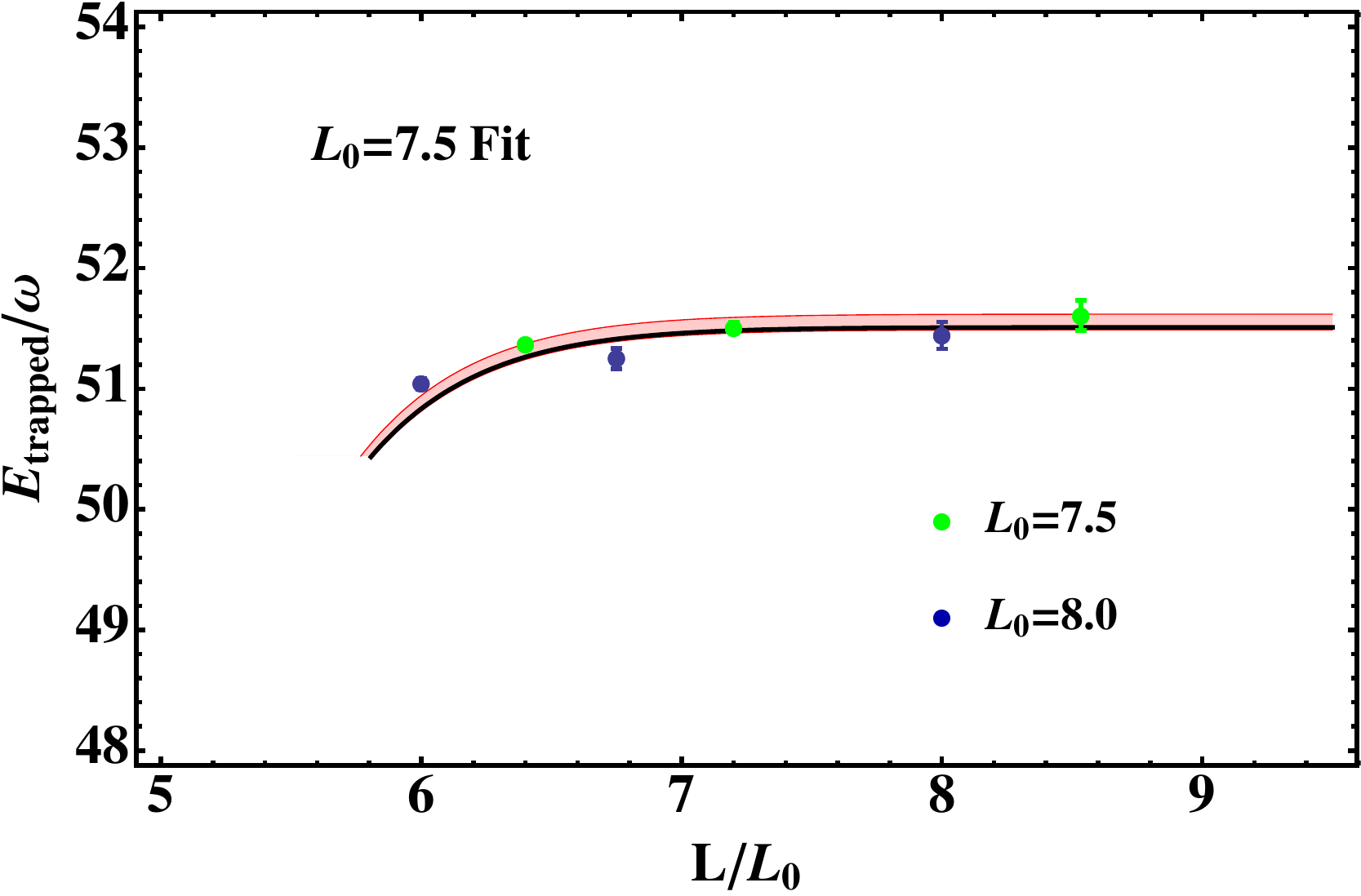}
\includegraphics[width=\figwidth]{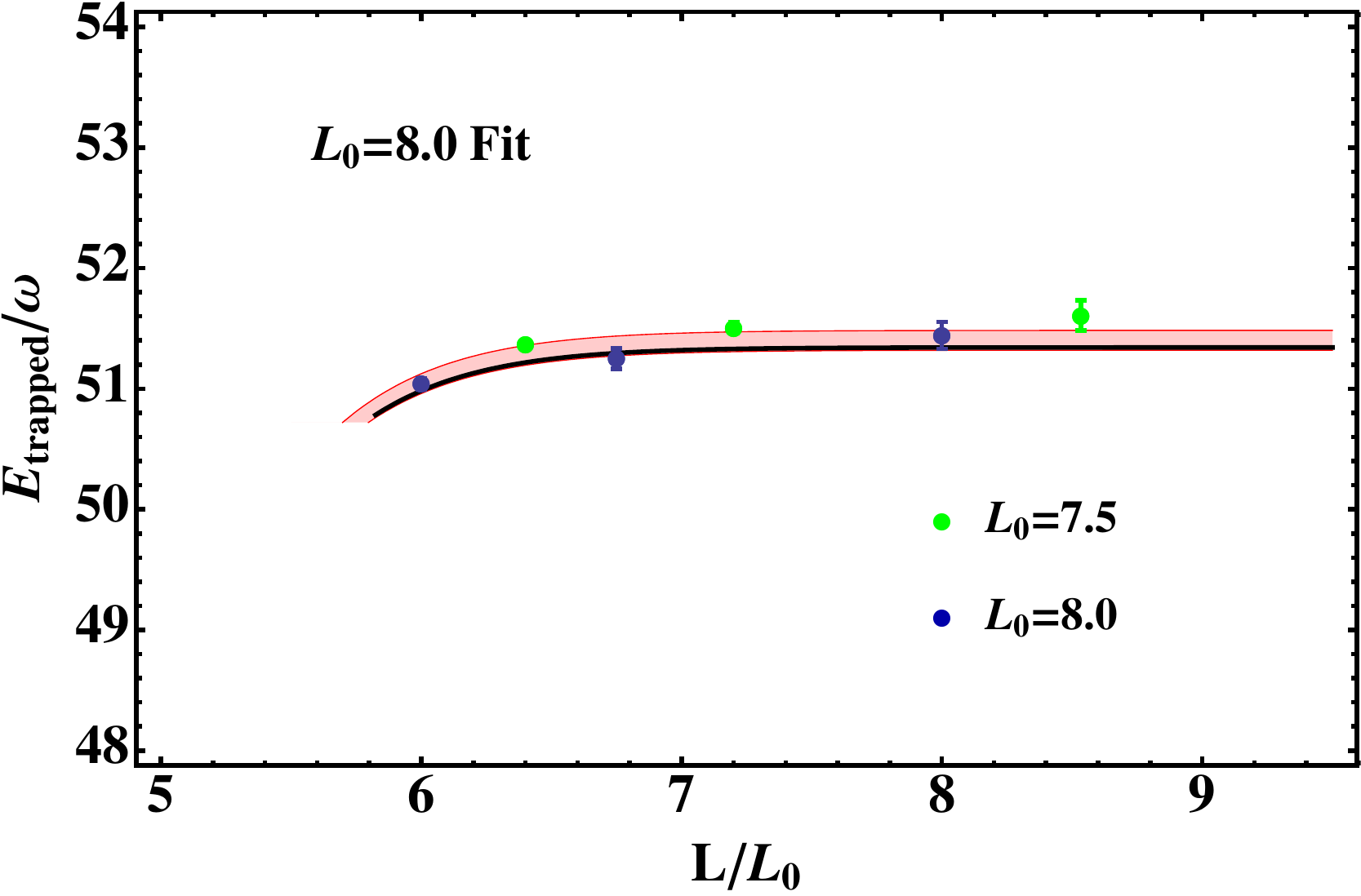}
\includegraphics[width=\figwidth]{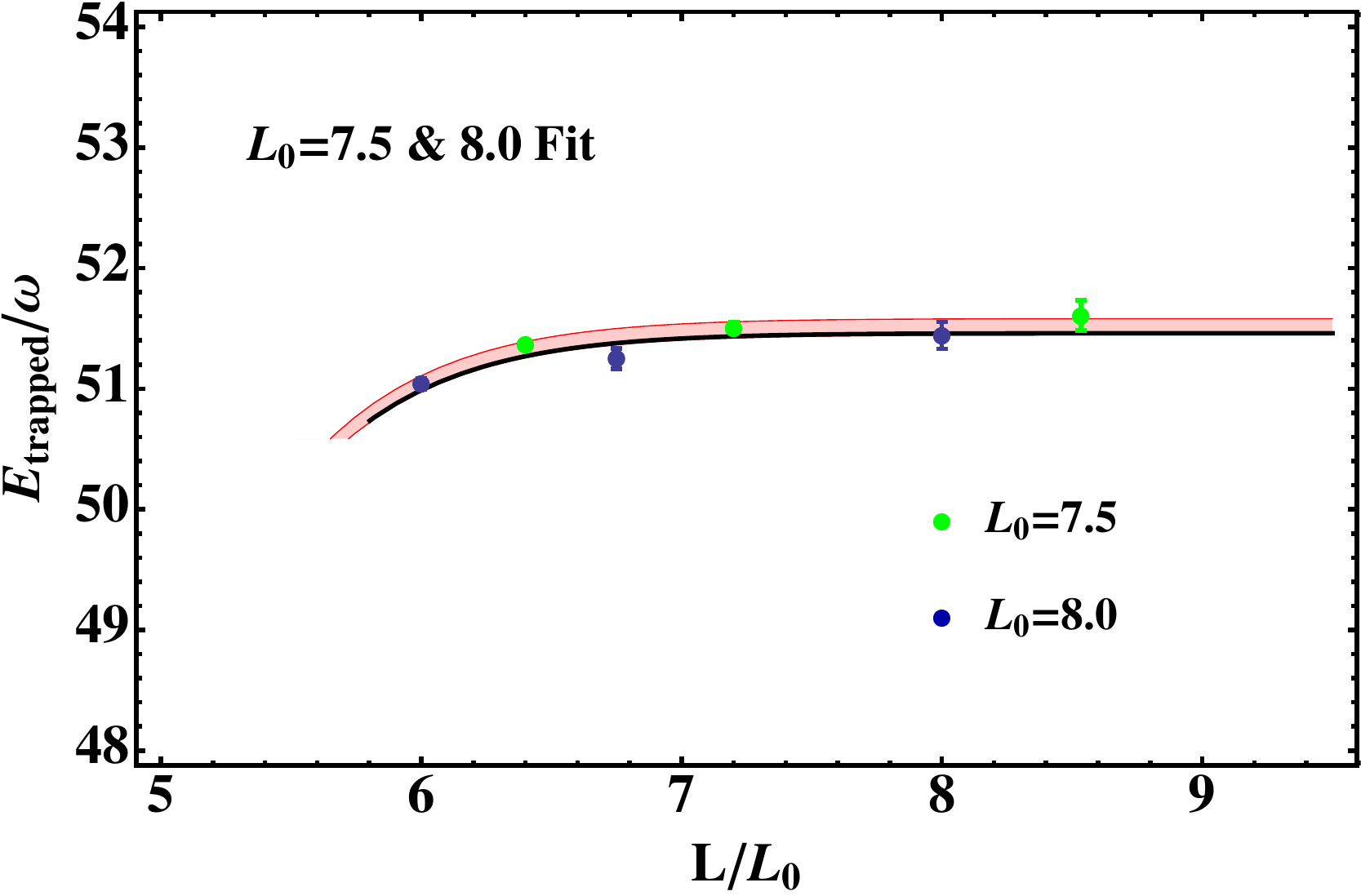}
\includegraphics[width=\figwidth]{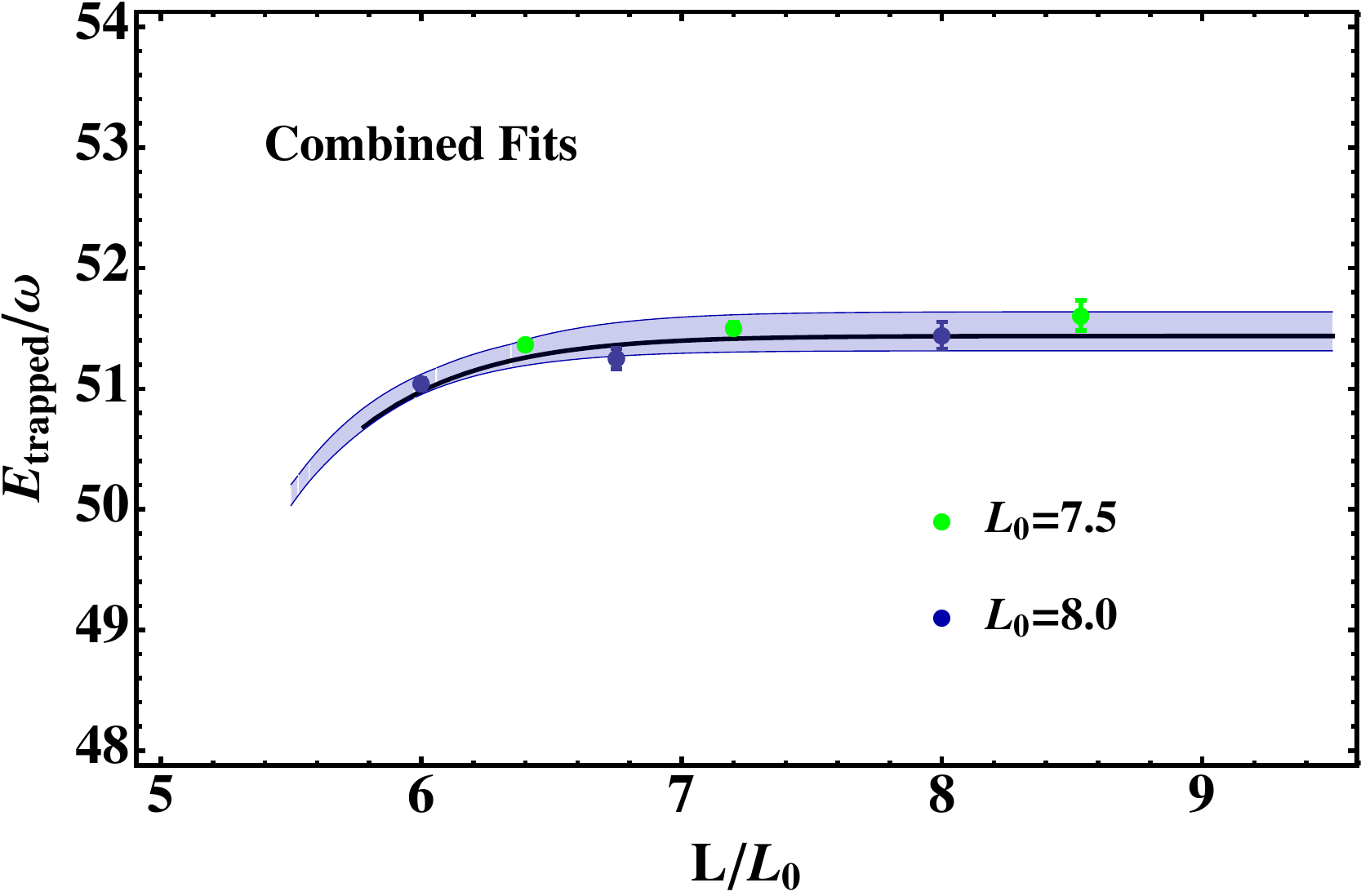}
\caption{%
\label{fig:extrap1}%
Volume dependence of the ground state energies (in units of $\omega$) for moderate $N$ ($N=24$).  The data points indicate the individual results for our six values of $L/L_0$. An infinite volume extrapolation is shown as a solid line, while a band represents the associated statistical and fitting systematic error bars of the extrapolation.
The upper plots show separate fits to the $L_0=7.5$ and $L_0=8.0$ points using \Eq{LoverL0_func}. The lower left plot shows a combined fit using both values of $L_0$, and the lower right plot shows the combined fit with a final error band obtained by combining the statistical and fitting systematic errors from all three extrapolations.
}
\end{figure}

\begin{figure}[h!]
\includegraphics[width=\figwidth]{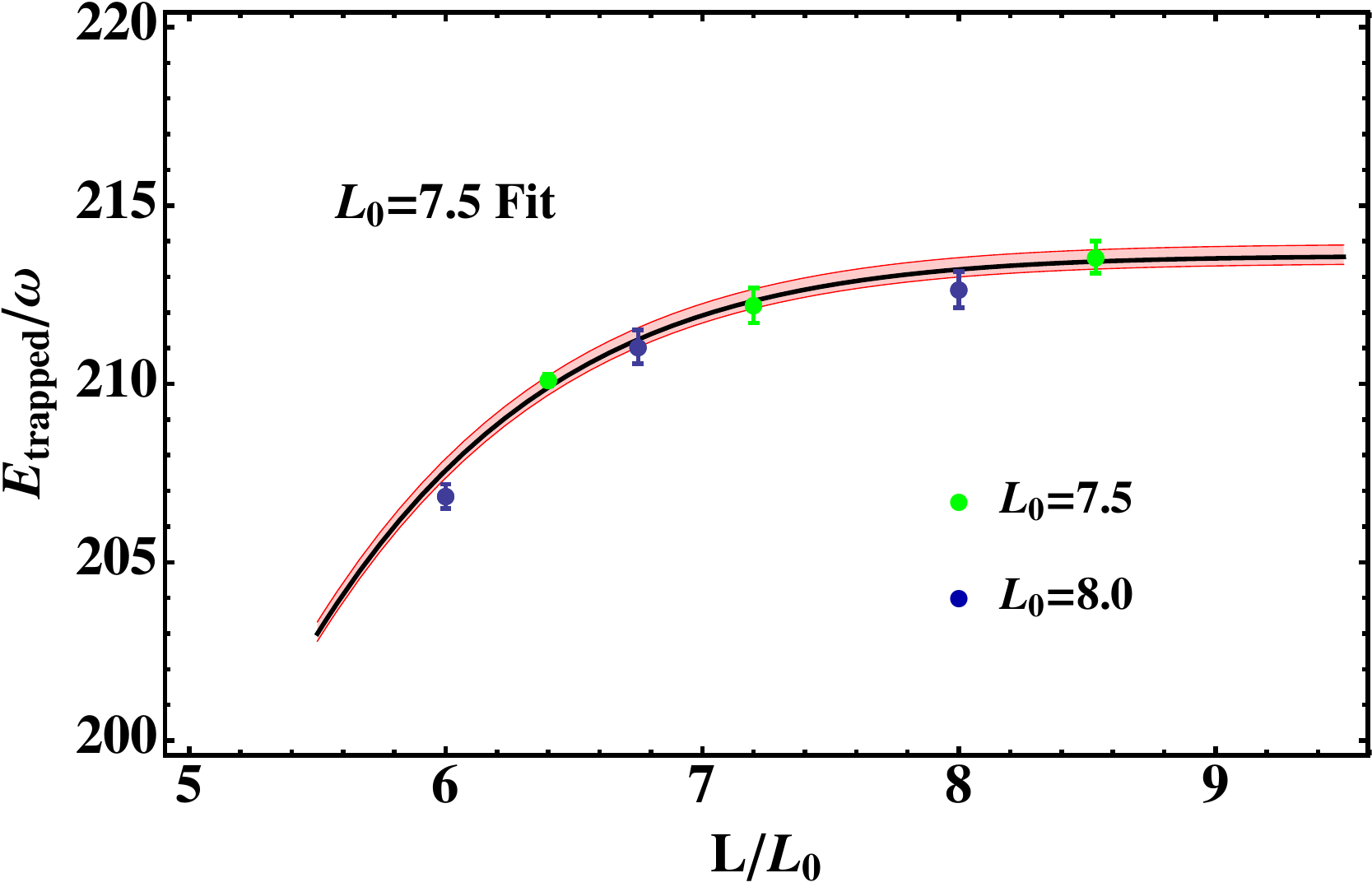}
\includegraphics[width=\figwidth]{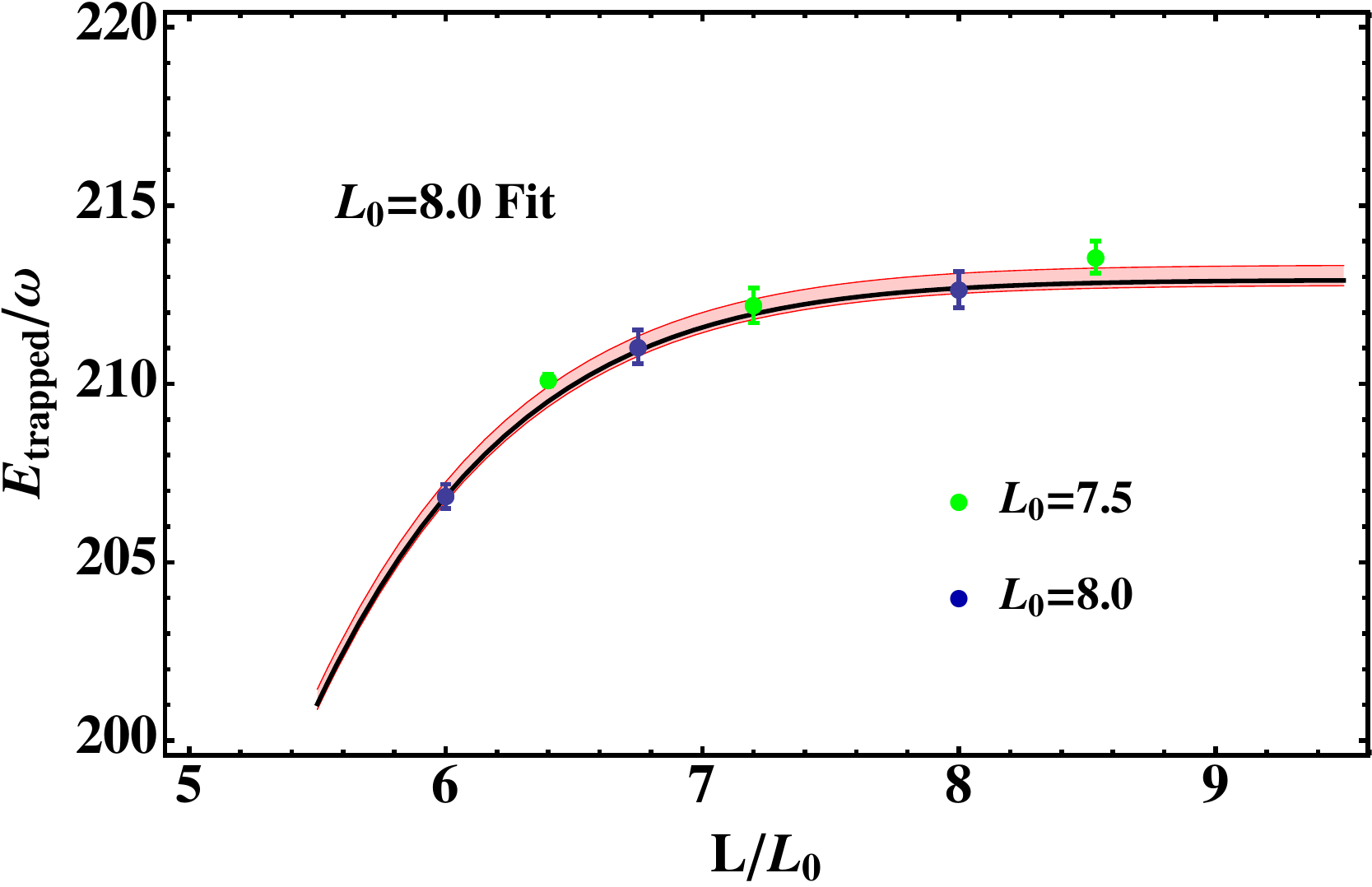}
\includegraphics[width=\figwidth]{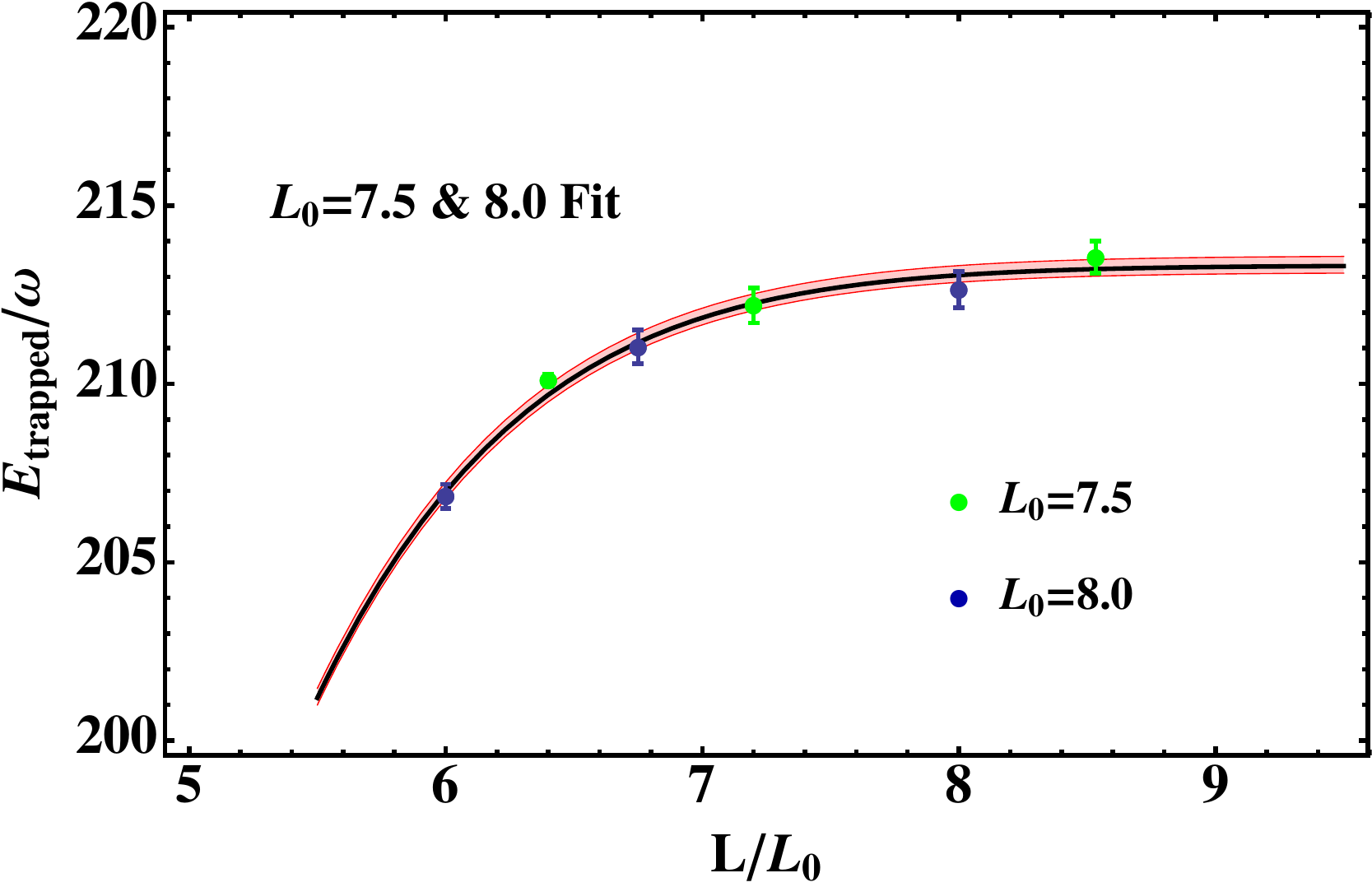}
\includegraphics[width=\figwidth]{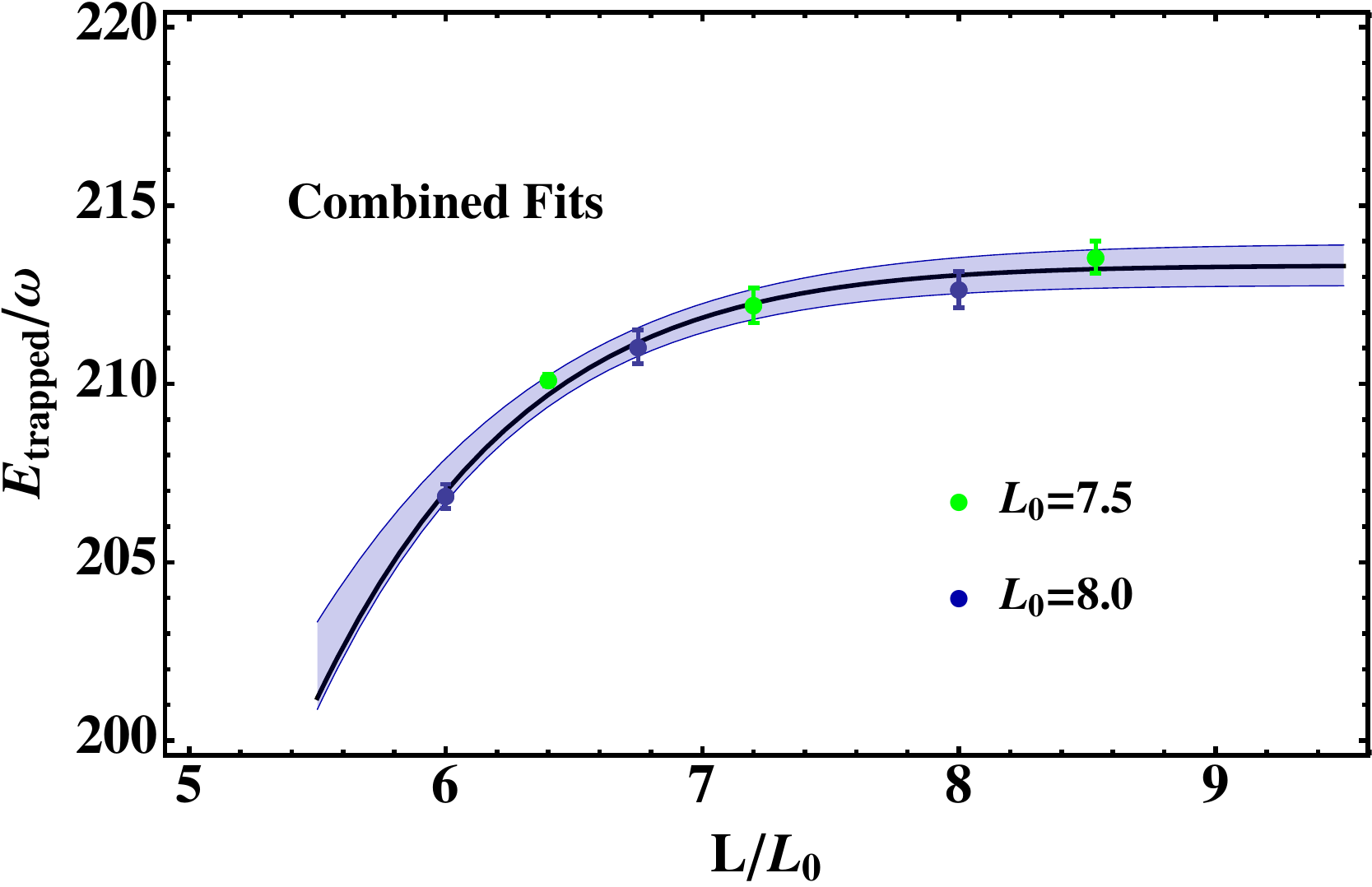}
\caption{%
\label{fig:extrap2}%
Volume dependence of the ground state energies (in units of $\omega$) for large $N$ ($N=70$).  The data points indicate the individual results for our six values of $L/L_0$. An infinite volume extrapolation is shown as a solid line, while a band represents the associated statistical and fitting systematic error bars of the extrapolation.
The upper plots show separate fits to the $L_0=7.5$ and $L_0=8.0$ points using \Eq{LoverL0_func}. The lower left plot shows a combined fit using both values of $L_0$, and the lower right plot shows the combined fit with a final error band obtained by combining the statistical and fitting systematic errors from all three extrapolations.
}
\end{figure}

To account for systematic errors arising from finite volume and lattice spacing effects, we have performed the calculation for three volumes ($L=48,54,64$) and at two values of the trap size ($L_0=7.5,8.0$).
We find that as more particles are added to the system, the discrepancies between results at different volumes grows.
The dependence on the lattice spacing is less clear, particularly due to the fact that changing $L_0$ changes not only the lattice spacing dependence ($b_s/L_0$), but also the finite volume dependence ($L_0/L$).
To separate these effects, an infinite volume extrapolation was performed for each value of $L_0$ using correlated fits of the data to
\begin{eqnarray}
f(L/L_0) = E_{\infty} \left(1-A e^{-B \left(L/L_0\right)^2}\right)
\label{eq:LoverL0_func}
\end{eqnarray}
over the plateau regions of each effective mass plot.
This form of the extrapolation function is a simplified version of the ansatz that finite volume errors depend on the probability,
\begin{eqnarray}
\calP(L/L_0) = \int_{L}^{\infty} | \psi(x/L_0) |^2 dx 
\end{eqnarray}
that the ground state wavefunction extends outside the box. 
We also make use of the fact that for unitary fermions, $\psi(x/L_0)$ is given asymptotically by a direct product of noninteracting harmonic oscillator wavefunctions.
We find that including wavefunctions from higher shells results in negligible change from the infinite volume extrapolations obtained using Gaussian fits.
These differences may ultimately be absorbed into the fitting coefficients $A,B$.

For the two different values of $L_0$ we find that the infinite volume extrapolations are consistent within error bars, indicating that spatial discretization errors are smaller than the combined statistical, fitting systematic, and infinite volume extrapolation errors.
For this reason, a third fit was also performed to all six data points simultaneously (see Figs.~\ref{fig:extrap1}, ~\ref{fig:extrap2}).
The spread between the three fits gives an approximation for any remaining spatial discretization errors.
For the final result, we added the statistical and fitting systematic errors from each fit in quadrature individually, and used the outer envelope to represent our total statistical, fitting systematic, extrapolation, and spatial discretization error.

\subsubsection{Final Results}
\label{sec:analysis_and_results.unitary_fermions_in_a_finite_trap.many-body_results.final_results}

Our results for the energies in units of $\omega$ and their corresponding errors are reported in \Tab{eall}. 
\begin{table}
\caption{%
\label{tab:eall}%
Ground state energies as a function of $N$ in units of $\omega$.
The error represents the combined statistical, fitting systematic, finite volume, and spatial discretization errors. See \Sec{analysis_and_results.unitary_fermions_in_a_finite_trap.many-body_results.additional_error} for possible additional systematic errors.
}
\begin{ruledtabular}
\begin{tabular}{|c|c|c|c|}
$N$ & $E_\text{trapped}/\omega$ &$N$ & $E_\text{trapped}/\omega$ \\
\hline
 4 & $5.071^{+0.032}_{-0.075}   $ & 38 & $94.34^{+0.33}_{-0.31}  $  \\
 6 & $8.347^{+0.080}_{-0.066}   $ & 40 & $100.50^{+0.26}_{-0.30} $ \\
 8 & $11.64^{+0.106}_{-0.124}   $ & 42 & $107.98^{+0.24}_{-0.33} $ \\
 10 & $16.05^{+0.031}_{-0.069}  $ & 44 & $115.41^{+0.31}_{-0.21} $ \\
 12 & $20.765^{+0.045}_{-0.093} $ & 46 & $122.94^{+0.36}_{-0.22} $ \\
 14 & $25.343^{+0.097}_{-0.081} $ & 48 & $130.45^{+0.38}_{-0.19} $ \\
 16 & $29.932^{+0.053}_{-0.093} $ & 50 & $137.98^{+0.39}_{-0.36} $ \\
 18 & $34.62^{+0.11}_{-0.08}    $ & 52 & $145.40^{+0.48}_{-0.17} $ \\
 20 & $39.31^{+0.11}_{-0.09}    $ & 54 & $152.97^{+0.46}_{-0.18} $ \\
 22 & $45.31^{+0.17}_{-0.10}    $ & 56 & $160.55^{+0.41}_{-0.29} $ \\
 24 & $51.44^{+0.20}_{-0.12}    $ & 58 & $168.16^{+0.42}_{-0.37} $ \\
 26 & $57.56^{+0.23}_{-0.13}    $ & 60 & $175.57^{+0.64}_{-0.31} $ \\
 28 & $63.65^{+0.25}_{-0.16}    $ & 62 & $183.16^{+0.53}_{-0.33} $ \\
 30 & $69.75^{+0.27}_{-0.12}    $ & 64 & $190.67^{+0.59}_{-0.06} $ \\
 32 & $75.89^{+0.31}_{-0.12}    $ & 66 & $198.19^{+0.64}_{-0.37} $ \\
 34 & $82.07^{+0.41}_{-0.31}    $ & 68 & $205.72^{+0.70}_{-0.26} $ \\
 36 & $88.05^{+0.46}_{-0.23}    $ & 70 & $213.26^{+0.68}_{-0.29} $ \\
\end{tabular}
\end{ruledtabular}
\end{table}
\begin{figure}[h!]
\includegraphics[width=0.6\linewidth]{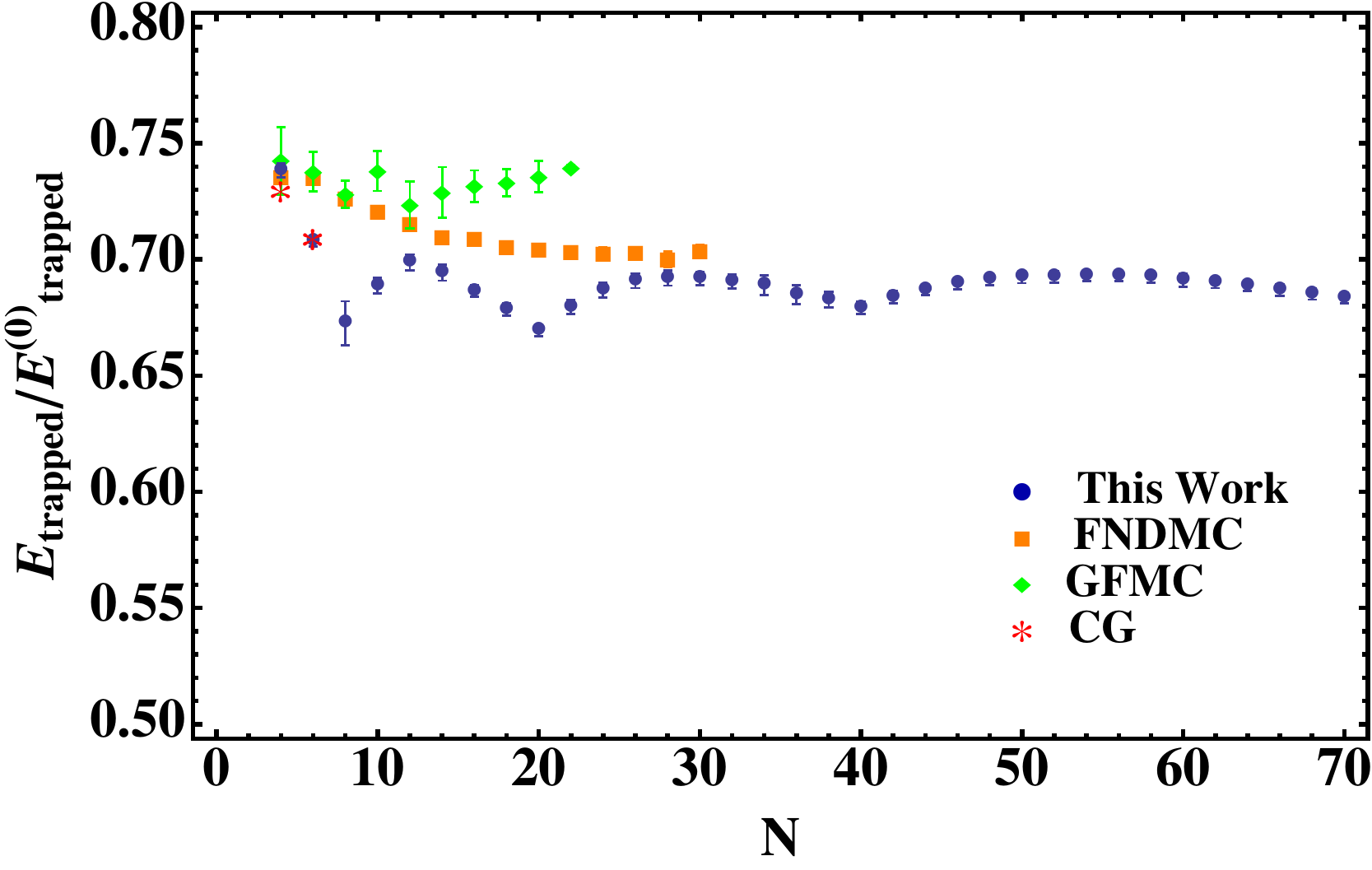}
\caption{%
Ground-state energies of $N$ trapped unitary fermions in units of the corresponding energies of $N$ trapped noninteracting fermions as a function of $N$. For comparison, we show results from GFMC \cite{Chang:2007zzd}, FN-DMC \cite{PhysRevLett.99.233201}, and CG \cite{Blume201186} methods.
}
\label{fig:bertschcomp}
\end{figure}
In \Fig{bertschcomp} we plot the results for the ground state energies in units of the energies for the corresponding noninteracting system, $E_\text{trapped}/E^{(0)}_\text{trapped}$.
For comparison, we also show the results from two fixed-node calculations: a Green's function Monte Carlo (GFMC) approach \cite{Chang:2007zzd} and a Diffusion Monte Carlo (FN-DMC) approach \cite{PhysRevLett.99.233201}.
By using the fixed-node constraint along with a variational principle, both of these methods provide upper bounds on the ground state energies.
We find that our energies are consistently lower than those obtained using both of these methods. Interestingly, fixed node calculations do not display the shell structure which is clearly present in our data. It is evident that this shell structure diminishes for large $N$, where eventually the thermodynamic limit should be reached.

\subsubsection{Possible additional sources of systematic error}
\label{sec:analysis_and_results.unitary_fermions_in_a_finite_trap.many-body_results.additional_error}

To calculate the error bars quoted in \Tab{eall} we have taken into account statistical, fitting systematic, extrapolation, and lattice errors. We note additionally that the spacing between the energy levels associated with breathing modes \cite{2006PhRvA..74e3604W}, $2 \omega b_\tau = 0.010$, is smaller than the inverse temporal extent of our lattice ($1/T \approx 0.017$), but larger than our quoted error bars. 
Furthermore, as an increasing number of particles are added to the system, a near continuum of different angular momentum states may result, also of $\mathcal{O}(\omega b_{\tau})$.

\begin{figure}[h!]
\includegraphics[width=\figwidth]{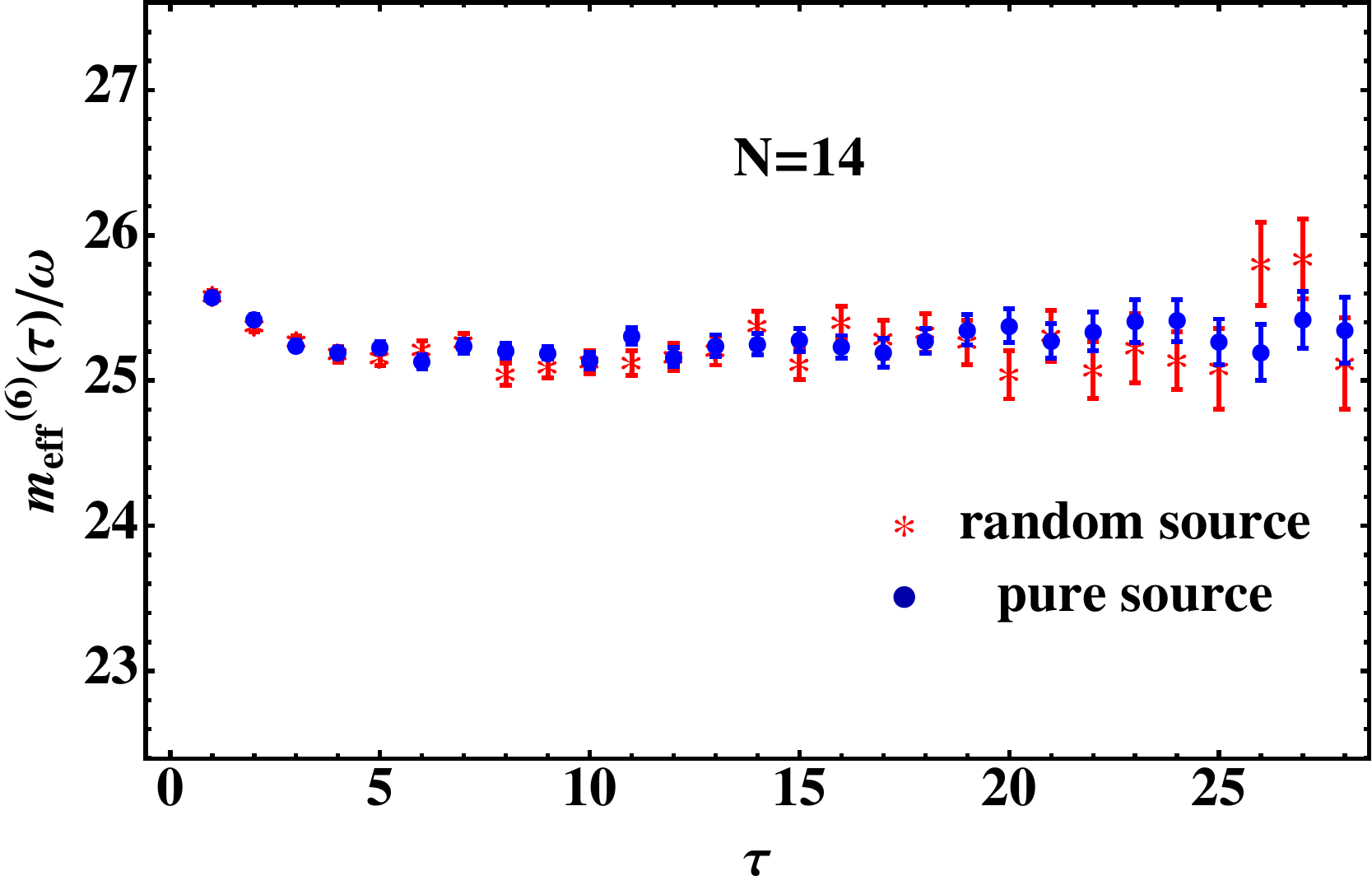}
\includegraphics[width=\figwidth]{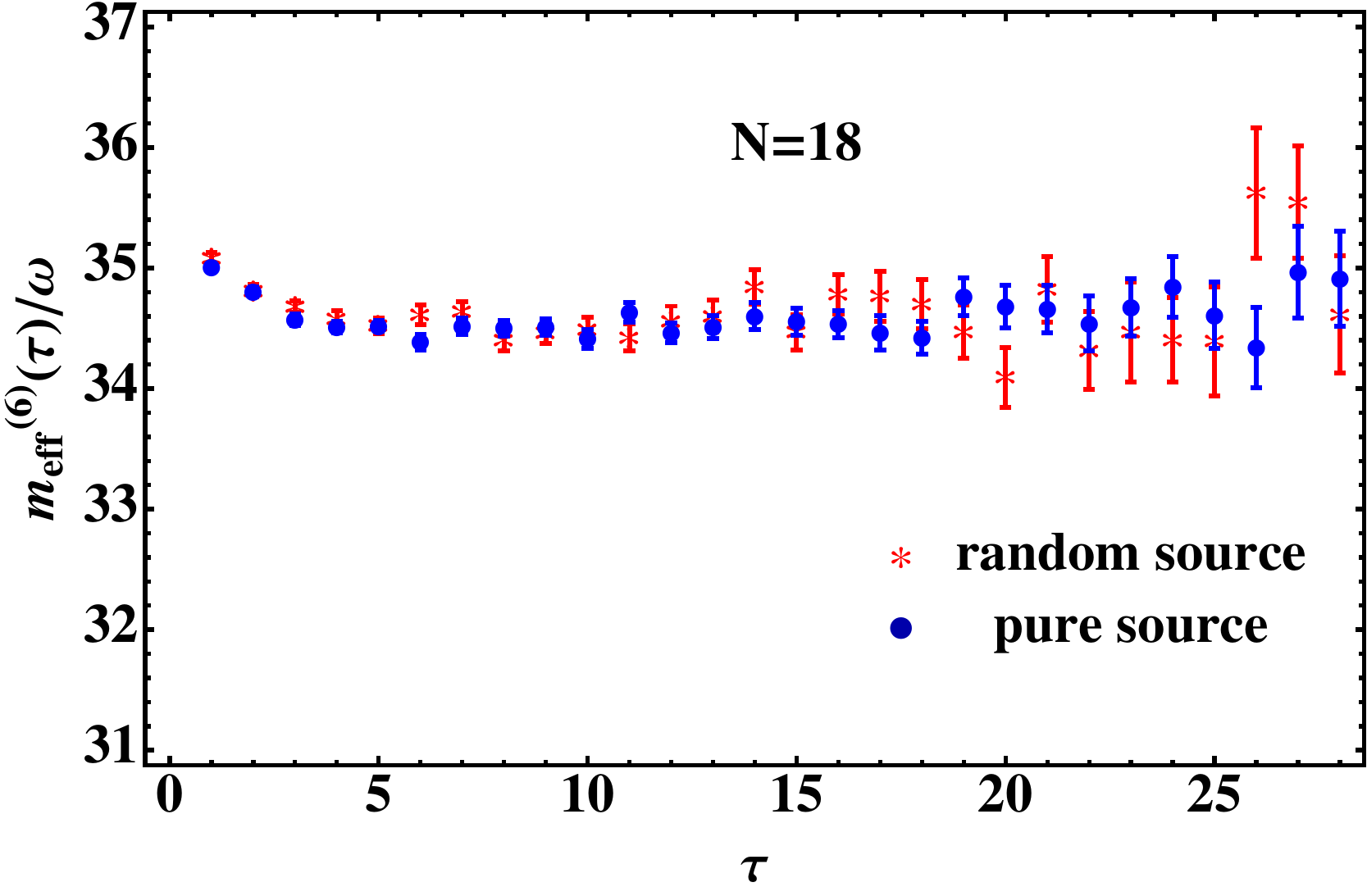}
\caption{%
\label{fig:2ndshell}%
Effective mass as a function of $\tau$ (lattice units) using two different sources for a half-filled shell ($N=14$) and a nearly closed shell ($N=18$) within the second shell ($L=48$, $L_0=8.0$).
The effective mass was calculated using the cumulant expansion with $N_\kappa=6$ using \Eq{cumulant_effm}.
The blue circles were generated using a source constructed by filling the single-particle states in the order given in \Tab{sources} while the red stars were generated using a source constructed of random linear combinations of the single-particle states within each shell.
}
\end{figure}

\begin{figure}[h!]
\includegraphics[width=\figwidth]{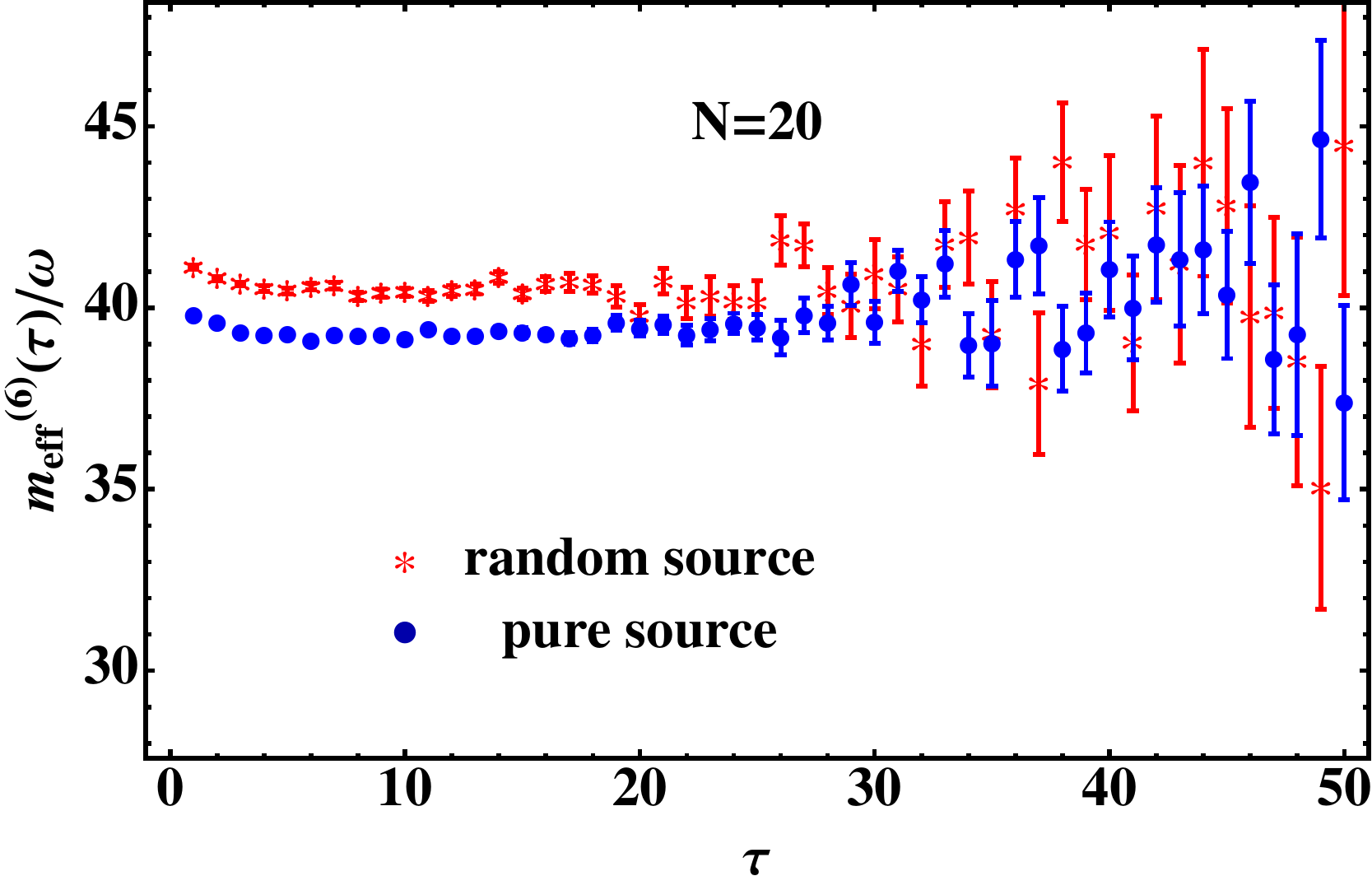}
\includegraphics[width=\figwidth]{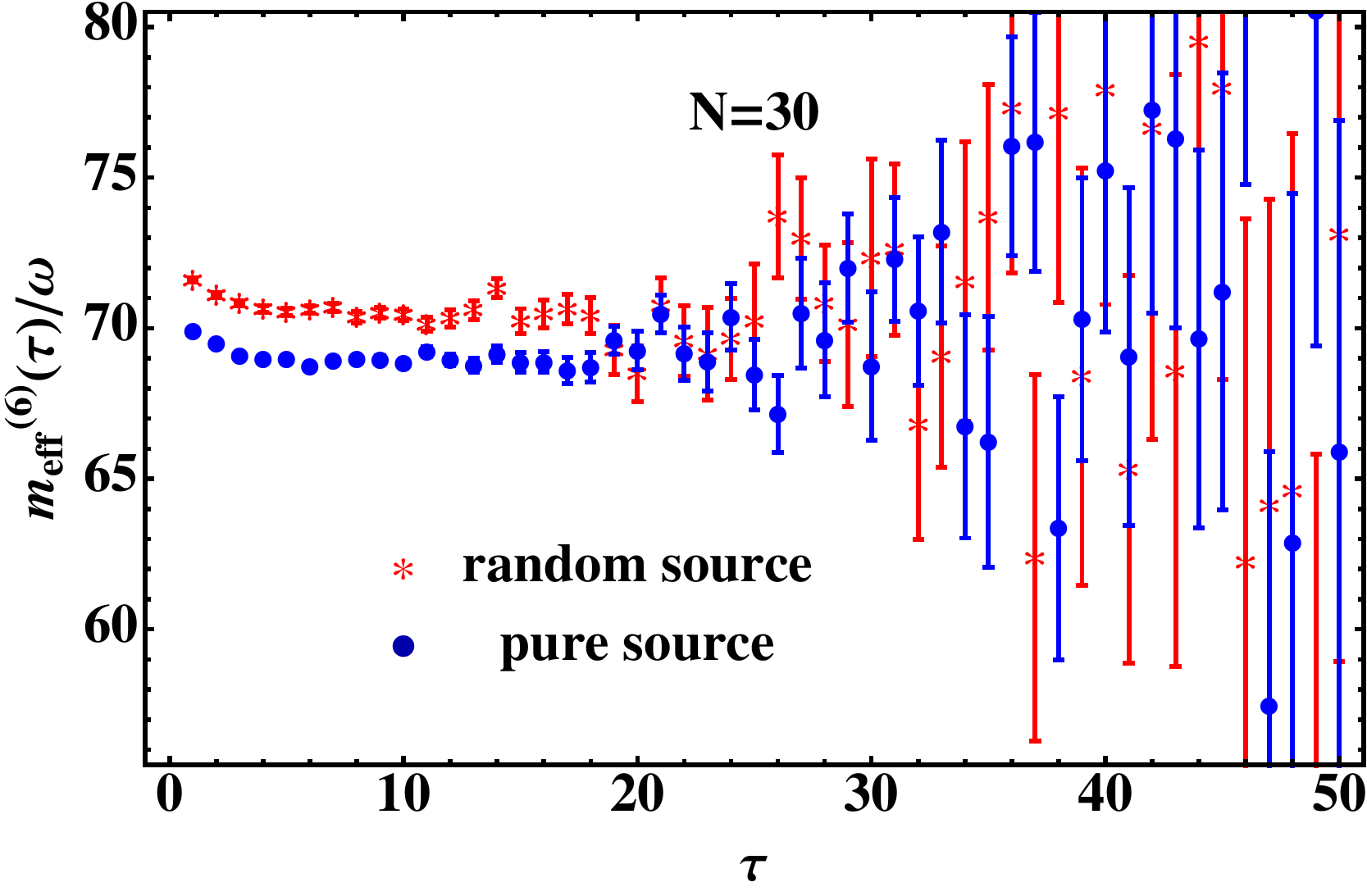}
\caption{%
\label{fig:3rdshell}%
Effective mass as a function of $\tau$ (lattice units) using two different sources for a closed shell ($N=20$) and a half-filled shell ($N=30$) within the third shell ($L=48$, $L_0=8.0$).
The effective mass was calculated using the cumulant expansion up to $N_\kappa=6$ using \Eq{cumulant_effm}.
The blue circles were generated using a source constructed by filling the single-particle states in the order given in \Tab{sources} while the red stars were generated using a source constructed of random linear combinations of the single-particle states within each shell.
}
\end{figure}

\begin{figure}[h!]
\includegraphics[width=\figwidth]{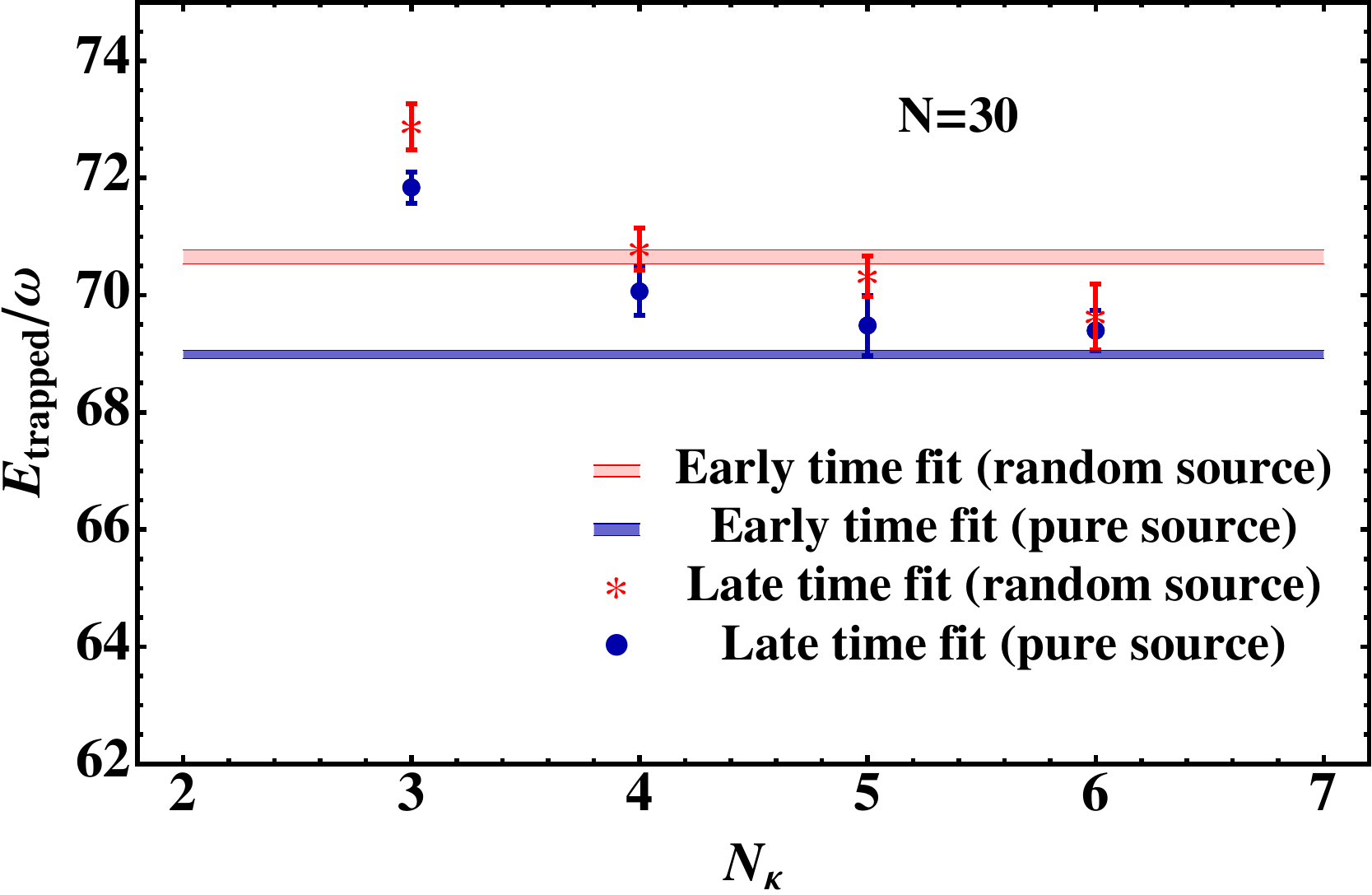}
\caption{%
\label{fig:compsourcefits}%
Fit results for the ground state energy using the two sources shown in \Fig{3rdshell} (right). The bands show results from fitting each source at early times ($\tau \sim 5$), while the data points show fit results as a function of the number of cumulants included for late times ($\tau \sim 30$). The late time fits from both sources are approaching the early time fit for the pure source, while the early and late time fits for the random source do not agree, indicating that the random source does not reach the ground state until later times.
}
\end{figure}

These excited state contributions could lead to systematic effects due to a failure to reach the ground state of the system. If excited state contamination is present in our results, it is possible that the overlap of our chosen sources and sinks (see Appendix ~\ref{sec:lattice_construction_observables}) with these excited states is shell dependent, causing our results to exhibit shell dependence even if this is not a property of the ground state. However, as noted in the beginning of \Sec{analysis_and_results}, we do not have any reason to believe we are near the thermodynamic limit, so it is quite conceivable that the shell structure we observe is a physical property of the ground state. 

The energy splittings are the same size for small $N$ as for large $N$, thus we might expect that if we are able to see the ground state within our time extent for small $N$, the same could be true for large $N$. Because our results for small $N$ agree with those from benchmark calculations, we can be assured that we have found the ground state in this case. This implies that the wavefunction overlap with excited states is very small.

To better quantify any possible effects from excited states, we may consider a correlation function whose long time behavior is dominated by two terms, the first corresponding to the ground state, the second to a breathing mode
\beq
\label{eq:excited}
\calC(\tau) \to Z_0 e^{-E_0 \tau} + Z_1 e^{-(E_0+2 \omega) \tau} \ ,
\eeq
where $Z_0$ and $Z_1$ represent the overlaps between our sources and sinks with the true ground and breathing mode states, respectively. Recall that the signs of $Z_0$ and $Z_1$ need not be positive due to the use of inequivalent sources and sinks. 

For large $N$ we typically find a plateau for time ranges $\tau \sim 5-30$. If we assume equal coupling of our sources to the ground state and breathing mode ($Z_0 = Z_1$), this would contribute to a drift in the effective mass plot of about $0.1 \omega$ for the time range considered. This is of approximately the same size as our statistical error bars in this region, so it is conceivable that such a drift would not be detected. 

One possible test to detect contamination from excited states is to repeat the calculation using a source that consists of random linear combinations of the states within each shell from \Tab{sources}. Using \Eq{excited}, one may show that the shift in the ground state energy for small $\omega \tau$ is approximately
\beq
E_0 + 2 \omega \frac{Z_1}{Z_0+Z_1} \ .
\eeq
If there is contamination from the second term and the new source changes the overlap with the excited state by at least $\mathcal{O}(1)$, this would give a shift in the effective mass plot of $\mathcal{O}(\omega)$ for all times considered ($\tau \leq 64$). 

We find that the effective mass plots produced using the random sources agree with those for our original (``pure") sources for $N$ within the first two shells (see \Fig{2ndshell}). For the third shell (\Fig{3rdshell}), the effective mass for the random source begins at higher values for both $N=20$ (closed shell) and $N=30$ (half-filled shell), however, the two sources begin to agree around $\tau \sim 30$.

By performing fits using the cumulant method, we find that in fact the random source plateaus at a later time than the pure source; results from fitting both sources at late times ($\tau \sim 30$) are consistent with each other (see \Fig{compsourcefits}). While the cumulant expansion converges too slowly at late times for us to extract a reliable ground state from these fits, it is clear that the results from both sources are approaching the early time ($\tau \sim 5$) fit for the pure source, giving us confidence in the energies extracted from this source. 

Thus, the random source test supports a lack of contamination from excited states in our quoted results. However, we do note that there is no guarantee that randomizing the source changes the $Z$-factors by at least $\mathcal{O}(1)$. Further analysis will be necessary to definitively establish this conjecture.

\section{Conclusion}
\label{sec:conclusion}

We have developed a new lattice method for studying large numbers of fermions at unitarity. The action is highly improved, so that our results require no extrapolation to zero range. In addition, we've applied a new method for calculating correlators from long-tailed distributions \cite{Endres:2011jm}, through which we are able to evade costly importance sampling. Our results agree with those from high precision solutions to the Schr\"odinger equation for $N \leq 6$ trapped fermions  \cite{Blume201186}, as well as with the energy of $N=3$ untrapped fermions calculated by Pricoupenko and Castin \cite{2007JPhA...4012863P}. Due to the low cost of the simulation we are able to then study up to $N=70$ trapped fermions, finding lower values than published results.  One feature we find is that shell effects persist at the $\sim 2\%$ above $N=40$ fermions, making it impossible to extract a reliable value for the Bertsch parameter $\xi$.  The shell effects we find are much more pronounced than what we see for untrapped fermions \cite{Endres:2011tba}.  
 
 In a future work we will present results for the homogeneous system of up to $N=66$ unitary fermions in a box, including our extraction of the Bertsch parameter, $\xi$, as well as data on the superfluid gap and integrated contact density for unitary fermions in a box. We believe this method could be applicable for a wide variety of nonrelativistic many-body systems, and these studies of unitary fermions, in addition to their inherent value, pave the way for investigations of more complex systems at zero temperature.

\acknowledgments

We have profited from discussions and communications with many people, including A. Bulgac, J. Carlson, Y. Castin, M. Forbes, S. Gandolfi, A. Gezerlis, M. Savage, D. Son, S. Tan.

This research utilized resources at the New York Center for Computational Sciences at Stony Brook University/Brookhaven National Laboratory which is supported by the U.S. Department of Energy under Contract No. DE- AC02-98CH10886 and by the State of New York.
Computations for this work were also carried out in part on facilities of the USQCD Collaboration, which are funded by the Office of Science of the U.S. Department of Energy.
This work was supported by U. S. Department of Energy grants DE-FG02-92ER40699 (M.G.E.) and DE-FG02-00ER41132 (D.B.K., J-W. L. and A.N.N.).
M.G.E is supported by the Foreign Postdoctoral Researcher program at RIKEN.

\appendix

\section{Tuning}
\label{sec:tuning}

The two particle transfer matrix is a linear function of the two-body couplings $C_{2n}$:
\begin{eqnarray}
\calT(\bfC) = \calT_{free} + \sum_{n=0}^{N_{\calO}-1} C_{2n} \calT_{2n}\ ,
\end{eqnarray}
where $\calT_{free}$ is the free fermion transfer matrix, and $\calT_{2n}$ contains contributions to the interaction $C$ starting at order $2n$ in momenta.
Eigenvalues $\lambda_k(\bfC)$ of this transfer matrix, however, are nonlinear functions of the couplings.
We may compute the derivative of these eigenvalues with respect to the couplings using the Feynman-Hellman theorem:
\begin{eqnarray}
W_{kn}(\bfC) \equiv \frac{\partial \lambda_k}{ \partial C_{2n} } = \langle \psi_k | \calT_{2n} | \psi_k \rangle\ ,
\end{eqnarray}
where $\calT(\bfC)  | \psi_k \rangle = \lambda_k(\bfC)  | \psi_k\rangle$ for $k=1,\ldots, \textrm{dim}(\calT)$ and the eigenstates $ | \psi_k \rangle$ implicitly depend on $\bfC$.
Tuned values for the $N_{\calO}$ couplings are defined as the values $C_{2n}$ ($n=0,\ldots,N_\calO-1$) for which $\chi^2(\bfC) = 0$, where
\begin{eqnarray}
\chi^2(\bfC) = \sum_{k=1}^{N_{\calO}} \delta\lambda_k(\bfC)^2
\end{eqnarray}
and $\delta\lambda_k(\bfC) = \lambda_k(\bfC) /\lambda^*_k-1$.
Starting from an initial guess for the couplings $C^0_{2n}$, we may iteratively search for the solution to $\chi(\bfC)=0$ using:
\begin{eqnarray}
\bfC^{r+1} = \bfC^r + \epsilon \tilde W(\bfC^r)^{-1} \delta\lambda(\bfC^r)\ ,
\end{eqnarray}
provided the inverse of the $N_{\calO}$-dimensional sub-matrix $\tilde W$ exists, where $\tilde W_{kn} = W_{kn}$ for $n=0,\ldots,N_{\calO}-1$ and $k=1,\ldots,N_{\calO}$.
The small parameter $\epsilon$ may be chosen adaptively in order to improve the convergence of the iterative procedure.

\section{Correlation Functions}
\label{sec:lattice_construction_observables}

%\section{Correlation Functions}
%\label{sec:lattice_construction.observables}

Multi-fermion sources may be constructed from direct products of single particle states $|\alpha^\sigma_i \rangle$, where $i=1,\ldots,N_\sigma$ labels each state with quantum number $\alpha$ and $\sigma = (\uparrow, \downarrow)$ labels the species.
In order to satisfy Fermi-Dirac statistics, fermions of the same species must have different quantum numbers.
As is well-known from quantum mechanics, a simple way to impose the proper anti-symmetrization requirements on multi-fermion states is to use Slater-determinants.
Thus correlation functions of $N = N_\uparrow + N_\downarrow$ fermions may be expressed as:
\begin{eqnarray}
\calC_{N_\downarrow, N_\uparrow}(\tau) = \langle \det{ S^{\downarrow}(\tau) } \det { S^{\uparrow}(\tau) } \rangle\ ,
\end{eqnarray}
where $S^\sigma$ is an $N_\sigma$-dimensional Slater matrix corresponding to the species $\sigma$, given by
\begin{eqnarray}
S^\sigma_{i,j}(\tau) = \langle \alpha^\sigma_{i} | K^{-1}(\tau,0) | \alpha^\sigma_{j} \rangle\ .
\label{eq:slater}
\end{eqnarray}
Although it is not a requirement, a convenient choice for the single particle states $|\alpha_i^\sigma \rangle$ is to use eigenstates of the noninteracting system.
For trapped fermions, they are SHO states ($\alpha=\bfn$) in the Cartesian basis.
A list of the sources used in our simulations is provided in \Tab{sources}.

\begin{table}
\centering
\caption{\label{tab:sources} Single fermion sources used in trapped ($\alpha=\mathbf{n}$) fermion calculations. }
\begin{tabular}{|c|ccc|c|ccc|}
\hline 
i &  $\mathbf{n}^\uparrow_i=\mathbf{n}^\downarrow_i$ & shell & deg & i & $\mathbf{n}^\uparrow_i=\mathbf{n}^\downarrow_i$ & shell & deg \\ \hline 
1 & (  0,  0,  0) & 0 &  1 &21  & (  0,  0,  4) & 4 & 15\\ \cline{2-4}
2& (  0,  0,  1) & 1 &  3 &22  & (  0,  1,  3) &   & \\ 
3 & (  0,  1,  0) &   &  &23  & (  0,  2,  2) &   &   \\
4  & (  1,  0,  0) &   &  &24    & (  0,  3,  1) &   &   \\             \cline{2-4}
5& (  0,  0,  2) &  2 & 6 &25   & (  0,  4,  0) &   &    \\
6  & (  0,  1,  1) &   &  &26  & (  1,  0,  3) &   &   \\
7  & (  0,  2,  0) &   & &27  & (  1,  1,  2) &   &   \\ 
8  & (  1,  0,  1) &   & &28  & (  1,  2,  1) &   &  \\
9 & (  1,  1,  0) &   &  &29   & (  1,  3,  0) &   &   \\    
10   & (  2,  0,  0) &   &  &30   & (  2,  0,  2) &   & \\    \cline{2-4}     
11 & (  0,  0,  3) & 3 & 10 &31  & (  2,  1,  1) &   & \\
12  & (  0,  1,  2) &   &  &32  & (  2,  2,  0) &   &    \\
13   & (  0,  2,  1) &   &  &33  & (  3,  0,  1) &   &    \\
14  & (  0,  3,  0) &   & &34  & (  3,  1,  0) &   &    \\
15  & (  1,  0,  2) &   &   &35  & (  4,  0,  0) &   &   \\  \cline{6-8}
16  & (  1,  1,  1) &   &   &&&& \\
17  & (  1,  2,  0) &   &    &&&&\\
18  & (  2,  0,  1) &   &    &&&&\\
19  & (  2,  1,  0) &   &    &&&&\\
20 & (  3,  0,  0) &   &    &&&&\\    
\hline
\end{tabular}
\end{table}

Typically multi-particle sources constructed from single particle states possess poor overlap with the unitary Fermi gas ground state.
This may easily be seen from the fact that at early times, where few interactions have occurred, the correlation function falls off exponentially like that of free fermions with a Z-factor near unity.
A better approach is to incorporate pairing correlations into the interpolating field by constructing sources and sinks out of two-fermion wave functions \cite{PhysRevLett.91.050401}.
In practice, such an approach may only be carried out at the sink, however, because our numerical approach requires that sources be separable functions; this is none-the-less adequate to achieve far superior overlap with the ground state.
A consequence of using sources and sinks that differ is that correlation functions and effective masses need not be monotically decreasing functions of time.
Thus when studying correlators of this form, care must be taken to distinguish shallow local minima in effective masses from true plateaus.

For $N^\uparrow = N^\downarrow = N/2$, these considerations lead us to study correlation functions of the form:
\begin{eqnarray}
\calC_{N_\downarrow,N_\uparrow}(\tau) = \langle \det{ S^{\downarrow\uparrow}(\tau)} \rangle\ ,
\label{eq:slater2}
\end{eqnarray}
where
\begin{eqnarray}
S^{\downarrow\uparrow}_{i,j}(\tau) = \langle \Psi | K^{-1}(\tau,0) \otimes  K^{-1}(\tau,0) | \alpha^\downarrow_i  \alpha^\uparrow_j \rangle
\label{eq:propagator2}
\end{eqnarray}
and $| \alpha^\downarrow \alpha^\uparrow \rangle = | \alpha^\downarrow \rangle \otimes | \alpha^\uparrow \rangle $.
In the coordinate basis, we consider two-fermion states $|\Psi\rangle$ of the form $\langle \bfx^{\downarrow} \bfx^{\uparrow} | \Psi \rangle = \Psi(\bfr_{rel})$ where $\bfr_{rel} = \bfx^\downarrow - \bfx^\uparrow$ is the relative coordinate of the two fermions.
It is helpful to express the two-particle wave functions as a Fourier transform: $\Psi(\bfr_{rel}) = \int d\bfp \tilde\Psi(\bfp) e^{- \bfp\cdot\bfr_{rel}}$, allowing \Eq{propagator2} to be written as 
\begin{eqnarray}
S^{\downarrow\uparrow}_{i,j}(\tau) = \sum_{\bfp} \tilde\Psi(\bfp) \langle \bfp | K^{-1}(\tau,0) | \alpha^\downarrow_i \rangle \langle -\bfp | K^{-1}(\tau,0) |\alpha^\uparrow_j \rangle\ .
\end{eqnarray}
Since the projection onto the sink involves only a single sum over momenta, evaluation of \Eq{propagator2} scales like $\calO(L^3)$ rather than the usual $\calO(L^6)$,

Numerical evidence suggests that the best choice for $\Psi(\bfr_{rel})$ is a lattice approximation to the two-particle s-wave solution to the continuum Schr\"odinger equation for unitary fermions, which possess a $1/|\bfr_{rel}|$ singularity.
We therefore consider a momentum space wave-function of the form
\begin{eqnarray}
\tilde\Psi(\bfp) = 
\frac{2\beta}{|\bfp|} d \left(\frac{  |\bfp|}{  2\beta } \right)\ ,
\label{eq:trapped_pair_source} % Keep both of these unless you are sure they aren't used!
\end{eqnarray}
where $d(x)$ is Dawson's integral function.
Note that the wave-function has a free parameter $\beta$ which may be tuned to maximize the overlap with the ground state.
Physically one expects $\beta \sim 1/\sqrt{2}L_0$, and this is what we use in practice.

For odd numbers of fermions, such as in our few-body studies, one may construct a mixed matrix built out of both single and two fermion wave functions.
In the case $N^\downarrow = N^\uparrow+1$, one may may construct such a Slater matrix by replacing row $i$ of $S^{\downarrow\uparrow}$ with the same row of $S^{\downarrow}$.
This replacement corresponds to a removal of the i-th single fermion state $\alpha^\uparrow_i$ from the source and thus also breaking the pair involving state $i$ at the sink.

\section{Measurement Strategy}
\label{sec:measurement_strategy}

Our studies have shown that for large numbers of fermions, the effective mass, 
\begin{eqnarray}
m_{eff}(\tau) = \frac{1}{\Delta \tau} \log{ \left[ \frac{ \calC(\tau) }{  \calC(\tau+\Delta\tau) } \right]   }\ ,
\label{eq:generalized_effm}
\end{eqnarray}
obtained from correlators measured in the conventional way, as a sample average of propagators measured on background $\phi$ configurations, often exhibits a distribution overlap problem.
Several indicators for this problem include: 1) lack of $1/\sqrt{N_{conf}}$ scaling in the error bars for correlators and measured quantities derived from them, 2) sudden jumps in estimated quantities and their associated error bars as a function of the sample size $N_{conf}$ and 3) continued growth of effective masses at late times, with no evidence for a plateau.
\Fig{overlap_problem} provides a mild demonstration of the third case for $N=4$ untrapped unitary fermions; plotted is the logarithm of the correlation function $\bar \calC(\tau)$ which has been estimated using ensembles of size ranging from $N_{conf} = 0.06M$ to $3.84M$ configurations.
An estimate of the effective mass at late times, quantified by the slope of $-\log{\bar \calC(t)}$ in this figure, appears to decrease as the size of the ensemble is increased.
As the number of configurations in the ensemble is increased by several orders of magnitude, $\bar \calC(\tau)$ eventually appears to converge to what is expected to be the true value of the correlator, indicated by the dashed line and estimated using a much larger ensemble of size $N_{conf}=2B$ configurations.

\begin{figure}
\includegraphics[width=\figwidth]{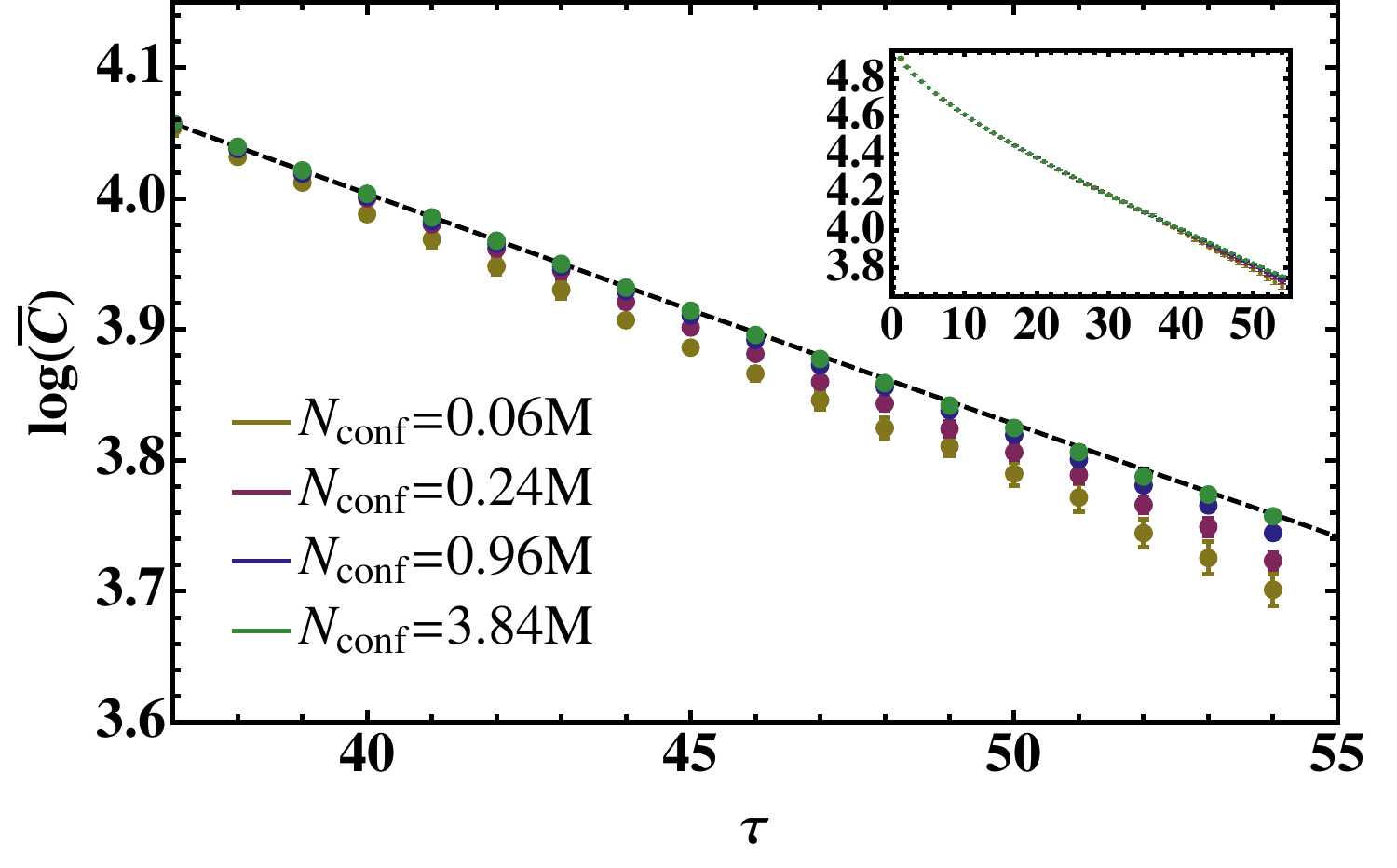}
\caption{%
\label{fig:overlap_problem}%
Natural logarithm of the N=4 fermion correlation function for untrapped unitary fermions of mass $M=5$ on an $L=10$ lattice as a function of sample size.
Dashed line indicates the result obtained using $N_{conf}=2B$ configurations.
}
\end{figure}

In order to understand this behavior better, it is instructive to study the distribution of the correlator as a function of $\tau$.
The distribution of an arbitrary operator $Y(\phi)$ measured on a background field configuration $\phi$ is given by:
\begin{eqnarray}
P(y) = \int [d\phi]\rho(\phi) \delta(Y(\phi)-y )\ .
\label{eq:corr_dist}
\end{eqnarray}
A plot of the correlator distribution, taking $Y(\phi)= \calC_\phi(\tau)$ and $y=c$, is shown in \Fig{correlator_distribution} for $N=4$ fermions at several values of $\tau$, and demonstrates the formation of a long tail in the late time limit.
Also shown is a corresponding plot of the distribution for the logarithm of the correlation function, taking $Y(\phi)=\log \calC_{\phi}(\tau)$ and $y = \log c$, along with the results from a Gaussian fit to the histograms.
The excellent agreement between the fit results and the measured log-correlator distribution suggests that the multi-fermion correlation function is log-normally distributed, or nearly so.
Such distributions are known to possess very long tails which dominate the distribution mean, and undersampling the tail can result in an underestimate in the correlation function, and thus an overestimate in the energy obtained from \Eq{generalized_effm} at large times, as was evident in \Fig{overlap_problem}.

\begin{figure}
\includegraphics[width=\figwidth]{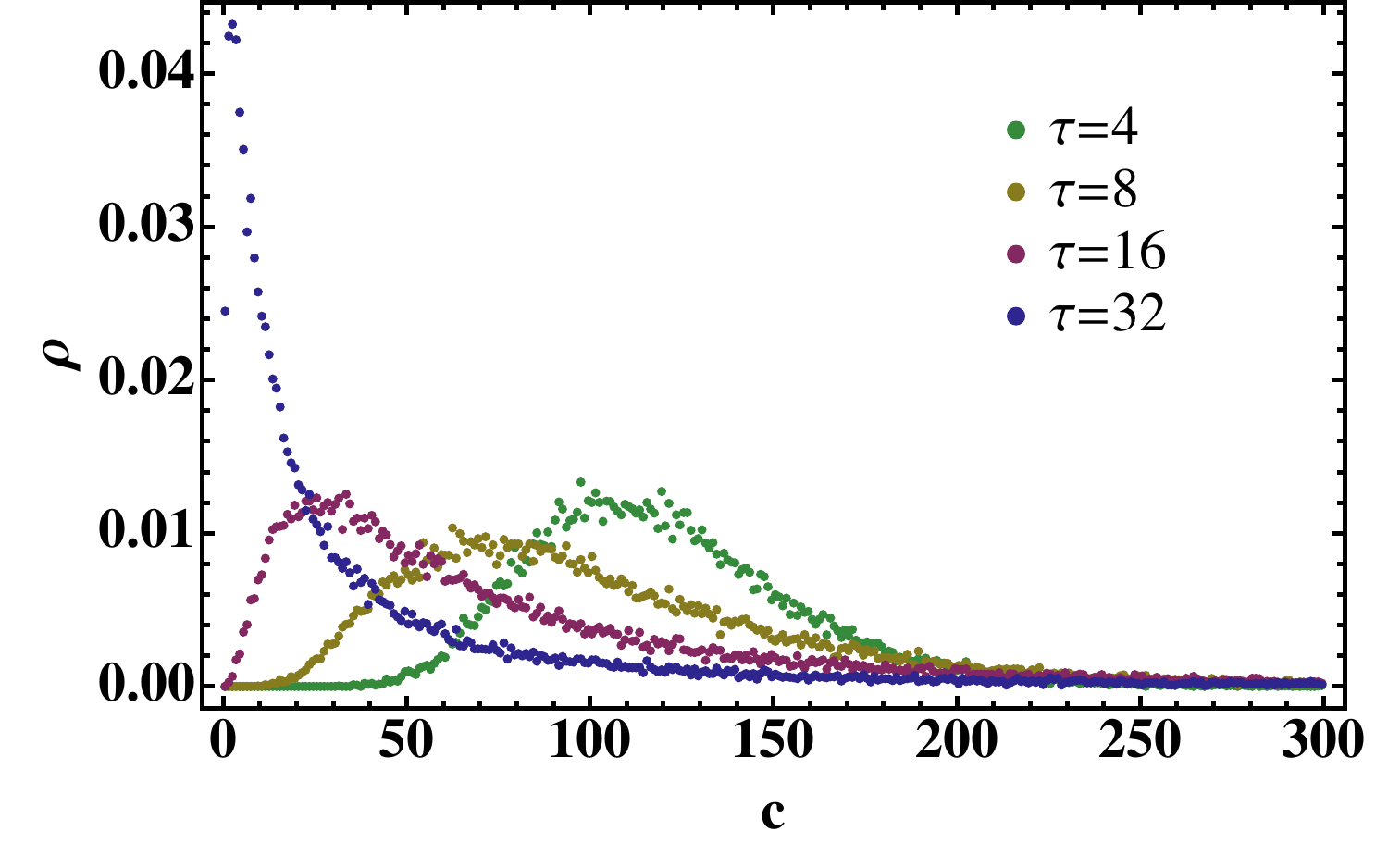}
\includegraphics[width=\figwidth]{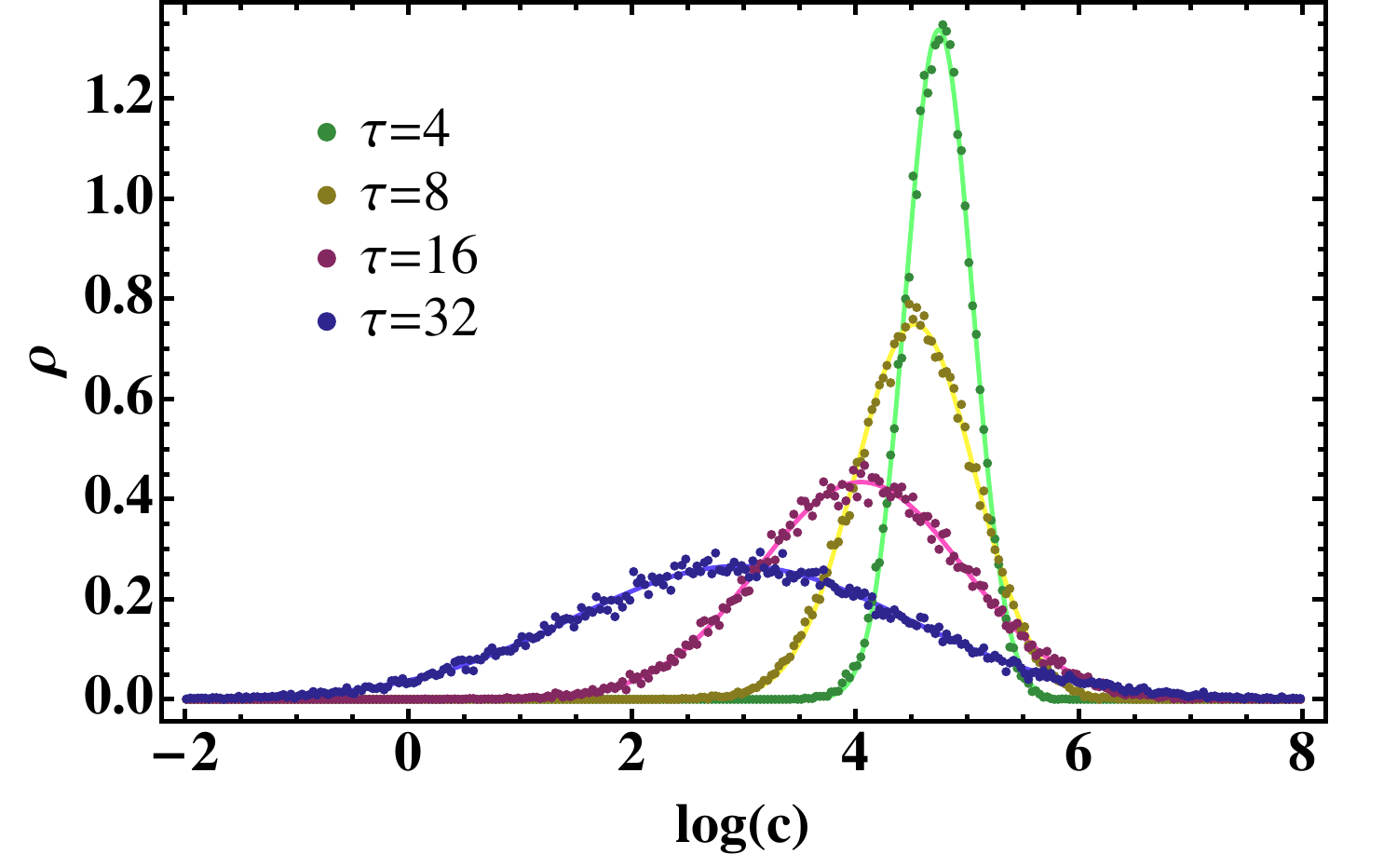}
\caption{%
\label{fig:correlator_distribution}%
$N=4$ fermion correlator and natural log-correlator distributions at various time separations $\tau$ for unitary fermions of mass $M=5$ on an $L=10$ lattice.
Solid curves in the log-correlator distribution plot correspond to Gaussian fits to the distribution.
}
\end{figure}

Provided we know the underlying distribution for the correlation function, we may estimate the number of configurations required such that the sample average $\bar \calC(\tau)$ is normally distributed.
Deviations of the sample mean from the normal distribution may result in an overlap problem and reflect the fact that the sample size is too small for the central limit theorem to apply.
In particular, if $\sigma$ and $\rho$ are the second and third central moments of the correlator distribution function, then by the Berry-Esseen theorem, one should show that the condition $N_{conf}\ll\rho^2/\sigma^3$ holds before invoking the central limit theorem.
This condition comes from quantifying the deviation in the cumulative distribution function for $\left( \bar \calC - \langle \calC \rangle \right) \sqrt{N_{conf}}/\sigma$ from that of the standard normal distribution, where $\bar \calC$ is the sample mean of the correlation function obtained from a sample size $N_{conf}$.
An example of the cumulative distribution function of this quantity for $N=4$ unitary fermions at several times is shown in \Fig{overlap_CDF}, and was obtained from 20K ensembles each of sample size $N_{conf}=100K$.
The true mean $\langle \calC \rangle$ was estimated using a sample of size $N_{conf}=2B$ configurations.
In this example, we find that for $\tau=24$ there is little deviation from the standard normal cumulative distribution function, whereas for $\tau=36$, significant deviation is evident.

In \cite{Endres:2011jm}, it was shown within mean field theory that the log-correlator distribution function defined by \Eq{corr_dist} is Gaussian with mean $\bar y=\log Z + E_0(N) \tau$ and variance $\sigma^2 = \frac{40}{9\pi} E_0(N)\tau$, where $E_0(N)$ is the free gas ground state energy for $N$ noninteracting fermions ($N/2$ fermions of each species) and $\log Z$ is the corresponding overlap between the ground state wave function and source and sink wave functions.
This in turn implies that the correlator distribution is log-normally distributed in mean field limit.
We may use the Berry-Esseen theorem along with our mean field result for the correlator distribution to estimate the minimal number of configurations required for a given value of $N$ and $\tau$.
The results is $N_{conf} \gg e^{3 \frac{40}{9\pi} E_0(N)\tau}$, scaling exponentially in the time and free gas energy.
Applying this result to the case $N=4$, we one finds that $N_{conf} \gg  3K$ configurations are required for $\tau=24$, $N_{conf} \gg 175K$ for $\tau=36$, and $N_{conf} \gg 10M$ for $\tau=48$.
The onset of an overlap problem around $\tau\sim32$ in \Fig{overlap_CDF} obtained from $N_{conf}=100K$ configurations is consistent with the prediction based on the application of Berry-Esseen theorem to our mean field calculation.

\begin{figure}
\includegraphics[width=\figwidth]{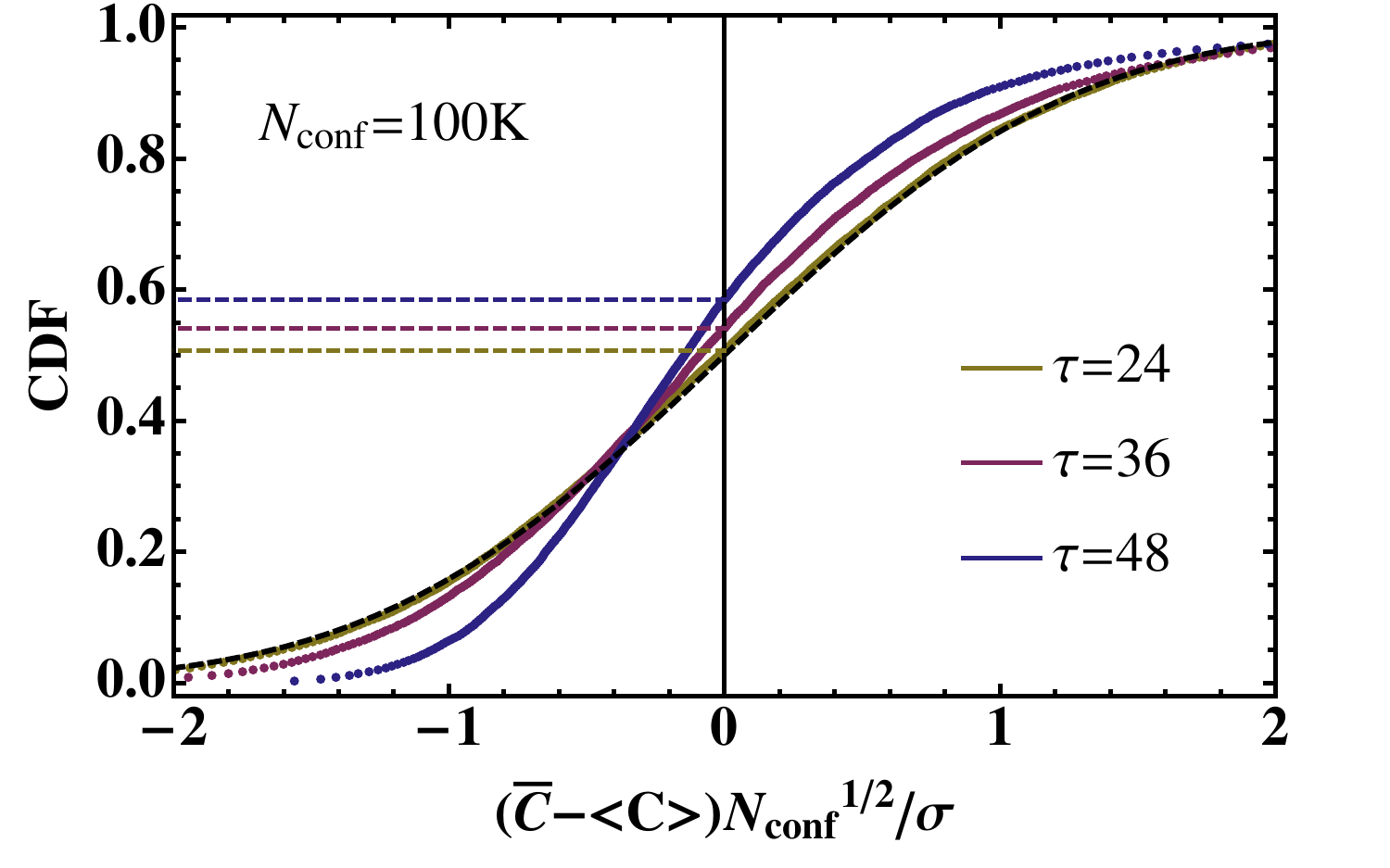}
\caption{%
\label{fig:overlap_CDF}%
Plot of the cumulative distribution function for $\left( \bar \calC - \langle \calC \rangle \right) \sqrt{N_{conf}}/\sigma$ for $\bar \calC$ estimated using $20K$ ensembles each of size  $N_{conf} = 100K$; $\langle \calC \rangle$ is estimated using $2 B$ configuration.
The dashed line is the cumulative distribution function for the standard normal distribution.
}
\end{figure}

The traditional technique for avoiding difficulties associated with distribution overlap problems is to use importance sampling in the Monte Carlo simulation.
In the case of large numbers of fermions, this might be achieved by reweighting the probability measure by either the correlation function at some late time, or some other carefully chosen weight factor.
In the former case, one might use the product $\rho(\phi) \calC_\phi(\tau_0)$ for an arbitrary but large value of $\tau_0$ as a probability measure for the auxiliary fields, and then measure ensemble averages of the ratio $\calC_\phi(\tau)/\calC_\phi(\tau_0)$ to estimate the correlator at times $\tau$.
\footnote{Since effective masses depend only on the ratio $\calC(\tau+1)/\calC(\tau)$, the overall normalization of correlation functions  determined from an ensemble average of $1/\calC_\phi(\tau_0)$ using  $\rho(\phi) \calC_\phi(\tau_0)$ as a probability measure is unimportant.}
In taking such an approach, however, the ensembles generated are typically only suitable for estimating a specfic operator (e.g., a single correlator at a specific value of $N$) or a small class of operators, and are inappropriate for most others.
Consequently, the simulation cost is enhanced by the number of operators being measured in addition to the difficulty of performing unquenched simulations using a far more complicated effective action for the auxiliary field, which generally will involve the logarithm of a correlation function.
This may be likened to performing a simulation in the Grand Canonical ensemble, where a different simulation must be performed at each value of chemical potential to achieve estimates of the energy as a function of density.

A far more efficient approach proposed in \cite{Endres:2011jm} is to find a better estimator for $\calC(\tau)$ that is free from the distribution overlap problem rather than rely on importance sampling.
Provided $\calC_\phi(\tau)>0$ for every $\phi$, \footnote{For the case $N^\uparrow=N^\downarrow$, one can show explicitly that correlators of the type defined in \Eq{slater} and \Eq{slater2} are positive for every background field configuration.} a systematic method for extracting useful information from an undersampled log-normal-like distribution may be devised by considering the cumulant expansion:
\begin{eqnarray}
\log \calC_{N_\kappa}(\tau) \equiv \sum_{n=1}^{N_{\kappa}} \frac{ \kappa_n(\tau) }{n!}\ ,
\label{eq:cumulant_expansion}
\end{eqnarray}
where $\kappa_n(\tau)$ is the $n$-th cumulant of the distribution for $\log \calC_\phi(\tau)$, which is presumed to be nearly normally distributed.
In this expansion, systematic uncertainties associated with the truncation of the series at order $N_{\kappa}$ are traded for statistical uncertainties associated with including increasing numbers of cumulants which have been estimated from an ensemble of finite size.
For a perfect log-normally distributed $\calC_\phi(\tau)$, \Eq{cumulant_expansion} is exact at $N_{\kappa}=2$, since all higher order cumulants vanish.
In practice, if the correlator distribution is not log-normal, deviations in the distribution would be quantified by the non-zero contributions to \Eq{cumulant_expansion} from $\kappa_n$ with $n>2$.
Such contributions--one would hope--are relatively small, allowing one to reliably obtain an estimate for $\log \calC(\tau)$ based on estimates of $\kappa_n$.

The generalized effective mass associated with each partial sum in \Eq{cumulant_expansion} may be expressed as:
\begin{eqnarray}
m_{eff}^{(N_\kappa)}(\tau) = \frac{1}{\Delta \tau} \sum_{n=1}^{N_\kappa} \frac{ 1}{n!} \left[  \kappa_n(\tau) -  \kappa_n(\tau + \Delta\tau)  \right]\ .
\label{eq:generalized_cumulant_effm}
\end{eqnarray}
By studying \Eq{generalized_cumulant_effm} as a function of $N_{\kappa}$, one may determine the ideal value $N^*_{\kappa}$ for which the statistical uncertainties and truncation errors become comparable.
Such an $N^*_{\kappa}$ then defines a best estimate value for the effective mass at a given time $\tau$.
Alternatively, we may define an energy $E_{N_\kappa} = \lim_{\tau\to\infty} m_{eff}^{(N_\kappa)}(\tau)$\footnote{Although we have not proved the convergence of $m_{eff}^{(N_\kappa)}(\tau)$ as a function of $\tau$, all of of our numerical evidence suggests that this quantity tends to a constant at late times.} and study its convergence as a function of $N_\kappa$.
In all of our studies, we use the latter approach.

Finally we comment on the applicability of the cumulant method to odd numbers of fermions.
In the case $N^\uparrow=N^\downarrow+1$, analysis using the cumulant expansion breaks down, since negative correlators $\calC_\phi(\tau)$ may exist.
For large numbers of fermions, we find that the fraction of negative correlators in a given ensemble is typically less than a few percent, however.
Furthermore, unlike the positive part of the distribution, the negative part exhibits no long tail at large $\tau$.
This suggests that the positive and negative parts of the distribution may be treated not only independently, but also differently: for the positive portion one may use the cumulant expansion technique, and for the negative portion a standard ensemble average, and the results may then be combined.
Although we do not consider odd numbers of fermions in this paper, we believe these considerations will be of importance in future studies of the pairing gap, which requires accurate estimates of the energy for both even and odd $N$.

\section{Simulation details}
\label{sec:simulation_details}

The lattice theory described in \Sec{lattice_construction} has been implemented on a number of clusters and massively parallel architectures.
As a result of the low memory footprint, which scales like $\calO(N\times L^3)$, and extremely fast character of the algorithm \cite{Endres:2010sq}, we use an embarrassingly parallel implementation: multiple streams are farmed out to many different cores and random generators associated with each core are seeded independently of each other.
Simulations were performed in double precision, and random numbers were generated using L\"uscher's Ranlux pseudo-random number generator \cite{Luscher:1993dy}.
Preliminary studies have shown no statistical advantage to using Gaussian auxiliary fields over $Z_2$ noise, and therefore all of our studies have been performed using the latter. 

Due to the extremely fast nature of our algorithm, it was necessary to perform ensemble-averages of multi-fermion correlator data on-line in order to eliminate bottlenecks associated with file I/O and to also reduce storage requirements.
Data was therefore ensemble averaged into $N_\calB$ blocks of size $N_{conf}/N_\calB$, where $N_{conf}$ is the total number of configurations generated.
$N_\calB$ was chosen small enough to avoid the I/O and storage issues, but also large enough to maintain adequate control over the statistical errors in our analysis.

We have checked our algorithm and implementation by comparing numerical predictions of the energies for several exactly soluble systems with their known solutions.
The four-fermion interaction, for instance, was checked using a high precision measurement ($N_{conf}=4.7B$ configurations) of the lowest and first excited state energies for two unitary fermions of mass $M=5$ in a finite box of size $L=8$.
The simulation was performed using $N_{\calO}=4$, with couplings $C_0=0.487259$, $C_2=0.298043, C_4=-0.211675$ and $C_6=0.0405311$.
The temporal extent of the lattice was chosen to be $T=64$, which is approximately three times larger than the predicted inverse energy difference in the ground and first excited states $\lambda_1^*$ and $\lambda_2^*$, given by L\"uscher's formula.
The lowest two measured energies, $\lambda_1$ and $\lambda_2$, were then determined by fitting the time-dependence of the effective mass (\Eq{generalized_effm} with $\Delta\tau=1$) to a constant plus exponential form.
The effective mass was fit over an interval $[\tau_{min}, \tau_{max}]$, where $\tau_{max}=T$ and $\tau_{min}$ was varied until a plateau was achieved in the fit values of $\lambda_1$ and $\lambda_2$.
We found that the results for the ground state and first excited state agreed with the theoretical result determined by L\"uscher's formula to within errors of about 0.007\% and 8\% percent, respectively.

The implementation of the external potential was checked by measuring the ground and excited state energies of a single fermion confined to a harmonic trap as a function of $L/L_0$.
In this case, since there are no auxiliary fields present, this check may be regarded as a numerical calculation, rather than a simulation.
We chose $L_0=2, 2.5, 3, 3.5$ and $4$, and $\omega=0.005$ in order to ensure negligible temporal and spatial discretization errors.
Sources with appropriate symmetry properties were constructed in order to extract the lowest seven energies of the SHO.
In all calculations, the temporal extant was chosen such that $T\gg 1/\omega$ in order to eliminate contamination from higher energy states in the single fermion correlation functions.
Numerical calculations obtained from spatial lattices $L=8, 16, 32, 64$ are shown in \Fig{sho_finite_vol}, and show good agreement with the continuum theory.

\bibliography{unitary}

\end{document}